\documentclass[longauth]{aa}

\usepackage[varg]{txfonts}
\usepackage{enumerate}
\usepackage[bookmarks=false]{hyperref}
\usepackage{graphicx}
\usepackage{color}
\usepackage[dvipsnames]{xcolor}
\usepackage{natbib}
\usepackage{longtable}
\usepackage{caption}
\usepackage{amssymb}
\usepackage{overpic}
\usepackage{float}
\usepackage{morefloats}

\usepackage{rotating}

\hypersetup{
    colorlinks,
    linkcolor={BlueViolet},
    citecolor={BlueViolet},
    urlcolor={BlueViolet}
}

\graphicspath{{./figures_small/}}
\bibpunct{(}{)}{;}{a}{}{,}

\begin{document}

\title{CLASH-VLT: A Highly Precise Strong Lensing Model of the \newline Galaxy Cluster RXC~J2248.7$-$4431 (Abell S1063) and \newline Prospects for Cosmography}

\author{G.~B.~Caminha       \inst{\ref{unife}}\thanks{e-mail address: \href{mailto:gbcaminha@fe.infn.it}{gbcaminha@fe.infn.it}}                    \and
        C.~Grillo           \inst{\ref{dark}}                           \and
        P.~Rosati           \inst{\ref{unife}}                          \and
        I.~Balestra         \inst{\ref{obs_munich},\,\ref{inaftrieste}} \and
        W.~Karman           \inst{\ref{Kapteyn}}                        \and
        M.~Lombardi         \inst{\ref{unimilano}}                      \and
        A.~Mercurio         \inst{\ref{inafcapo}}                       \and
        M.~Nonino           \inst{\ref{inaftrieste}}                    \and
        P.~Tozzi            \inst{\ref{inafflorence}}                   \and
        A.~Zitrin           \inst{\ref{caltec},\,\ref{hf}}              \and
        A.~Biviano          \inst{\ref{inaftrieste}}                    \and
        M.~Girardi          \inst{\ref{inaftrieste},\,\ref{unitrieste}} \and
        A.~M.~Koekemoer     \inst{\ref{stsi}}                           \and
        P.~Melchior         \inst{\ref{oh_1},\,\ref{oh_2}}              \and
        M.~Meneghetti       \inst{\ref{inafbologna},\,\ref{infnbologna}}\and
        E.~Munari           \inst{\ref{inaftrieste},\,\ref{unitrieste}} \and
        S.~H.~Suyu          \inst{\ref{sinica}}                         \and
        K.~Umetsu           \inst{\ref{sinica}}                         \and
        M.~Annunziatella    \inst{\ref{inaftrieste},\,\ref{unitrieste}} \and
        S.~Borgani          \inst{\ref{inaftrieste},\,\ref{unitrieste}} \and
        T.~Broadhurst       \inst{\ref{basque}}                         \and
        K.~I.~Caputi        \inst{\ref{Kapteyn}}                        \and
        D.~Coe              \inst{\ref{stsi}}                           \and
        C.~Delgado-Correal  \inst{\ref{unife}}                          \and
        S.~Ettori           \inst{\ref{inafbologna},\,\ref{infnbologna}}\and
        A.~Fritz            \inst{\ref{inafmilano}}                     \and
        B.~Frye             \inst{\ref{uniarizona}}                     \and
        R.~Gobat            \inst{\ref{labparis}}                       \and
        C.~Maier            \inst{\ref{univienna}}                      \and
        A.~Monna            \inst{\ref{obs_munich},\,\ref{mpi}}         \and
        M.~Postman          \inst{\ref{stsi}}                           \and
        B.~Sartoris         \inst{\ref{unitrieste}}                     \and
        S.~Seitz            \inst{\ref{obs_munich},\,\ref{mpi}}         \and
        E.~Vanzella         \inst{\ref{inafbologna}}                    \and
        B.~Ziegler          \inst{\ref{univienna}}                      
        }
        
\institute{
Dipartimento di Fisica e Scienze della Terra, Universit\`a degli Studi di Ferrara, Via Saragat 1, I-44122 Ferrara, Italy\label{unife}\and
Dark Cosmology Centre, Niels Bohr Institute, University of Copenhagen, Juliane Maries Vej 30, DK-2100 Copenhagen, Denmark\label{dark}\and
University Observatory Munich, Scheinerstrasse 1, 81679 Munich, Germany\label{obs_munich}\and
INAF - Osservatorio Astronomico di Trieste, via G. B. Tiepolo 11, I-34143, Trieste, Italy\label{inaftrieste}\and
Kapteyn Astronomical Institute, University of Groningen, Postbus 800, 9700 AV Groningen, The Netherlands \label{Kapteyn} \and
Dipartimento di Fisica, Universit\`a  degli Studi di Milano, via Celoria 16, I-20133 Milano, Italy\label{unimilano} \and
INAF - Osservatorio Astronomico di Capodimonte, Via Moiariello 16, I-80131 Napoli, Italy\label{inafcapo}\and
INAF - Osservatorio Astrofisico di Arcetri, Largo E. Fermi, I-50125, Firenze, Italy\label{inafflorence}\and
Cahill Center for Astronomy and Astrophysics, California Institute of Technology, MS 249-17, Pasadena, CA 91125, USA\label{caltec}\and
Hubble Fellow\label{hf} \and
Dipartimento di Fisica, Universit\`a  degli Studi di Trieste, via G. B. Tiepolo 11, I-34143 Trieste, Italy\label{unitrieste}\and
Space Telescope Science Institute, 3700 San Martin Drive, Baltimore, MD 21208, USA\label{stsi}\and
Center for Cosmology and Astro-Particle Physics, The Ohio State University, Columbus, OH 43210, USA\label{oh_1}\and
Department of Physics, The Ohio State University, Columbus, OH 43210, USA\label{oh_2}\and
INAF - Osservatorio Astronomico di Bologna, Via Ranzani 1, I- 40127 Bologna, Italy\label{inafbologna}\and
INFN - Sezione di Bologna, viale Berti Pichat 6/2, 40127 Bologna, Italy\label{infnbologna}\and
Institute of Astronomy as Astrophysics, Academia Sinica, P.O.Box 23-141, Taipei 10617, Taiwan \label{sinica}\and
Ikerbasque, Basque Foundation for Science, Alameda Urquijo, 36-5 Plaza Bizkaia, E-48011, Bilbao, Spain\label{basque}\and
INAF - Istituto di Astrofisica Spaziale e Fisica cosmica (IASF) Milano, via Bassini 15, I-20133 Milano, Italy\label{inafmilano}\and
Department of Astronomy/Steward Observatory, University of Arizona, 933 North Cherry Avenue, Tucson, AZ 85721, USA\label{uniarizona}\and
Laboratoire AIM-Paris-Saclay, CEA/DSM-CNRS-Universit\`e Paris Diderot, Irfu/Service d'Astrophysique, CEA Saclay, Orme des Merisiers, F-91191 Gif sur Yvette, France\label{labparis}\and
University of Vienna, Department of Astrophysics, T\"urkenschanzstr. 17, A-1180, Wien, Austria\label{univienna}\and
Max Planck Institute for Extraterrestrial Physics, Giessenbachstrasse, 85748 Garching, Germany\label{mpi}
}

\abstract
{We perform a comprehensive study of the total mass distribution
  of the galaxy cluster RXC~J2248.7$-$4431 ($z=0.348$) with a
  set of high-precision strong lensing models, which take advantage of
  extensive spectroscopic information on many multiply lensed
  systems. In the effort to understand and quantify inherent
  systematics in parametric strong lensing modelling, we explore a
  collection of 22 models where we use different samples of
  multiple image families, different parametrizations of the mass
  distribution, as well as cosmological parameters.
 As input
  information for the strong lensing models, we use the CLASH \emph{HST}
  imaging data and spectroscopic follow-up observations, carried out
  with the VIMOS and MUSE spectrographs on the VLT, to identify and
  characterize bona-fide multiple image families, and measure their
  redshifts down to $m_{F814W}\simeq 26$. A total of 16 background
  sources, over the redshift range $1.0-6.1$, are multiply lensed into
  47 images, 24 of which are spectroscopically confirmed and belong to 10
  individual sources. These also include a multiply lensed Lyman-$\alpha$ blob at $z=3.118$.
  The cluster total mass distribution and underlying cosmology
  in the models are optimized by matching the observed positions of the multiple
  images on the lens plane. MCMC techniques are used to quantify
  errors and covariances of the best-fit parameters.
 We show that with
  a careful selection of a large sample of \emph{ spectroscopically
    confirmed} multiple images, the best-fit model can reproduce their
  observed positions with a rms scatter of $0\arcsec.3$
  in a fixed flat $\Lambda$CDM cosmology, whereas the lack of
  spectroscopic information or the use of inaccurate photometric
  redshifts can lead to biases in the values of the model parameters. We find
  that the best-fit parametrisation for the cluster total mass
  distribution is composed of an elliptical pseudo-isothermal mass
  distribution with a significant core for the overall cluster halo
  and truncated pseudo-isothermal mass profiles for the cluster
  galaxies. We show that by adding bona-fide photometric-selected multiple images to the
  sample of spectroscopic families one can further, although slightly,
  improve constraints on the model parameters. In particular, we find that
  the degeneracy between the lens total mass distribution and the underlying geometry of the Universe,
  probed via angular diameter distance ratios between the lens and the sources and the observer and the sources,  
  can be partially removed.
  Allowing cosmological parameters to vary together with the cluster
  parameters, we find (at $68\%$ confidence level)
  $\Omega_m=0.25^{+0.13}_{-0.16}$ and $w=-1.07^{+0.16}_{-0.42}$ for a
  flat $\Lambda$CDM model, and $\Omega_m=0.31^{+0.12}_{-0.13}$ and
  $\Omega_\Lambda=0.38^{+0.38}_{-0.27}$ for a universe with $w=-1$ and
  free curvature.  Finally, using toy models mimicking the overall
  configuration of multiple images and cluster total mass distribution, we estimate
  the impact of the line of sight mass structure on the positional rms to
  be $0\arcsec.3\pm 0\arcsec.1$. We argue that the apparent sensitivity of 
  our lensing model to cosmography is due to the combination
  of the regular potential shape of RXC~J2248, a large number of bona-fide
  multiple images out to $z=6.1$, and a relatively modest
  presence of intervening large-scale structure, as revealed by our
  spectroscopic survey.}


\titlerunning{A Highly Precise Strong Lensing Model of the Cluster RXC~J2248 and Prospects for Cosmography}

\authorrunning{G.~B.~Caminha et al.}

\maketitle

\section{Introduction}
\label{sec:introduction}

Different cosmological probes agree on finding that the
total energy density of the present Universe is composed of ordinary
baryonic matter for less than 5\%, a poorly understood form of
non-relativistic matter, called dark matter, for approximately 20\%,
and an enigmatic constituent with negative pressure (i.e., with an
equation of state of the form $P = w \rho$, where $P$ and $\rho$ are
the pressure and the density, respectively, and $w$ is a negative
quantity), called dark energy, for more than 70\%.
The latter component can account for the current epoch of accelerating expansion of the Universe \citep[e.g.,][]{1998AJ....116.1009R, 1999ApJ...517..565P, 2002MNRAS.330L..29E, 2005ApJ...633..560E, 2011ApJS..192...18K, 2014A&A...571A..16P}.

The combination of both geometrical probes and statistics depending on
the cosmic growth of structure, e.g. the cluster mass
function or the matter power spectrum, has long been recognized as
critical in the effort to measure the global geometry of the Universe and test at the
same time theories of gravity. In this context, gravitational lensing is
a powerful astrophysical tool that can be used to investigate the
global structure of the Universe. The matter distribution at different
scales and different cosmic epochs can be probed with cosmic shear
techniques. Both weak and strong lensing methods are very effective
in measuring the mass distribution of dark matter halos on galaxy and
cluster scales. In addition, the observed positions and time-delays of
 multiple images of strongly lensed sources are sensitive to the
geometry of the Universe. In fact, these observables depend on the
angular diameter distances between the observer, lens, and source,
thereby one can in principle constrain cosmological parameters, as
a function of redshift, which describe the relative contributions to
the total matter-energy density \citep[see, for
example,][]{1992grle.book.....S}.

 On galaxy scales, detailed strong lensing models of background, multiply-imaged
quasars \citep[e.g.,][]{2010ApJ...711..201S, 2013ApJ...766...70S} and
sources at different redshifts \citep[e.g.,][]{2014MNRAS.443..969C},
and analyses of statistically significant samples of strong lenses
\citep[e.g.,][]{2008A&A...477..397G, 2010ApJ...708..750S} have shown
promising results that can  complement those of other
cosmographic probes and test their possible unknown systematic
effects.  On galaxy cluster scales, only recently it has been possible
to exploit the observed positions of spectroscopically confirmed
families of multiple images to obtain precise measurements of the
total mass distributions in the core of these lenses
\citep[e.g.,][]{2008A&A...481...65H, 2015ApJ...800...38G} and the
first constraint on cosmological parameters
\citep[e.g.,][]{2010Sci...329..924J, 2015ApJ...813...69M}. In the last few years, there has
been a significant improvement in the strong lensing modelling of galaxy
clusters, based on both \emph{Hubble Space Telescope} (hereafter \emph{HST}) multi-color imaging, to identify
and measure with high precision the angular positions of the multiple
images, and deep spectroscopy, to secure the redshifts of the
lensed sources and cluster members.

The \emph{HST} Multi-Cycle Treasury
Program Cluster Lensing And Supernova survey with Hubble \citep[CLASH;
P.I.: M. Postman;][]{2012ApJS..199...25P} and the Director
Discretionary Time program Hubble Frontier
Fields\footnote{\url{http://www.stsci.edu/hst/campaigns/frontier-fields/}}
(HFF; P.I.: J. Lotz) have led to the identification of tenths of 
hundreds of multiple images, 
deflected and distorted by the gravitational fields of massive galaxy
clusters. Their apparent positions have been measured with an accuracy lower than an arcsecond and morphologies well characterized. The
spectroscopic redshifts of many of these systems have been obtained as
part of a separate Very Large Telescope (VLT) spectroscopic follow-up
campaigns with the VIsible Multi-Object Spectrograph
\citep[VIMOS;][]{2003SPIE.4841.1670L} and the Multi Unit Spectroscopic
Explorer \citep[MUSE;][]{2010SPIE.7735E..08B}.  In particular, the ESO
Large Programme 186.A$-$0798 \citep[P.I.:
P. Rosati;][]{2014Msngr.158...48R}, the so-called CLASH-VLT project (hereafter just CLASH-VLT),
has provided an extensive spectroscopic data set on several of these
galaxy cluster lenses.

In this paper, we focus on the HFF cluster RXC~J2248.7$-$4431 (or
Abell S1063; hereafter RXC~J2248), which was part of the CLASH survey.
The clusters sample selection and the observations are presented in \citet{2008ApJS..174..117M} and
\citet{2009MNRAS.392.1509G}. We take advantage of our CLASH multiband
\emph{HST} data and extensive spectroscopic information that we have
collected on the cluster members and background, lensed sources in
this galaxy cluster with the VIMOS and MUSE instruments at the VLT
(see, \citealt{2013A&A...559L...9B},
\citealt{2015A&A...574A..11K}). Combining the \emph{HST} and VLT data sets, we
develop a highly accurate strong lensing model, which is able to
constrain the mass distribution of the lens in the inner region and at
the same time provides interesting constraints on the cosmological
parameters, which are ultimately limited by the intervening large
scale structure along the line of sight and the model assumptions on the
mass distribution.

When not specified, the computations were made considering a flat $\rm
\Lambda CDM$ cosmology with $\Omega_m = 0.3$ and $H_0 = 70\, {\rm
  km/s/Mpc}$.  In this cosmology, $1''$ corresponds to a physical scale
of $4.92\, {\rm kpc}$ at the cluster redshift ($z_{lens} = 0.348$).
In all images North is top and East left.

\section{RXC~J2248}
\label{sec:rxj2248}

RXC~J2248 is a rich galaxy cluster at $z_{lens}=0.348$ and was first
identified as Abell S1063 in \citet{1989ApJS...70....1A}. The high mass and redshift of RXC~J2248 make it a powerful gravitational lens creating
several strong lensing features, such as giant arcs, multiple image
families and distorted background sources. As detailed in this
article, a total of 16 multiple image families, ten of which
are spectroscopically confirmed, have been securely
identified to date over an area of 2 arcmin$^2$.
RXC~J2248 was one of the 25 clusters observed
within CLASH \citep{2012ApJS..199...25P} in 16 filters, from the UV through
the NIR, with the ACS and WFC3 cameras onboard \emph{HST}.
The full-depth, distortion-corrected HST mosaics in each filter were all produced using procedures similar to those described in \citet{2011ApJS..197...36K}, including additional processing beyond the default calibration pipelines, and astrometric alignment across all filters to a precision better than a few milliarcseconds.

\begin{figure*}[!ht]
  \centering
  \includegraphics[clip, width = 1.0\textwidth]{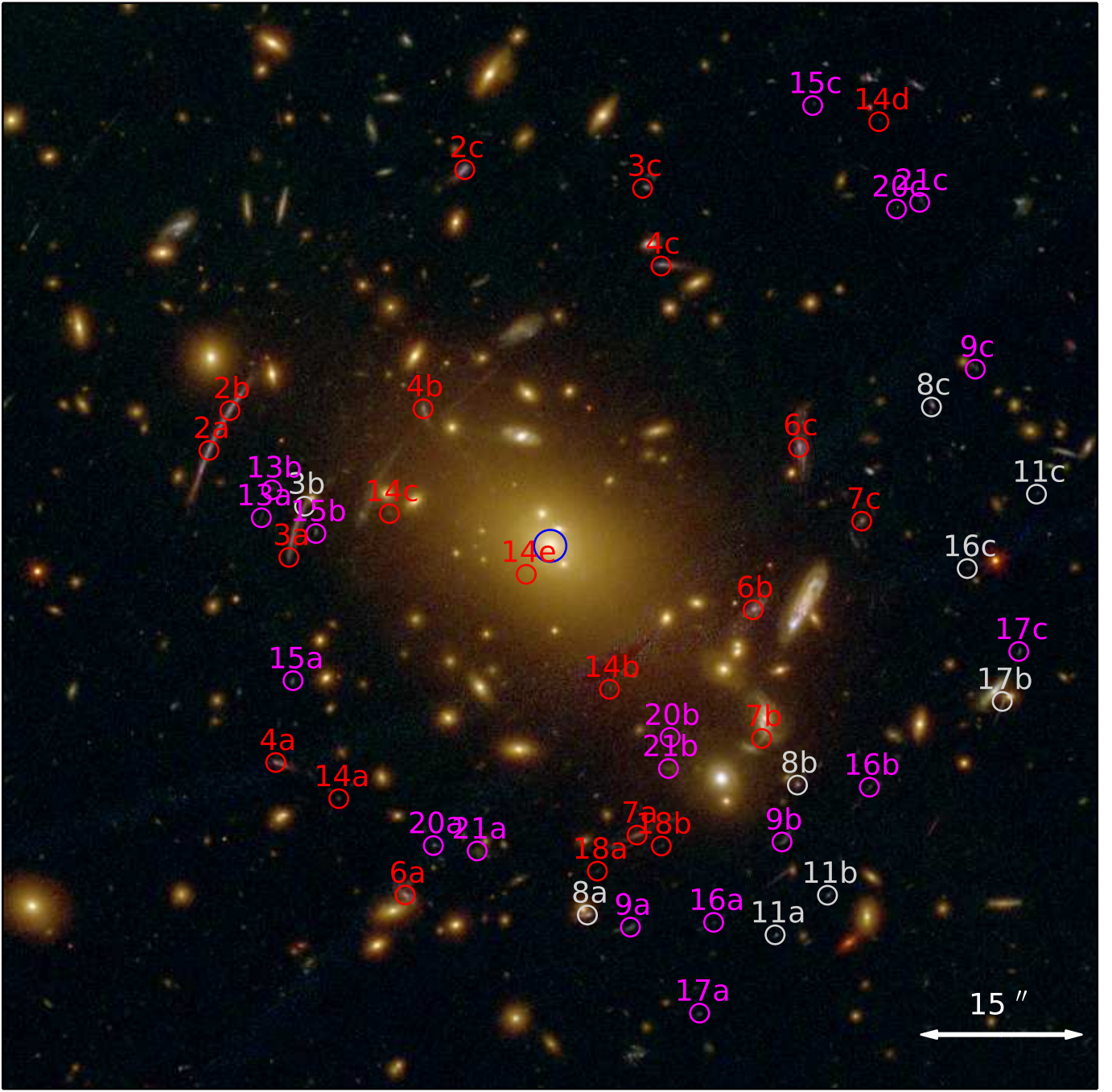}
  \caption{Colour composite image of RXC~J2248 obtained using the 16
    \emph{HST}/ACS and WFC3 filters. Spectroscopically confirmed multiple
    images are indicated in red, multiple images with no
    spectroscopic redshift in magenta. White circles indicate sources
    close to a cluster member, or possibly lensed by line of sight structures, or
    with no secure counter images. These last images are not used in the lens
    model. More information is provided in Table
    \ref{tab:families}. The blue circle shows the position of the
    BCG. We remark that the multiple image ID 14e is used only in the model F1-5th, see Table \ref{tab:summary_bf}.}
  \label{fig:multiple_image_systems}
\end{figure*}

In Figure
\ref{fig:multiple_image_systems}, we show a colour image of RXC~J2248
obtained from the combination of the CLASH ACS and WFC3
filters.  The red circles indicate the position of the multiple images
with spectroscopic redshift, the magenta circles the families with no
spectroscopic confirmation, while the white circles indicate sources close
to cluster members or possibly lensed by line of sight mass structures or,
not secure counter images.  We remark that the positions of the multiple
images are uniformly distributed around the cluster core, providing constraints on the overall cluster mass distribution.
Most of the families are composed of two or three multiple images, except for the
family at redshift 6.111 (ID 14), which is composed of five identified images
\citep[see][]{2013A&A...559L...9B, 2014MNRAS.438.1417M}.
After the submission of this paper, deeper \emph{HST} imaging from the HFF program became available, allowing us to detect the fifth, faint image (ID 14e) close to the BCG (see Figure \ref{fig:src_lens_planes}).
The spectroscopic confirmation of the redshift of this multiple image will be given in Karman et al. (in prep.).
Due to the late identification of image 14e we include it in only one strong lensing model, labeled as F1-5th in Table \ref{tab:summary_bf}.
We anticipate that the high redshift of this
source and its multiple image configuration, similar to an Einstein's
cross, will play an important role in constraining the cluster total mass distribution
and the relation between angular diamater distances and redshifts (for
more details see Section \ref{sec:results_sl}).

\begin{figure}[!ht]
  \centering
  \includegraphics[width = .5\textwidth]{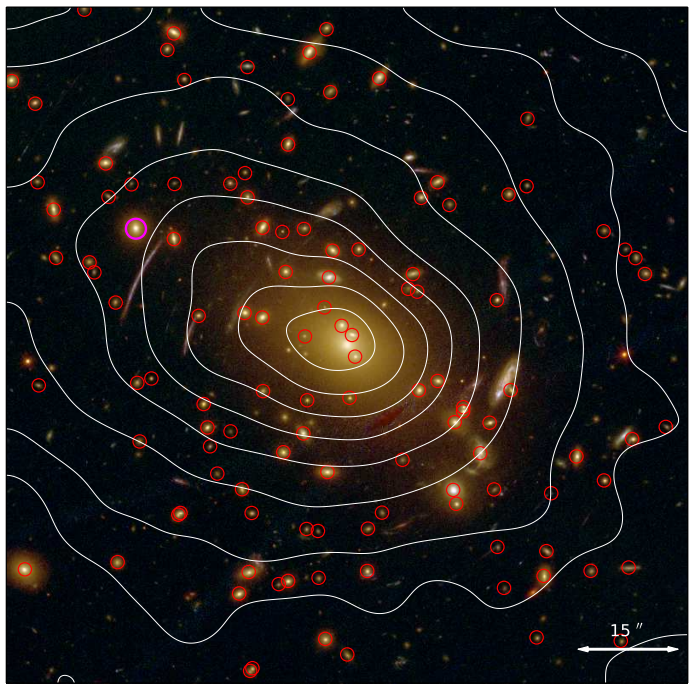}
  \caption{Colour composite image of RXC~J2248 overlaid with the
    \emph{Chandra} X-ray contours in white
    \citep{2012AJ....144...79G}. Red circles indicate the selected cluster members (see Section
    \ref{subsec:cluster_members}). The magenta circle shows the second brightest cluster
    member, used as the reference for the normalization of the
    mass-to-light ratio of the cluster members, i.e. $L_0$ in
    the Equation (\ref{eq:mass_to_light}).}
  \label{fig:members_xray}
\end{figure}

The total mass distribution of RXC~J2248 has been studied using different probes, such as X-ray emission \citep{2012AJ....144...79G}, strong \citep{2014MNRAS.438.1417M, 2014ApJ...797...48J, 2014MNRAS.444..268R, 2015ApJ...801...44Z} and weak lensing analyses \citep{2013MNRAS.432.1455G, 2014ApJ...795..163U, 2015ApJ...806....4M, 2015MNRAS.449.2219M} with generally good agreement between these different techniques.
\citet{2012AJ....144...79G} indicates that the galaxy cluster has undergone a recent off-axis merger, and \citet{2015MNRAS.449.2219M} find the cluster to be embedded in a filament with corresponding orientation. However, moderately deep X-ray Chandra observations show an elongated but regular shape, with no evidence of massive substructures in the inner region (see Figure \ref{fig:members_xray}).

Previous strong lensing analyses \citep{2014MNRAS.438.1417M,
  2014ApJ...797...48J, 2014MNRAS.444..268R} have shown that the
cluster total mass distribution of RXC~J2248 can be well represented by
a single elliptical dark matter halo, with the addition of the galaxy
cluster members. These studies have suggested that the dark matter
halo has a significantly flat core of $\approx
17$\arcsec.
The influence of the BCG during the cluster merging process \citep[e.g.,][]{2015MNRAS.451.1177L} and baryonic physics effects \citep[e.g.,][]{2015arXiv150703590T} can account for the formation of a core in the dark-matter density distribution of clusters and galaxies.
However, more simulations should be explored in order to better characterize these effects in objects with different formation histories and mass scales.

The regular shape and lens efficiency of RXC~J2248, in combination with
the high quality multi-colour imaging and extensive spectroscopy measurements, makes
it a very suitable system for testing high precision strong lensing
modelling of the mass distribution of galaxy clusters, with appreciable
leverage on the underlying geometry of the Universe.

Upcoming deeper observations of this cluster by the Grism Lens-Amplified Survey from Space
\citep[GLASS, GO-13459, P.I.: T. Treu,][]{2015ApJ...812..114T}, the HFF campaign and using MUSE, are expected to further increase the number of identified multiple image families and spectroscopic confirmations.

\begin{table*}[!ht]
\small
\centering
\caption{\label{tab:families} Multiple image systems}
\begin{tabular}{ccclcrccc}
\hline\hline
ID & RA & DEC & $z_{spec}$ & $z_{model}$ & $\mu$ & $mag_{814}^{obs}$ & $mag_{814}^{unlensed}$ \\
\hline
2a & 342.19559 & $-44.52839$ & 1.229$^{a,b,d}$ & ---             & $29.8_{-1.8}^{+3.1}$  & $22.59 \pm 0.01$ & $26.3_{-0.1}^{+0.1}$ \\
2b & 342.19483 & $-44.52735$ & 1.229$^{a,b,d}$ & ---             & $-23.5_{-1.6}^{+1.1}$ & $22.89 \pm 0.01$ & $26.3_{-0.1}^{+0.1}$ \\
2c & 342.18631 & $-44.52107$ & 1.229$^{a,b}$ & ---             & $5.4_{-0.2}^{+0.2}$   & $22.91 \pm 0.02$ & $24.7_{-0.1}^{+0.1}$ \\
\hline
3a     & 342.19269 & $-44.53118$ & 1.260$^{a,b}$ & ---         & $18.4_{-0.6}^{+0.8}$  & $24.56 \pm 0.04$ & $27.7_{-0.1}^{+0.1}$ \\
3b$^*$ & 342.19212 & $-44.52984$ & 1.260$^{a,b}$ & ---         & $^{\dagger}-21.2_{-2.6}^{+2.2}$ & $23.78 \pm 0.02$ & $27.1_{-0.1}^{+0.1}$ \\
3c     & 342.17986 & $-44.52156$ & 1.260$^{d}$ & ---                    & $3.3_{-0.1}^{+0.1}$   & $24.62 \pm 0.04$ & $25.9_{-0.1}^{+0.1}$ \\
\hline
4a & 342.19317 & $-44.53652$ & --- & ---                     & $3.5_{-0.1}^{+0.1}$   & --- & --- \\
4b & 342.18782 & $-44.52730$ & 1.398$^{a,b}$ & ---             & $-4.8_{-0.2}^{+0.2}$  & $22.65 \pm 0.04$ & $24.3_{-0.1}^{+0.1}$ \\
4c & 342.17919 & $-44.52358$ & 1.398$^{a,b,d}$ & ---             & $4.7_{-0.1}^{+0.1}$   & $23.81 \pm 0.03$ & $25.5_{-0.1}^{+0.1}$ \\
\hline
6a & 342.18847 & $-44.53998$ &1.428$^{a,b,e}$ & ---             & $3.9_{-0.2}^{+0.2}$   & $22.59 \pm 0.01$ & $24.1_{-0.1}^{+0.1}$ \\
6b & 342.17585 & $-44.53254$ &1.428$^{e}$ & ---             & $-8.1_{-0.5}^{+0.6}$  & $22.19 \pm 0.02$ & $24.5_{-0.1}^{+0.1}$ \\
6c & 342.17420 & $-44.52831$ &1.428$^{d,e}$ & ---             & $7.8_{-0.2}^{+0.3}$   & $22.51 \pm 0.01$ & $24.7_{-0.1}^{+0.1}$ \\
\hline
7a & 342.18006 & $-44.53842$ & 1.035$^{e}$ & ---             & $7.3_{-0.1}^{+0.2}$ & $23.68 \pm 0.04  $ & $25.8_{-0.1}^{+0.1}$   \\
7b & 342.17554 & $-44.53590$ & 1.035$^{e}$ & ---             & $-13.2_{-0.7}^{+0.9}$ & ---              &  ---\\
7c & 342.17191 & $-44.53023$ & 1.035$^{e}$ & ---             & $5.7_{-0.1}^{+0.2}$   & $24.05 \pm 0.04$ & $25.9_{-0.1}^{+0.1}$ \\
\hline
8a$^*$ & 342.18006 & $-44.53842$ & 1.837$^{a}$ & ---           & $^{\dagger}8.8_{-1.4}^{+1.8}$  & $23.73 \pm 0.01$ & $26.1_{-0.2}^{+0.2}$ \\
8b$^*$ & 342.17554 & $-44.53590$ & 1.837$^{a}$ & ---           & $^{\dagger}-6.8_{-0.4}^{+0.4}$  & $24.55 \pm 0.04$ & $26.6_{-0.1}^{+0.1}$ \\
8c$^*$ & 342.17191 & $-44.53023$ & --- & ---                    & $^{\dagger}4.1_{-0.2}^{+0.2}$   & $24.71 \pm 0.04$ & $26.2_{-0.1}^{+0.1}$ \\
\hline
9a & 342.18030 & $-44.54082$ & --- & $2.48_{-0.05}^{+0.05}$     & $8.3_{-0.2}^{+0.3}$   & $24.63 \pm 0.06$ & $26.9_{-0.1}^{+0.1}$ \\
9b & 342.17480 & $-44.53860$ & --- & ''                         & $-6.6_{-0.3}^{+0.2}$  & $25.01 \pm 0.08$ & $27.1_{-0.1}^{+0.1}$ \\
9c & 342.16779 & $-44.52627$ & --- & ''                         & $3.5_{-0.1}^{+0.1}$   & $25.79 \pm 0.11$ & $27.1_{-0.1}^{+0.1}$ \\
\hline
11a$^{*}$ & 342.17505 & $-44.54102$ & 3.116$^{e}$   & ---    & $^{\dagger}16.8_{-1.5}^{+1.6} $  & $25.85 \pm 0.08$ & $28.9_{-0.1}^{+0.1}$ \\
11b$^{*}$ & 342.17315 & $-44.53999$ & 3.116$^{a,b,e}$   & ---    & $^{\dagger}-17.3_{-1.9}^{+1.6}$ & $25.54 \pm 0.06$ & $28.6_{-0.1}^{+0.1}$ \\
11c$^{*}$ & 342.16557 & $-44.52953$ & ---      & ---           & $^{\dagger}4.1_{-0.2}^{+0.2}$    & $27.26 \pm 0.20$ & $28.8_{-0.2}^{+0.2}$ \\
\hline
13a & 342.19369 & $-44.53014$ & --- & $1.27_{-0.03}^{+0.03}$    & $32.5_{-2.7}^{+9.1}$  & $26.40 \pm 0.17$ & $30.2_{-0.2}^{+0.4}$ \\
13b & 342.19331 & $-44.52942$ & --- & ''                        & $-30.1_{-10.1}^{+1.4}$& ---              & --- \\
\hline
14a & 342.19088 & $-44.53747$ & 6.112$^{b,c,e}$ & ---                & $6.4_{-0.2}^{+0.3}$   & $25.74 \pm 0.08$ & $27.8_{-0.1}^{+0.1}$ \\
14b & 342.18106 & $-44.53462$ & 6.111$^{b,c,e}$ & ---                & $-7.8_{-0.8}^{+0.9}$  & $25.47 \pm 0.24$ & $27.7_{-0.3}^{+0.3}$ \\
14c & 342.18904 & $-44.53004$ & ---         & ---                & $-12.3_{-6.0}^{+2.1}$ & $25.11 \pm 0.07$ & $27.8_{-0.5}^{+0.2}$ \\
14d & 342.17129 & $-44.51982$ & 6.111$^{b,c}$& ---                       & $2.6_{-0.1}^{+0.1}$   & $27.85 \pm 0.60$ & $28.9_{-0.6}^{+0.6}$ \\
14e & 342.18408 & $-44.53162$ & ---& ---                       & ---   & --- & --- \\
\hline
15a & 342.19254 & $-44.53439$ & --- & $3.14_{-0.10}^{+0.09}$   & $10.9_{-0.3}^{+0.7}$  & $25.27 \pm 0.07$ & $27.9_{-0.1}^{+0.1}$ \\
15b & 342.19171 & $-44.53055$ & ---& ''                        & $-10.5_{-0.5}^{+0.5}$ & $25.56 \pm 0.08$ & $28.1_{-0.1}^{+0.1}$ \\
15c & 342.17369 & $-44.51940$ & ---& ''                        & $2.6_{-0.1}^{+0.1}$   & $26.94 \pm 0.16$ & $28.0_{-0.2}^{+0.2}$ \\
\hline
16a     & 342.17728 & $-44.54069$ & --- &$1.43_{-0.02}^{+0.02}$& $7.7_{-0.3}^{+0.2}$   & $26.72 \pm 0.30$ & $28.9_{-0.3}^{+0.3}$ \\
16b     & 342.17163 & $-44.53717$ & ---& ''                    & $-15.7_{-1.5}^{+1.0}$ & $26.16 \pm 0.21$ & $29.1_{-0.2}^{+0.2}$ \\
16c$^*$ & 342.16894 & $-44.53256$ & ---& ''                   & ---                       & $26.42 \pm 0.31$ & ---                     \\
\hline
17a     & 342.17779 & $-44.54306$ & --- &$2.39_{-0.06}^{+0.05}$& $5.4_{-0.2}^{+0.1}$   & $25.79 \pm 0.14$ & $27.6_{-0.2}^{+0.1}$ \\
17b$^*$ & 342.16681 & $-44.53493$ & ---& ''                    & ---                      & ---           & ---    \\
17c     & 342.16621 & $-44.53363$ & ---& ''                    & $12.0_{-0.5}^{+0.7}$  & $25.52 \pm 0.07$ & $28.2_{-0.1}^{+0.1}$ \\
\hline
18a & 342.18150 & $-44.53936$ & 4.113$^{e}$ & ---              & $30.0_{-2.6}^{+2.5}$  & $26.88 \pm 0.26$ & $30.6_{-0.3}^{+0.3}$ \\
18b & 342.17918 & $-44.53870$ & 4.113$^{e}$ & ---              & $-25.3_{-2.5}^{+1.7}$ & $27.37 \pm 0.23$ & $30.9_{-0.3}^{+0.2}$ \\
\hline
20a & 342.18745 & $-44.53869$ &$3.118^{a}$&$3.11_{-0.10}^{+0.11}$& $6.4_{-0.2}^{+0.3}$   & $25.40 \pm 0.07$ & $27.4_{-0.1}^{+0.1}$ \\
20b & 342.17886 & $-44.53587$ &$3.118^{a}$& ''                   & $-5.3_{-0.3}^{+0.3}$  & $25.94 \pm 0.07$ & $27.7_{-0.1}^{+0.1}$ \\
20c & 342.17065 & $-44.52209$ & ---   & ''                     & $2.8_{-0.1}^{+0.1}$   & $26.13 \pm 0.13$ & $27.2_{-0.2}^{+0.1}$ \\
\hline
21a & 342.18586 & $-44.53883$ & --- & $3.49_{-0.12}^{+0.13}$   & $8.8_{-0.3}^{+0.5}$      & $25.91 \pm 0.11$ & $28.2_{-0.1}^{+0.1}$ \\
21b & 342.17892 & $-44.53668$ & --- & ''                       & $-6.6_{-0.3}^{+0.3}$     & ---             & ---  \\
21c & 342.16981 & $-44.52192$ & --- & ''                       & $2.7_{-0.1}^{+0.1}$      & $25.91 \pm 0.13$ & $27.0_{-0.1}^{+0.1}$ \\
\hline \hline
\end{tabular}
\tablefoot{
  Properties of the multiple images. The coordinates correspond to the luminosity peak used in the strong lensing models. Best-fit redshifts and magnifications with 68\% confidence level errors (columns $z_{model}$ and $\mu$, respectively) are computed using the model F2 (see Section \ref{sec:results_sl}). Observed magnitudes in the F814W filter ($mag_{814}^{obs}$) are Kron magnitudes measured with \emph{SExtractor}. In the last column, the unlensed magnitudes $mag_{814}^{unlensed}$ are shown.\\
  \tablefoottext{*}{Multiple images close to a cluster member, or possibly lensed by LOS structures, or not secure counter image (not used in the model).}
  \tablefoottext{$\dagger$}{Magnification given by the model ID F2a. For family 11, we quote the values at the model-predicted positions (see text for details).}
  \tablefoottext{a}{This work.}
  \tablefoottext{b}{\citet{2013A&A...559L...9B}. Independent redshift measuremenst by:}
  \tablefoottext{c}{\citet{2013A&A...559L...1B};}
  \tablefoottext{d}{\citet{2014MNRAS.444..268R};}
  \tablefoottext{e}{\citet{2015A&A...574A..11K}.}
}
\end{table*}

\subsection{VIMOS observations and data reduction}
\label{vimos_obs}

As part of the CLASH-VLT Large program, the cluster RXC~J2248 was observed
with the VIMOS spectrograph between June 2013 and May 2015.
The VIMOS slit-masks were designed in sets of four pointings with one of the
quadrants centered on the cluster core and the other three
alternatively displaced in the four directions (NE, NW, SE, SW). A
total of 16 masks were observed, 12 with the low-resolution (LR) blue
grism and 4 with the intermediate resolution (MR) grism; each mask was
observed for either $3 \times 20$ or $3\times15$ minutes (15 hours
exposure time in total).  Therefore, the final integration times for
arcs and other background galaxies varied between 45 minutes and 4
hours.  A summary of our VIMOS observations is presented in Table
\ref{logobs}.  We used $1''$-slits.  The LR-blue grism has a spectral
resolution of approximately $28 \AA$ and a wavelength coverage of
$3700-6700 \AA$, while the MR grism has a spectral resolution of
approximately $13 \AA$ and it covers the wavelength range $4800-10000
\AA$.

\begin{table}
\small
\caption{Log of VIMOS observations of the Frontier Fields cluster 
RXC~J2248, taken as part of the CLASH-VLT spectroscopic campaign.}
\label{logobs}
\begin{center}
\begin{tabular}{l c c}
\hline
\hline
Mask ID               & Date      & Exp. Time (s) \\
(1)                   &  (2)      & (3)  \\
\hline
Low-resolution masks               &                   &                        \\
MOS\_R2248\_LRB\_1\_M1     & Jun 2013  & $3\times1200$ \\
MOS\_R2248\_LRB\_2\_M1     & Jun 2013  & $3\times1200$ \\
MOS\_R2248\_LRB\_3\_M1     & Jul 2013  & $3\times1200$ \\
MOS\_R2248\_LRB\_4\_M1     & Jul 2013  & $3\times1200$ \\

MOS\_R2248\_LRB\_1\_M2     & Oct 2013  & $3\times900$ \\
MOS\_R2248\_LRB\_2\_M2     & Oct 2013  & $3\times900$ \\
MOS\_R2248\_LRB\_3\_M2     & Oct 2013  & $3\times900$ \\
MOS\_R2248\_LRB\_4\_M2     & Oct 2013  & $3\times900$ \\

MOS\_R2248\_LRB\_1\_M3     & Aug 2014  & $4\times900$ \\
MOS\_R2248\_LRB\_4\_M3     & Aug 2014  & $4\times900$ \\
MOS\_R2248\_LRB\_3\_M4     & Sep 2014  & $4\times900$ \\
MOS\_R2248\_LRB\_2\_M4     & May 2015  & $4\times900$ \\

Medium-resolution masks         &                   &                        \\
MOS\_R22248\_MR\_1\_M1      & Jul 2013  & $3\times1200$ \\
MOS\_R22248\_MR\_2\_M1      & Jul 2013  & $3\times1200$ \\
MOS\_R22248\_MR\_4\_M1      & Jul 2013  & $3\times1200$ \\
MOS\_R22248\_MR\_3\_M1      & Jul 2013  & $3\times1200$ \\
\hline
\end{tabular}
\\
\end{center}
\textbf{Notes.} Columns list the following information: (1) mask identification number, 
(2) date of the observations, and (3) number of exposures and integration time of single
exposures.
\end{table}

We define four quality classes by assigning a quality flag (QF) to
each redshift measurement, which indicates the reliability of a
redshift estimate. The four quality classes are defined as follows:
``secure'' (QF=3), ``likely'' (QF=2), ``insecure'' (QF=1), and based
on single emission line (QF=9).  Duplicate observations of hundreds of
sources across the whole survey allow us to quantify the reliability
of each quality class as follows: redshifts with QF=3 are correct with
a probability of $>99.99$\%, QF=9 with $\sim92$\% probability, QF=2
with $\sim75$\% probability, and QF=1 with $<40$\% probability.  In
this paper we will only consider redshifts with QF=3, 2, or 9.  A
total of 3734 reliable redshifts were measured over a field $\sim25$
arcmin across, where 1184 are cluster members and 2425 are field
galaxies (125 are stars).  For a complete description of the data
acquisition and reduction see Balestra et al. (in prep.) and Rosati et al. (in prep.).

In Figure \ref{fig:1a_spec}, we show the spectra of the multiply imaged
sources. On the left, we show the \emph{HST} cutout with the position of the $1\arcsec$-wide
slit of VIMOS. On the right, the one and two-dimensional spectra with the estimated
redshifts and quality flags are shown. All spectra
present clear emission lines, ensuring reliable redshifts for most of the
measurements, i.e. QF$=3$, with the exception of the low S/N spectrum
of image 8a (QF$=9$), however, its redshift of $z=1.837$ is confirmed
by MUSE observations (see Section \ref{subsec:muse_redshift}).

The positions and the redshift values ($z_{spec}$) of all multiple image families are given in Table \ref{tab:families}, and of magnified sources that are not multiply lensed are given in Table \ref{tab:high_mag_src}.
We also quote the Kron
observed magnitudes of each source measured with \emph{SExtractor}
\citep{1996A&AS..117..393B} in the F814W filter.
We use our strong lensing model F2 (see Table \ref{tab:summary_bf} and Section \ref{sec:results_sl}) to
compute the best-fit value of the redshift ($z_{model}$) of all multiply imaged sources not spectroscopically confirmed and to compute the magnification factors ($\mu$).
The value of $\mu$ is computed for a point-like object at the position of the images.
Since the model F2 is not suitable to compute the magnification of the families 8, 11 and multiple image 3b, we quote the magnifications values given by the model that includes all spectroscopically  confirmed multiple images (model ID F1a, see Table \ref{tab:summary_bf}).
For family 11, which presents two multiple images very close to the tangential critical line and the highest offset between the observed and model-predicted positions ($\Delta_i \approx 1\arcsec.4$), we quote the magnification values at the model-predicted positions.
This values are less sensitive to the systematic effects affecting this family and discussed in Section \ref{subsec:multple_image_selection}.

Magnitudes corrected by the magnification factor ($mag_{814}^{unlensed}$) are also estimated.
The apparent disagreement in the values of $mag_{814}^{unlensed}$ of some multiple image families can be due to the evaluation of $\mu$, which is computed in a point and not integrated over the extended image, and the difficulties in the photometric measurement of highly extended images, such as the multiple images of the families 2 and 3.
The high magnification efficiency of RXC~J2248 allows us to probe spectroscopically the very faint end of the galaxy luminosity function at high redshift, with intrinsic unlensed magnitudes extending down to  $M_{1600} \approx -15$, i.e. approximately 5 magnitudes bellow $M^*$ \citep{2011ApJ...737...90B}.

\begin{table*}
\centering
\caption{\label{tab:high_mag_src} Magnified but not multiply imaged sources}
\begin{tabular}{ccccccc}
\hline\hline
ID & RA & DEC & $z_{spec}$ & $\mu$ & $mag_{814}^{obs}$ & $mag_{814}^{unlensed}$ \\
\hline
B1  & 342.17404 & -44.53247 & 0.607 & $3.7_{-0.1}^{+0.1}$ &  $18.92\pm0.01$ & $20.3_{-0.1}^{+0.1}$ \\
B2  & 342.18442 & -44.53961 & 0.652 & $2.1_{-0.1}^{+0.1}$ &  $24.05\pm0.04$ & $24.9_{-0.1}^{+0.1}$ \\
B3  & 342.17925 & -44.54219 & 0.698 & $2.2_{-0.1}^{+0.1}$ &  $27.04\pm0.20$ & $27.9_{-0.2}^{+0.2}$ \\
B4  & 342.18402 & -44.52522 & 0.730 & $4.9_{-0.3}^{+0.1}$ &  $24.35\pm0.02$ & $26.1_{-0.1}^{+0.1}$ \\
B5  & 342.15632 & -44.54563 & 0.941 & $2.1_{-0.1}^{+0.1}$ &  $21.97\pm0.01$ & $22.8_{-0.1}^{+0.1}$ \\
B6  & 342.17554 & -44.54559 & 1.269 & $2.7_{-0.1}^{+0.1}$ &  $23.60\pm0.03$ & $24.7_{-0.1}^{+0.1}$ \\
B7  & 342.17241 & -44.54121 & 1.270 & $6.3_{-0.4}^{+0.2}$ &  $22.88\pm0.01$ & $24.9_{-0.1}^{+0.1}$ \\
B8  & 342.19929 & -44.51339 & 1.428 & $2.6_{-0.2}^{+0.1}$ &  $22.28\pm0.01$ & $23.3_{-0.1}^{+0.1}$ \\
B9  & 342.15719 & -44.54515 & 1.437 & $2.7_{-0.2}^{+0.1}$ &  $22.90\pm0.01$ & $24.0_{-0.1}^{+0.1}$ \\
B10 & 342.17695 & -44.54633 & 1.477 & $2.6_{-0.1}^{+0.1}$ &  $25.29\pm0.09$ & $26.3_{-0.1}^{+0.1}$ \\
B11 & 342.16109 & -44.53823 & 2.578 & $5.8_{-0.3}^{+0.2}$ &  $25.24\pm0.07$ & $27.2_{-0.1}^{+0.1}$ \\
B12 & 342.21712 & -44.52960 & 2.641 & $2.0_{-0.1}^{+0.1}$ &  $24.46\pm0.03$ & $25.2_{-0.1}^{+0.1}$ \\
B13 & 342.16214 & -44.53822 & 3.117 & $8.2_{-0.6}^{+0.4}$ &  $24.92\pm0.08$ & $27.2_{-0.1}^{+0.1}$ \\
B14 & 342.17392 & -44.54124 & 3.228 &$29.4_{-5.4}^{+13.6}$&  $25.64\pm0.10$ & $29.3_{-0.2}^{+0.5}$ \\
B15 & 342.16260 & -44.54296 & 3.240 & $6.2_{-0.3}^{+0.3}$ &  $23.80\pm0.02$ & $25.8_{-0.1}^{+0.1}$ \\
B16 & 342.20533 & -44.51552 & 3.542 & $4.2_{-0.2}^{+0.1}$ &  $23.84\pm0.02$ & $25.4_{-0.1}^{+0.1}$ \\
\hline
\end{tabular}

\tablefoot{List of significantly magnified sources ($\mu > 2$) with reliable redshift measurements (QF $>1$). The magnifications and errors are computed from 20000 random realizations of the model with fixed cosmology (ID F2, see Table \ref{tab:summary_bf}).\\
 }

\end{table*}

\begin{figure*}[!ht]
  \centering
  \begin{picture}(100,220)
    \put(0,0){\includegraphics[trim = 0px -50px 0px 0px, clip=true, width = .19\textwidth]{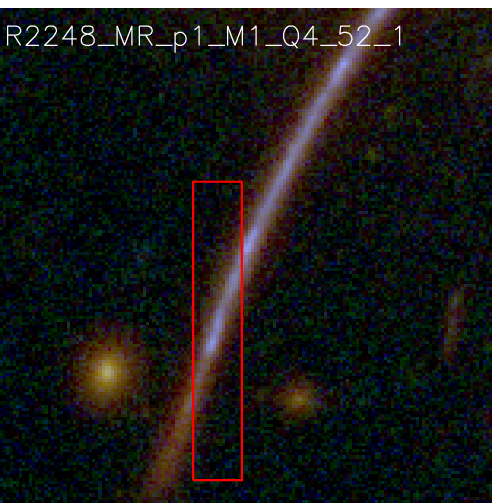}}
    \put(73, 40){{\color{white}\bf \large 2a/2b}}
  \end{picture}
  \begin{picture}(300,220)
    \put(0,0){\includegraphics[trim = 0px 0px 0px 25px, clip=true, width = .6\textwidth]{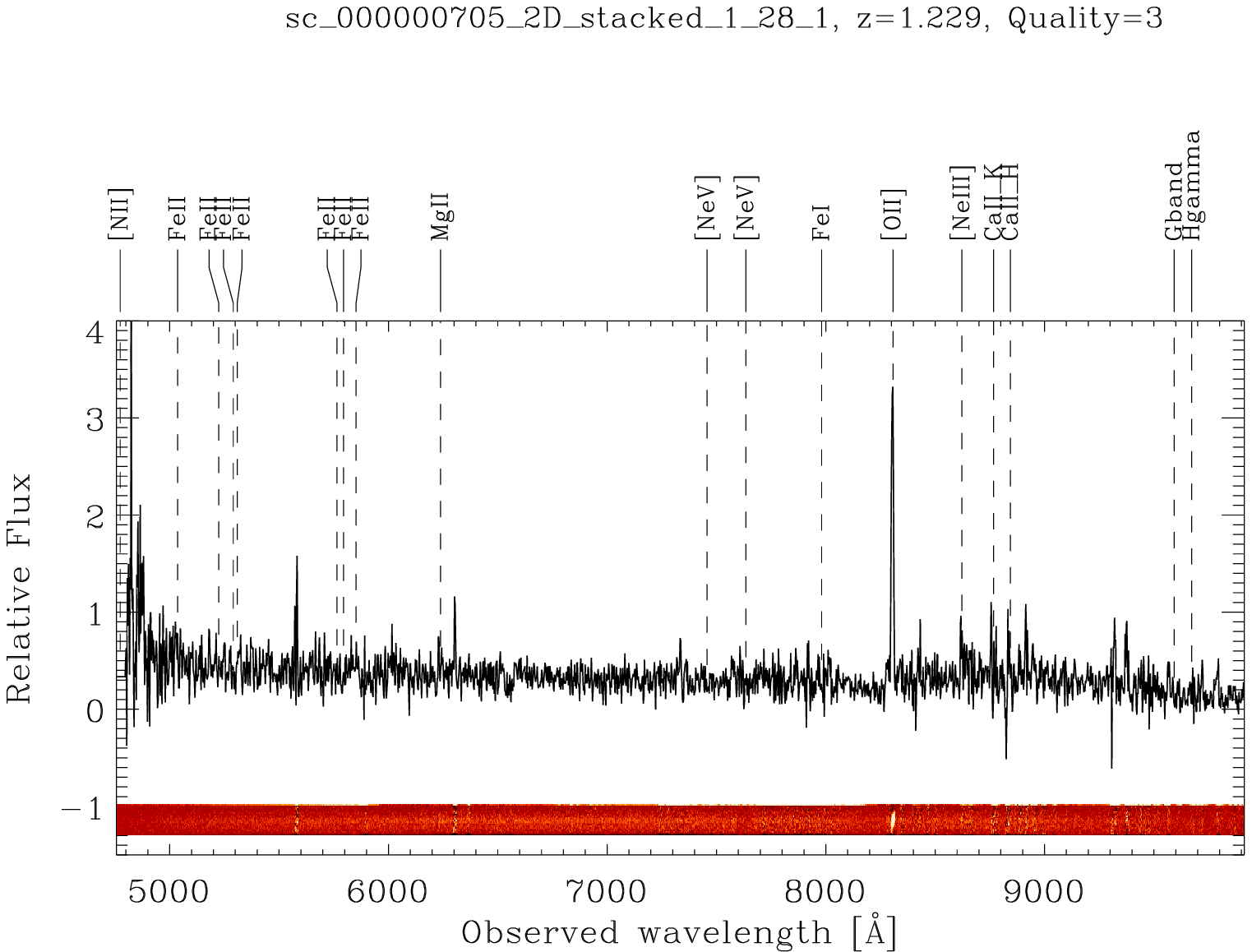}}
    \put(50, 40){{ \large z=1.229, QF=3}}
  \end{picture}

  \begin{picture}(100,220)
    \put(0,0){\includegraphics[trim = 0px -50px 0px 0px, clip=true, width = .19\textwidth]{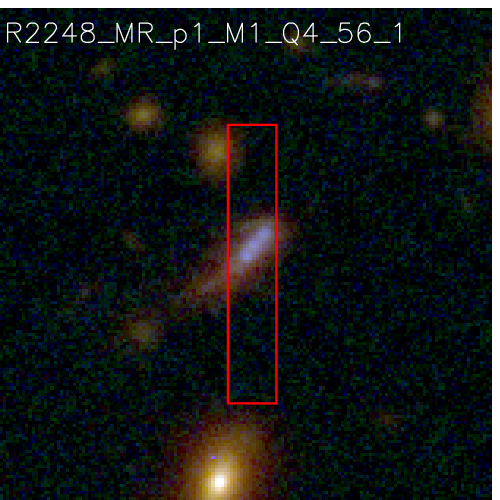}}
    \put(85, 40){{\color{white}\bf \large 2c}}
  \end{picture}
  \begin{picture}(300,220)
    \put(0,0){\includegraphics[trim = 0px 0px 0px 25px, clip=true, width = .6\textwidth]{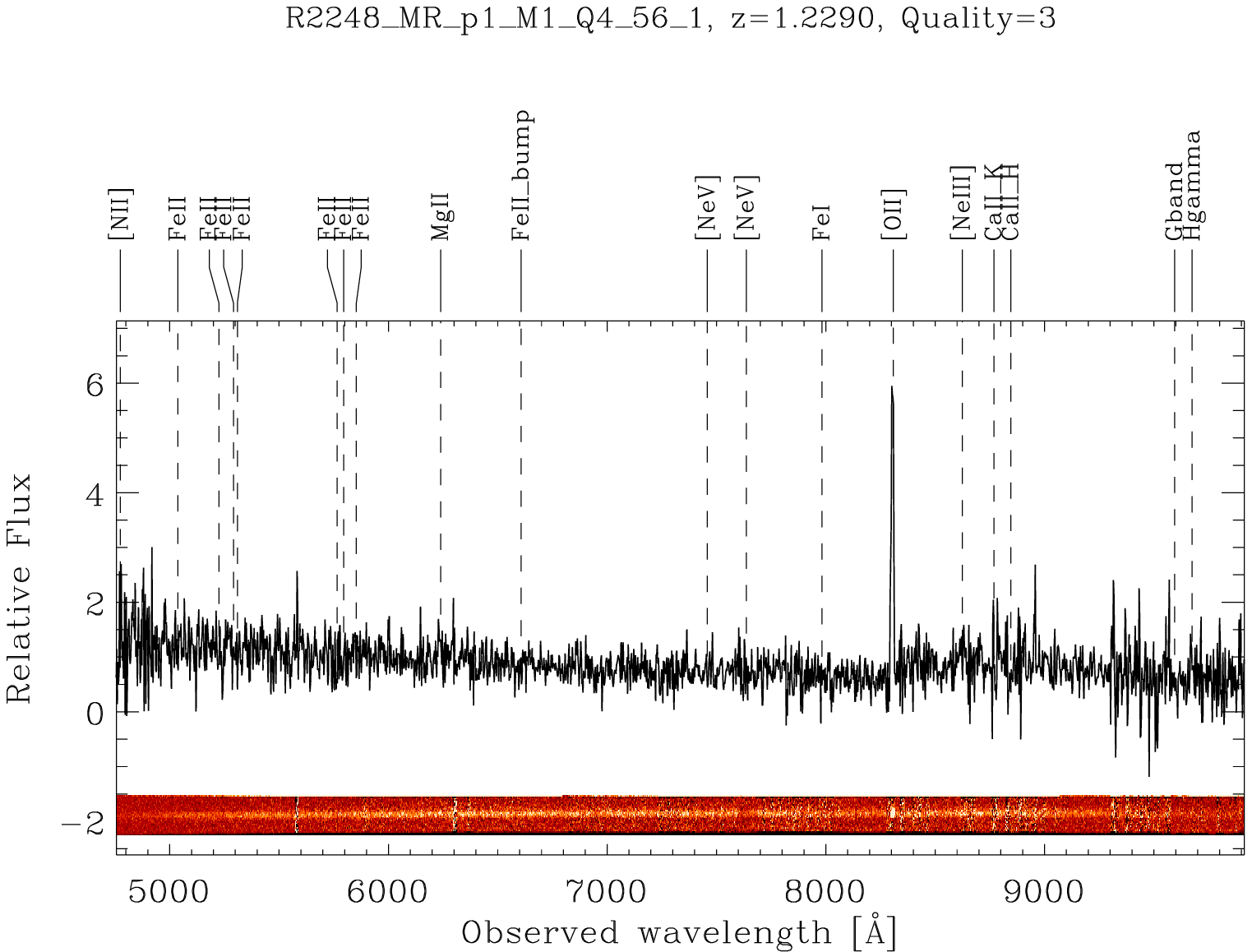}}
    \put(50, 40){{ \large z=1.229, QF=3}}
  \end{picture}

  \begin{picture}(100,220)
    \put(0,0){\includegraphics[trim = 0px -50px 0px 0px, clip=true, width = .19\textwidth]{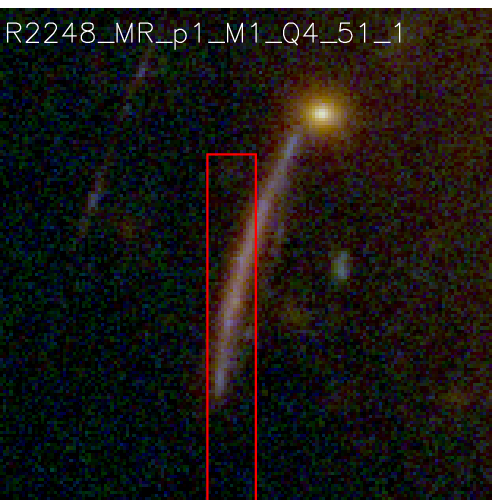}}
    \put(73, 40){{\color{white}\bf \large 3a/3b}}
  \end{picture}
  \begin{picture}(300,220)
    \put(0,0){\includegraphics[trim = 0px 0px 0px 25px, clip=true, width = .6\textwidth]{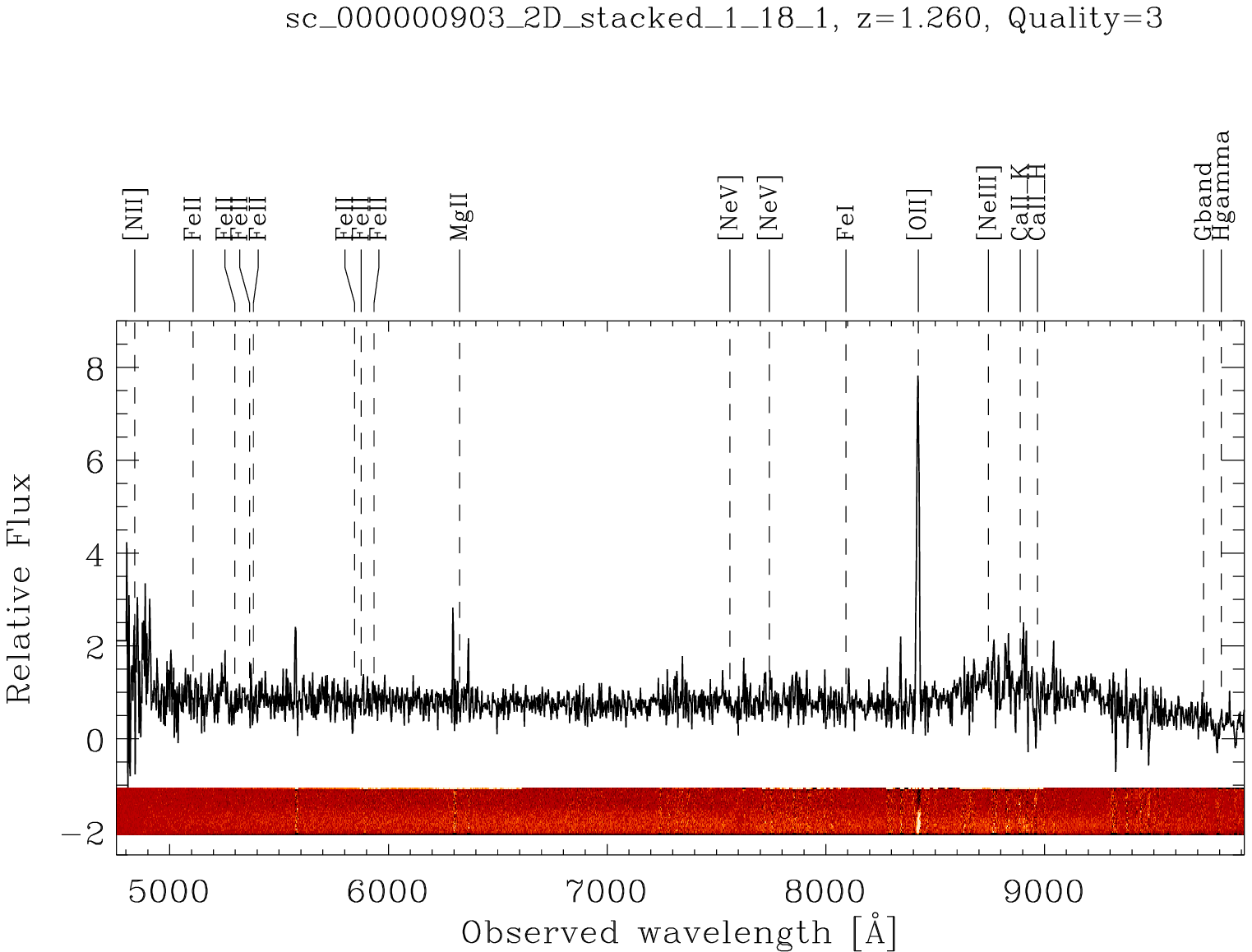}}
    \put(50, 40){{ \large z=1.260, QF=3}}
  \end{picture}

  \caption{VLT/VIMOS spectra of the multiple image
    systems. Left panels show \emph{HST} cutouts, 10\arcsec\ across with the position of the VIMOS \mbox{$1''$-wide} slits and the image ID from Table
    \ref{tab:families}. One- and
    two-dimensional spectra are shown on the right with measured redshifts and 
    quality flags, including typical emission and absorption lines.}
  \label{fig:1a_spec}
\end{figure*}

\begin{figure*}[!Ht]
  \centering
  \ContinuedFloat
  \begin{picture}(100,220)
    \put(0,0){\includegraphics[trim = 0px -50px 0px 0px, clip=true, width = .19\textwidth]{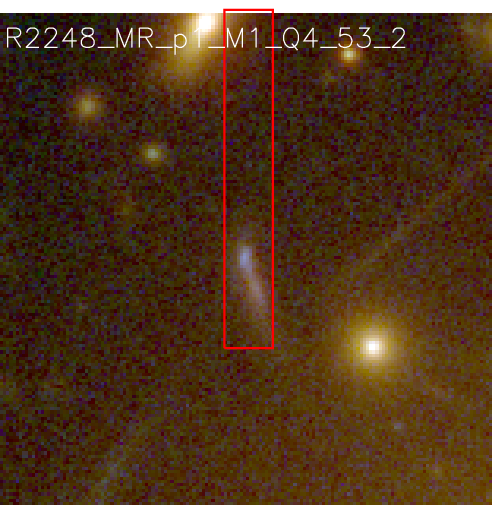}}
    \put(85, 40){{\color{white}\bf \large 4b}}
  \end{picture}
  \begin{picture}(300,220)
    \put(0,0){\includegraphics[trim = 0px 0px 0px 25px, clip=true, width = .6\textwidth]{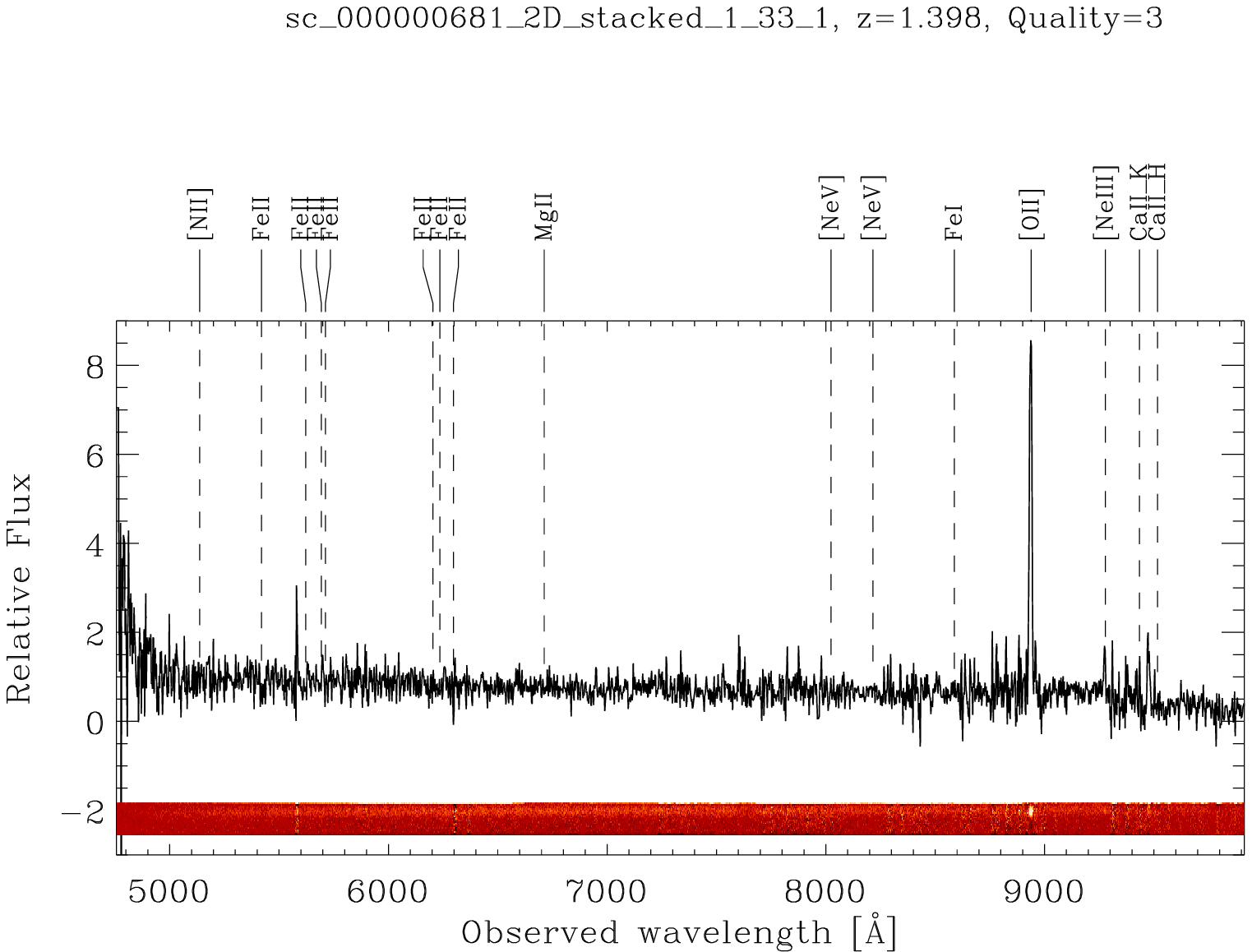}}
    \put(50, 40){{ \large z=1.398, QF=3}}
  \end{picture}

  \begin{picture}(100,220)
    \put(0,0){\includegraphics[trim = 0px -50px 0px 0px, clip=true, width = .19\textwidth]{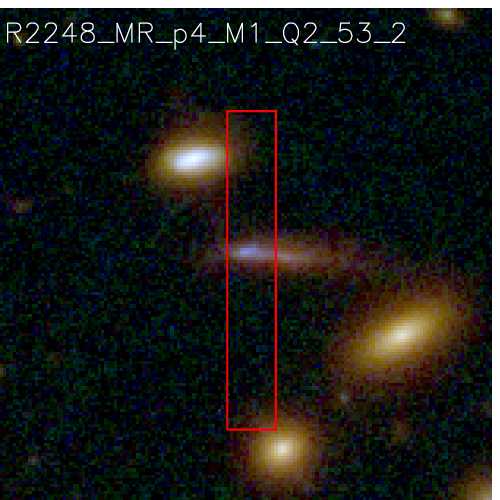}}
    \put(85, 40){{\color{white}\bf \large 4c}}
  \end{picture}
  \begin{picture}(300,220)
    \put(0,0){\includegraphics[trim = 0px 0px 0px 25px, clip=true, width = .6\textwidth]{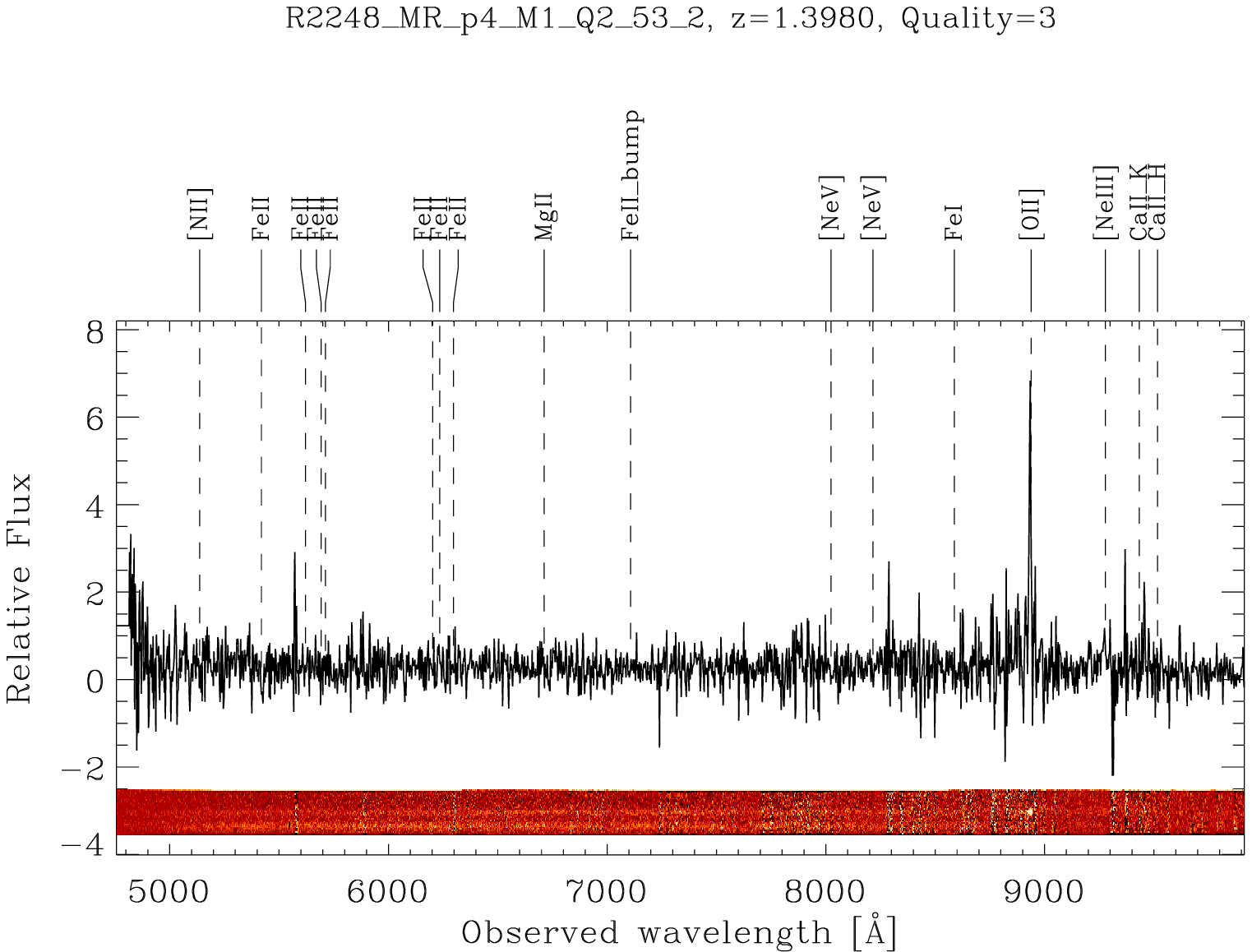}}
    \put(50, 40){{ \large z=1.398, QF=3}}
  \end{picture}

  \begin{picture}(100,220)
    \put(0,0){\includegraphics[trim = 0px -50px 0px 0px, clip=true, width = .19\textwidth]{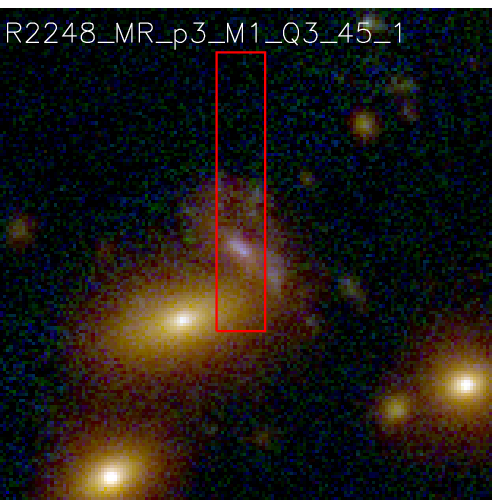}}
    \put(85, 40){{\color{white}\bf \large 6a}}
  \end{picture}
  \begin{picture}(300,220)
    \put(0,0){\includegraphics[trim = 0px 0px 0px 25px, clip=true, width = .6\textwidth]{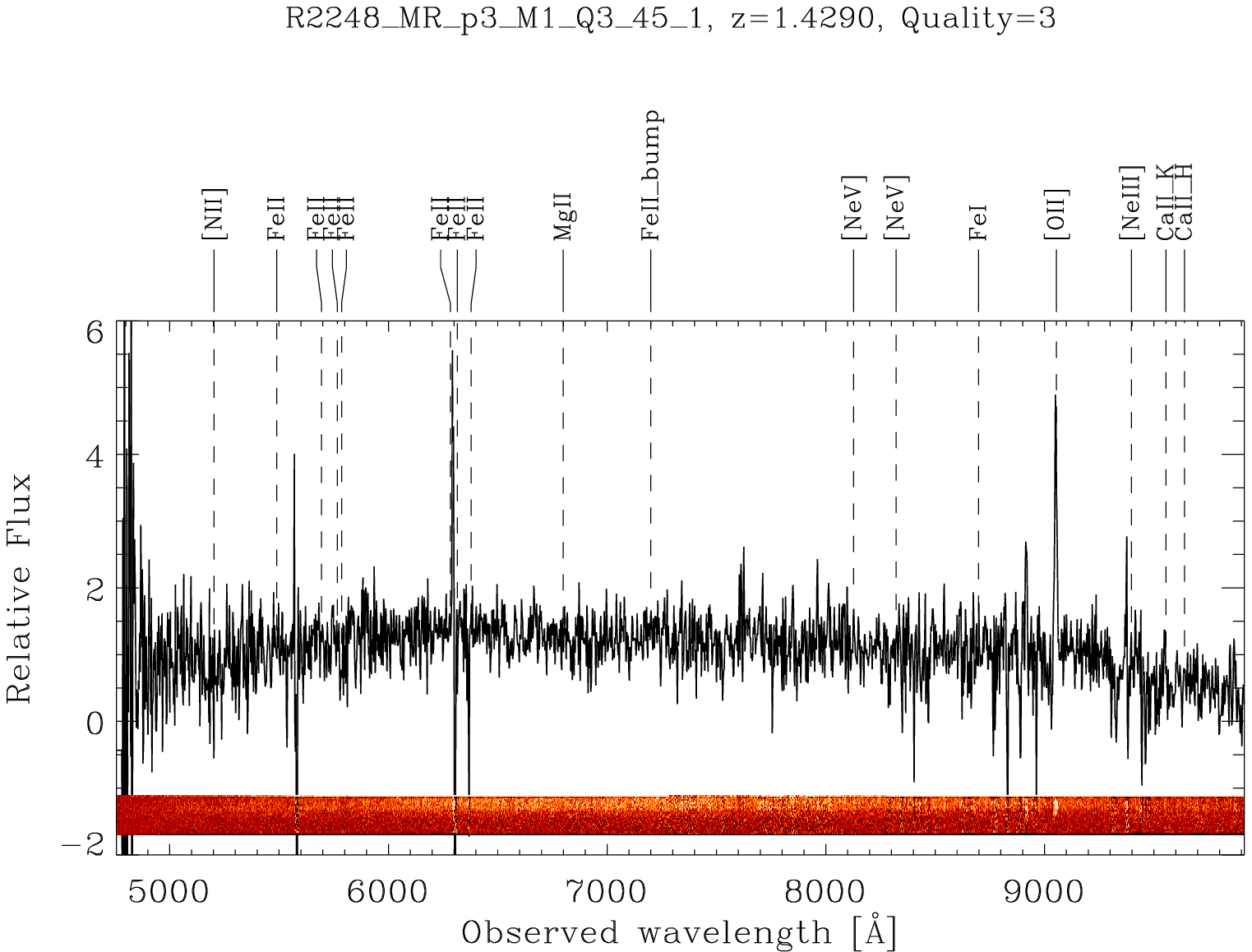}}
    \put(50, 40){{ \large z=1.428, QF=3}}
  \end{picture}

  \caption{(Continued)}
  \label{fig:1a_spec}
\end{figure*}

\begin{figure*}[!ht]
  \centering
  \ContinuedFloat
  
  \begin{picture}(100,220)
    \put(0,0){\includegraphics[trim = 0px -50px 0px 0px, clip=true, width = .19\textwidth]{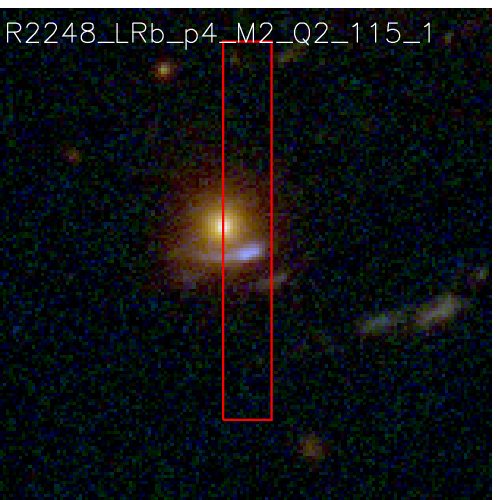}}
    \put(85, 40){{\color{white}\bf \large 8a}}
  \end{picture}
  \begin{picture}(300,220)
    \put(0,0){\includegraphics[trim = 0px 0px 0px 25px, clip=true, width = .6\textwidth]{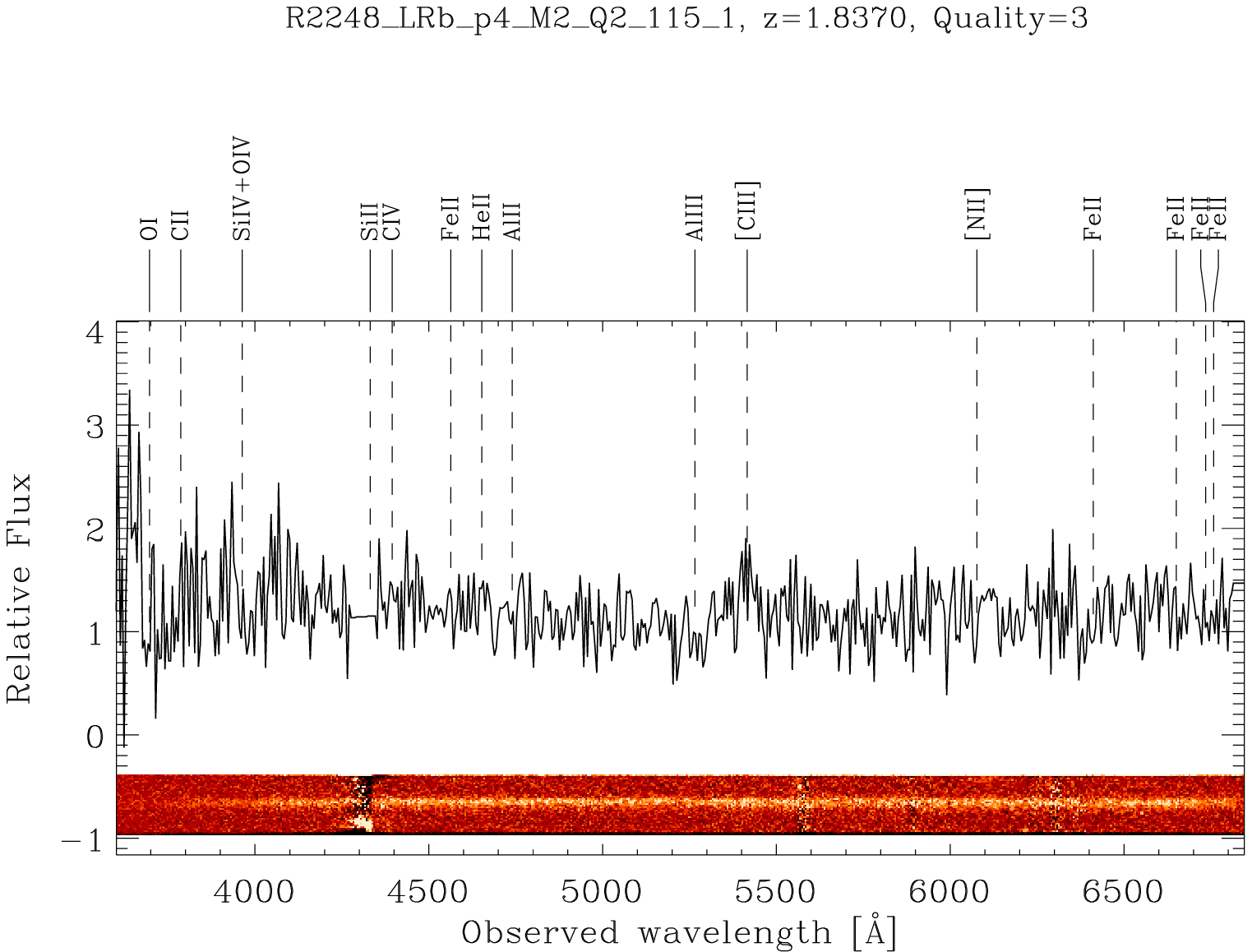}}
    \put(50, 45){{ \large z=1.837, QF=9}}
  \end{picture}

  \begin{picture}(100,220)
    \put(0,0){\includegraphics[trim = 0px -50px 0px 0px, clip=true, width = .19\textwidth]{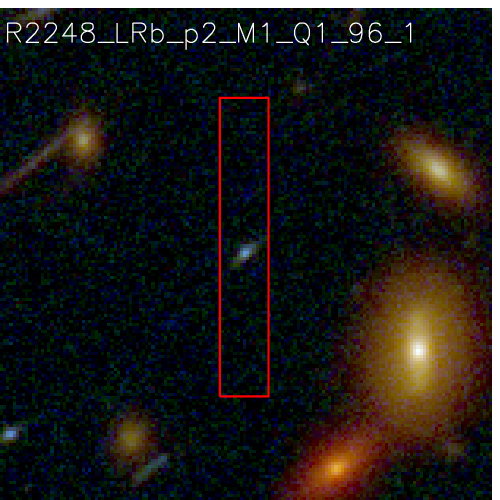}}
    \put(80, 40){{\color{white}\bf \large 11b}}
  \end{picture}
  \begin{picture}(300,220)
    \put(0,0){\includegraphics[trim = 0px 0px 0px 25px, clip=true, width = .6\textwidth]{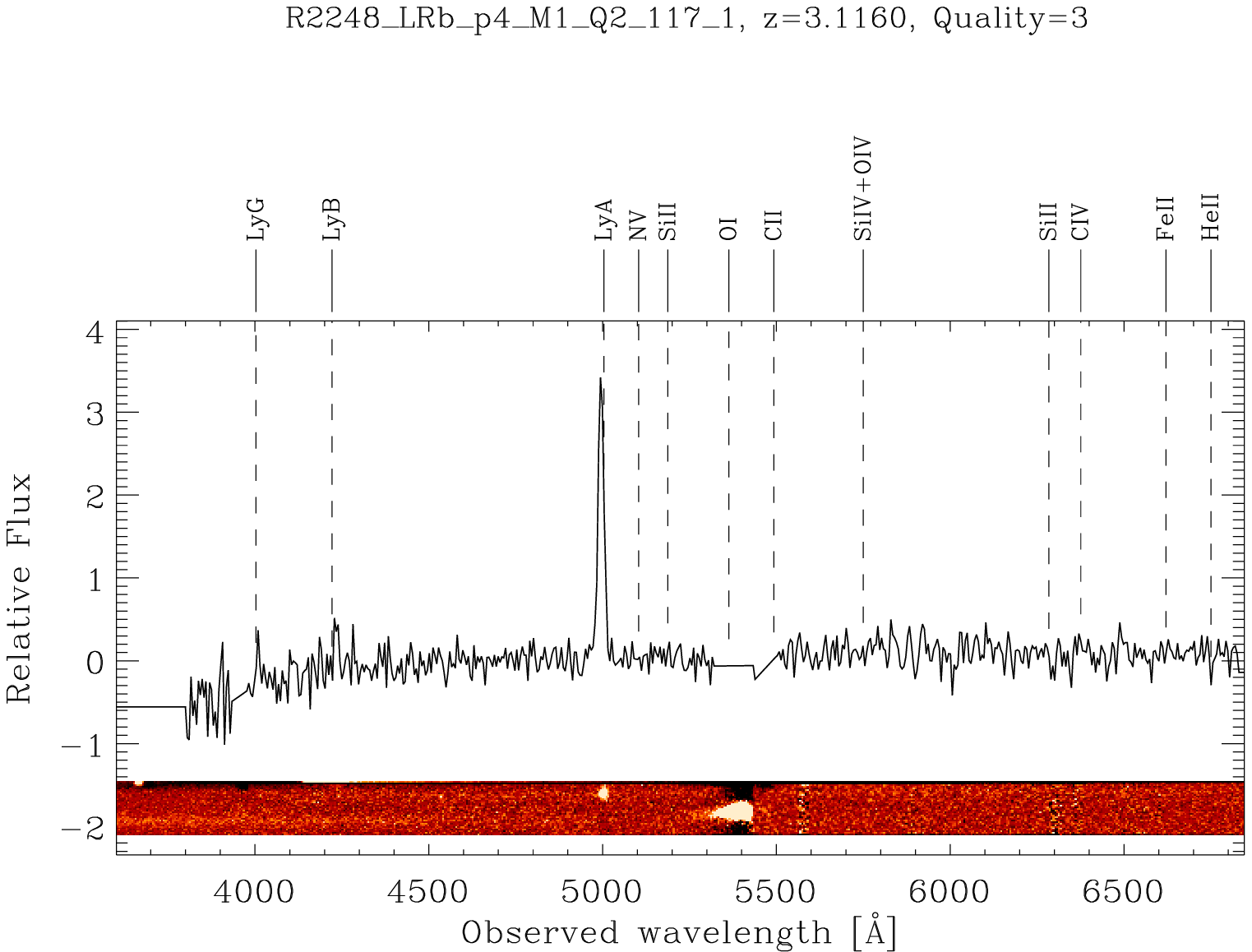}}
    \put(50, 42){{ \large z=3.116, QF=3}}
  \end{picture}

  \caption{(Continued)}
  \label{fig:1a_spec}
\end{figure*}

\subsection{MUSE redshift measurements}
\label{subsec:muse_redshift}
\begin{figure}[!ht]
  \centering
  \includegraphics[trim = 0px 0px 0px 0px, width = .475\textwidth, clip=true]{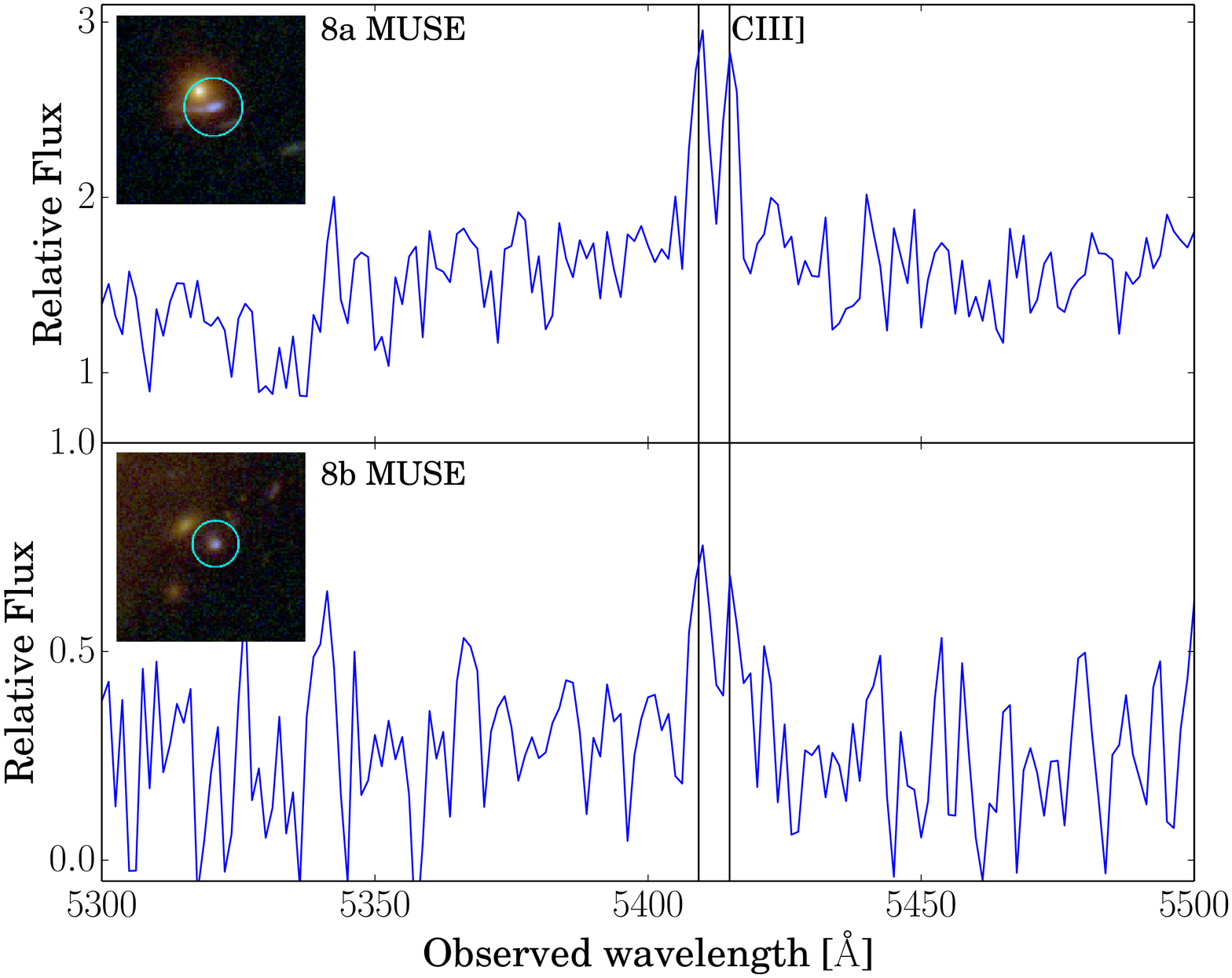}
  \caption{MUSE 1D spectra of the multiple images 8a and 8b. The
    vertical lines indicate the CIII] doublet emission wavelengths of a
    source at redshift 1.837. The small panels show the circles with
    $1\arcsec$ (top) and $0\arcsec.8$ (bottom) radius used to extract the two
    spectra. The flux is rescaled by a factor of $10^{-18}\rm erg
    /s/cm^2/\AA$.}
  \label{fig:family_8_spec}
\end{figure}

\begin{figure}[!ht]
  \centering
  \includegraphics[trim = 0px 0px 0px 0px, width = .475\textwidth, clip=true]{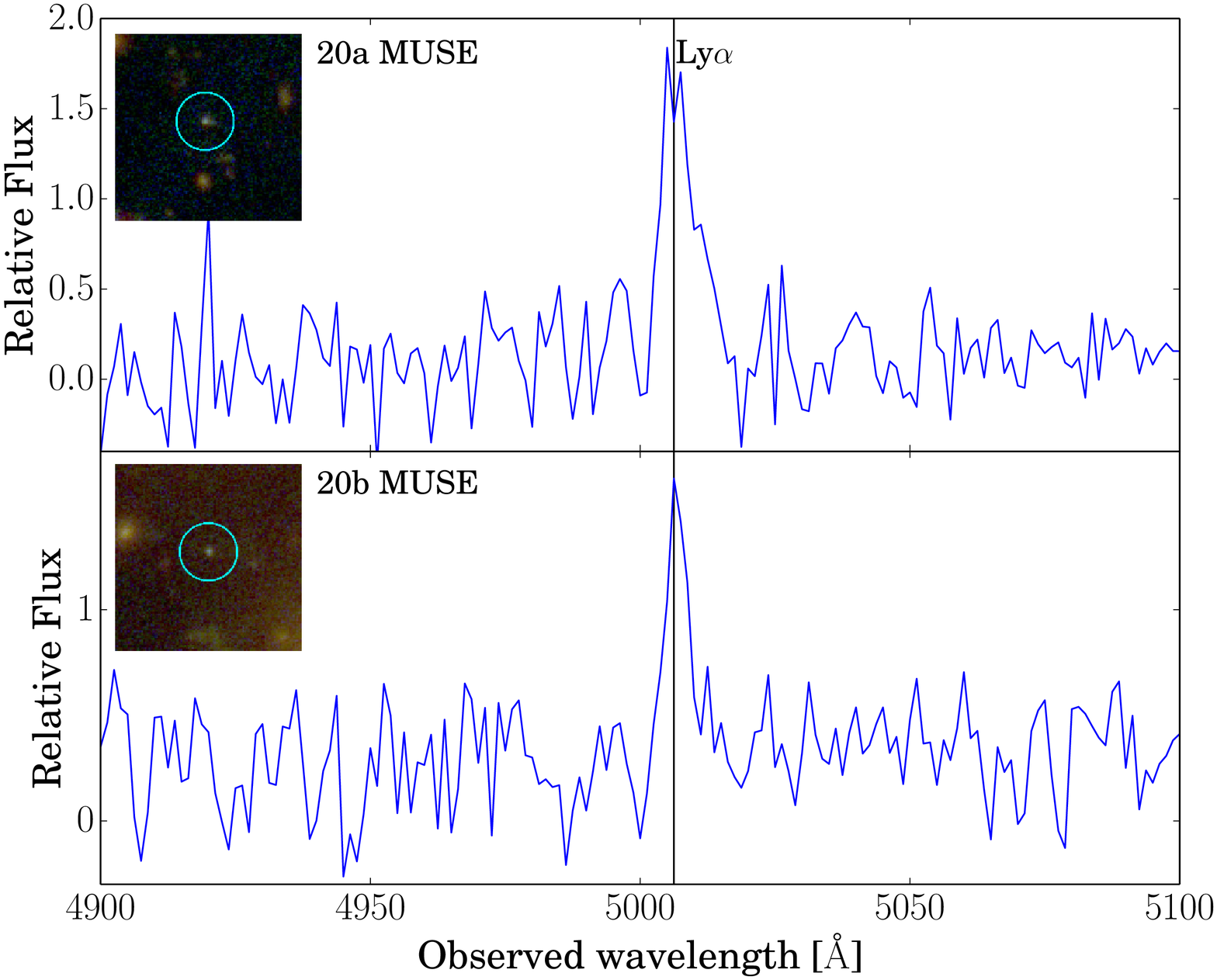}
  \caption{MUSE 1D spectra of the multiple images 20a and 20b. The
    vertical line indicates the Ly$\alpha$ emission wavelength of a source at
    redshift 3.118. The small panels show the circles with $1\arcsec$ radius
    used to extract the two spectra. The flux is rescaled by a factor
    of $10^{-18}\rm erg /s/cm^2/\AA$.}
  \label{fig:family_20_spec}
\end{figure}

Observations with the new integral-field spectrograph MUSE on the VLT
were conducted in the South-West part of the cluster as part of the
MUSE science verification program (ID 60.A-9345, P.I.: K. Caputi). A 8520 second total exposure
was obtained in June 2014 with a seeing of $\approx1\arcsec$. The MUSE data
cube covers $1\times 1$ arcmin$^2$, with a pixel size of 0\arcsec.2, over the
wavelength range $4750-9350\,\AA$ and with a spectral resolution of $\approx3000$ and a
dispersion of $1.25\,\AA$/pixel. Details on data reduction and results
can be found in \citet{2015A&A...574A..11K}.  We extracted 1D spectra
of the strong lensing features within circles with radius ranging from
$0\arcsec.5$ to $2\arcsec$, in order to minimize the contamination of
nearby objects and maximize the signal-to-noise.  In this work, guided
by the strong lensing model predictions, we revisited several spectra
and measured redshifts of two additional multiple image families not
included in the CLASH-VLT data and \citet{2015A&A...574A..11K}. In the
Figures \ref{fig:family_8_spec} and \ref{fig:family_20_spec}, we show
the MUSE spectra of the multiple images 8a/b and 20a/b. 
The spectra of both sources 8a and 8b show a pair of emission lines at the same wavelengths, which we identify as the resolved CIII] doublet ($1906.7, 1908.7 \AA$) at $z=1.837$. The fact that our lensing models predict these sources to be multiple images at $z_{model}=1.81\pm0.03$ lends strong support to this interpretation. We note that also the CLASH-VLT spectrum of the source 8a (see Figure \ref{fig:1a_spec}, 7th panel) shows an (unresolved) emission line although with low S/N (QF=9). 

In Figure \ref{fig:family_20_spec} we show the spectra extracted from apertures around the multiple images 20a/b.
The existence of an asymmetric emission line at $5007.5\,\AA$ is clear, which we identify with a $\rm Ly\alpha$ emission at redshift 3.118.
Inspection of the MUSE data cube around this wavelength reveals an extended low-surface brightness emission
around each image.
The excellent agreement between the modelled redshift of the compact sources (see Table \ref{tab:families}) and the extended emission shows that both are related.
Based on our lensing model, we interpret this diffuse double emission as two multiply imaged Ly$\alpha$ blobs.
LABs are commonly found in deep narrow band image surveys
\citep{1999MNRAS.305..849F, 2000ApJ...532..170S, 2001ApJ...554.1001F,
  2006A&A...452L..23N}, their Ly$\alpha$ luminosities are in the range
$10^{43}-10^{44}\, {\rm erg/s}$, with sizes up to $\sim\! 100$ kpc.
Although several mechanisms have been proposed to explain their large
luminosities, there is still no consensus on the physical nature of
these sources \citep[][and references
therein]{2015arXiv150201589P}.
A detailed study of this source will be presented in a forthcoming paper (Caminha et al. in prep.).

\section{Strong Lens Modelling}
\label{sec:strong_lens_modelling}

Here we use the strong lensing observables to reconstruct the total mass
distribution of RXC~J2248.  The positions of the multiple images, from
a single background source, depend on the relative distances (observer, lens and
source) and on the total mass distribution of the intervening lens. We describe below
our methodology to determine the mass distribution of the cluster from
the observed positions of the identified multiple images.

First, we visually identify the multiple images on the colour
composite \emph{HST} image (Figure \ref{fig:multiple_image_systems}).
We revisit all the previously suggested multiple image systems and explore new ones in this identification.
Using colour and morphological information of these objects, as well as the
expected parity from basic principles of gravitational lensing theory, we select luminosity peaks.
In a second step, we refine the measurements using the
stacked images of the optical (F435W, F606W, F625W, F775W, F814W and
F850lp) and near infra-red (F105W, F110W, F125W, F140W and F160W) filters,
depending on the colour of the multiple images. However, we do not use different stacked images to measure the luminosity peaks of different multiple images belonging to the same family. We draw different iso-luminosity contours around each peak and determine the position of the centroid of the innermost contour enclosing a few pixels ($\approx 5$, or 0.1 square arcsec).  With this procedure, we ensure that we consider the peaks of the light distribution of different multiple images which correspond to the same position on the source plane, thus avoiding systematics which are often introduced by automated measurements. The measured positions of multiple images are listed in Table \ref{tab:families}.

  Note that some extended arcs show multiple peaks or knots
  (e.g. families 2, 4 and 9), thus in principle allowing us to split these families
  in two subsets, as done in
  \citet{2014MNRAS.438.1417M}.  This technique can improve the
  constraints on the critical lines close to the multiple images,
  however it does not introduce any extra constraint in the overall best-fitting model.
  In this work, we choose to use only one peak for
  each family to avoid any possible systematic effect in the
  modelling and to save computational time.

\subsection{Mass model components}
\label{subsec:mass_model_components}

The optical and X-ray images of the cluster (see Figures \ref{fig:multiple_image_systems} and \ref{fig:members_xray}) indicate a regular elliptical shape, with no evident large asymmetries or massive substructures in the region where the multiple images are formed.
In view of its regular shape, we consider three main components for the total mass distribution in the lens modelling: 1) a smooth component describing the extended dark matter distribution, 2) the mass distribution of the BCG and 3) small scale halos associated to galaxy members.

We also check whether the presence of an external shear term associated to two
mass components in the north-east and south-west of the cluster
could improve the overall fit.
In these two regions (outside the field of view shown in Figure \ref{fig:multiple_image_systems}), we notice the presence of bright cluster galaxies that could contribute to the cluster total mass with additional massive dark matter halo terms.
However, we do not find any significant improvement in the reconstruction of the observed positions of the multiple images and these components are completely unconstrained.
Moreover, we test a model including an extra mass component in the core of the cluster ($R\lesssim 300$ kpc).
Also in this case the fit does not improve significantly to justify the increase in the number of free parameters, for which we obtain best-fitting values physically not very plausible (for example, we find an extremely high value for the mass ellipticity of this new term).

\subsubsection{Dark matter component}

For the smooth mass component (intra cluster light, hot gas and, mainly, dark matter) we adopt a Pseudo Isothermal Elliptical Mass Distribution \citep[hereafter PIEMD,][]{1993ApJ...417..450K}.
The projected mass density distribution of this model is given by:
\begin{equation}
\Sigma(R) = \frac{\sigma_v^2}{2 G}\left( \frac{1}{\sqrt{R^2(\varepsilon) + r_{core}^2}}\right),
\label{eq:sigma_piemd}
\end{equation}
where $R(x, y, \varepsilon)$ is an elliptical coordinate on the lens
plane and $\sigma_v$ is the fiducial velocity dispersion.  The
ellipticity $\varepsilon$ is defined as $\varepsilon \equiv 1 - b/a$,
where $a$ and $b$ are the semi-major and minor axis respectively.
There are 6 parameters describing this model: the centre position
($x_0$ and $y_0$), the ellipticity and its orientation angle ($\varepsilon$
and $\theta$, where the horizontal is the principal axis and the angle is counted counterclockwise), the fiducial velocity dispersion ($\sigma_v$) and
the core radius ($r_{core}$).  The PIEMD parametrization has been
shown to be a good model to describe cluster mass distributions in
strong lensing studies and sometimes provides a better fit than the
canonical Navarro-Frenk-White \citep[hereafter
NFW,][]{1996ApJ...462..563N, 1997ApJ...490..493N} mass distribution.
\citet{2015ApJ...800...38G}, using a similar high-quality data set,
found for example that the dark matter components of the HFF galaxy
cluster MACS~J0416.1$-$2403 are better described by PIEMD models.

To test the dependence of our main results on a specific mass
parametrization, we also consider an NFW distribution for the main
dark matter component.  In this work, we use a NFW model with
elliptical potential \citep[hereafther
PNFW][]{1993ApJ...417..450K,2002sgdh.conf...50K, 2002A&A...390..821G},
which significantly reduces computing time of the deflection angle
across the image in the \emph{lenstool} implementation.
For this model, the free parameters are the characteristic radius
$r_s$ and density $\rho_s$, besides the potential ellipticity,
orientation angle and the centre position ($\varepsilon$, $\theta$,
$x_0$ and $y_0$).  The main differences between these two models are
the presence of a core radius in the PIEMD model, while the PNFW has a
central cusp, and different slopes at large radii.

\subsubsection{Cluster members and BCG}
\label{subsec:cluster_members}

Membership selection is performed following the method adopted in
\citet{2015ApJ...800...38G} (see Section 3.3.1 of the referenced
paper).  Specifically, we investigate the loci, in a multidimensional
color space, of a large sample of spectroscopically confirmed cluster
members and field galaxies.  We define confirmed cluster members the galaxies within the spectroscopic range of $0.348\pm0.0135$,
corresponding to a velocity range of $\pm3000\rm\, km/s$ in the cluster
rest-frame. We thus find 145 members out of the 254 galaxies with
measured redshifts in the \emph{HST} field of view.  We then model the
probability density distributions (PDFs) of cluster member and field
galaxy colours as multidimensional gaussians, with means and
covariances determined using a robust method \citep[Minimum Covariance
Determinant,][]{Rousseeuw84}.  This ensures that a small fraction of
outliers in color space (due for example to inaccurate photometry,
contamination from angularly close objects, or presence of
star-forming regions) does not perturb significantly the measured
properties of the PDFs.  The color distribution is traced by all
independent combinations of available bands from the 16 CLASH filters,
with the exclusion of F225W, F275W, F336W, and F390W bands, which
often do not have adequate signal-to-noise ratio for red cluster
galaxies.  For all galaxies, we compute the probability of being a cluster member or a field galaxy, using the determined PDFs in
a Bayesian hypotheses inference.  We then classify galaxies using a
probability threshold which is a good compromise between purity and
completeness, and thus select 159 additional cluster members with no
spectroscopic redshifts.

In the strong lensing model, we consider only (spectroscopic and photometric) cluster members that are within $1\arcmin$ radius from the BCG centre, which encloses all the identified multiple images.
In this way, we save computational time by not computing the deflection angle of members in the outer regions of the cluster that are not expected to affect significantly the position of the multiple images.
Thus, we include 139 cluster members in the model, 64 of which are spectroscopically confirmed. 
 
Each cluster member is modelled as dual pseudoisothermal elliptical mass distribution \citep[dPIE,][]{2007arXiv0710.5636E, 2010A&A...524A..94S} with zero ellipticity and core radius, and a finite truncation radius $r_{cut}$.
The projected mass density distribution of this model is given by:
\begin{equation}
\Sigma (R) = \frac{\sigma_v^2}{2 G}\left( \frac{1}{\sqrt{R^2(\varepsilon=0)}} - \frac{1}{\sqrt{R^2(\varepsilon=0) + r_{cut}^2}}\right).
\label{eq:sigma_members}
\end{equation}
Following a standard procedure in cluster-scale strong lensing analyses \citep[e.g.,][]{2006MNRAS.372.1425H, 2007NJPh....9..447J, 2015ApJ...800...38G}, to reduce the number of free parameters describing the cluster members, we use the following velocity dispersion-luminosity and truncation radius-luminosity scaling relations:
\begin{equation}
\sigma^{gals}_{v,i} = \sigma^{gals}_v\left( \frac{L_i}{L_{0}} \right)^{0.25},\;r^{gals}_{cut,i} = r^{gals}_{cut}\left( \frac{L_i}{L_{0}} \right)^{0.5},
\label{eq:mass_to_light}
\end{equation}
where $L_{0}$ is a reference luminosity associated to the second most
luminous cluster member, indicated by the magenta circle in the Figure
\ref{fig:members_xray}.  Given the adopted relations, the total
mass-to-light ratio of the cluster members is constant and we reduce
the free parameters of all the member galaxies to only two parameters:
the reference velocity dispersion $\sigma^{gals}_v$ and truncation
radius $r^{gals}_{cut}$.  We measure the luminosities ($L_i$) in the
F160W band, the reddest available filter, in order to minimize the
contamination by blue galaxies around cluster members and to obtain a
good estimate of galaxy stellar masses.

Due to a generally different formation history, the BCG is often
observed to significantly deviate from these scaling relations
\citep{2012ApJ...756..159P}. We therefore introduce two additional
free parameters associated to the BCG ($\sigma_v^{BCG}$ and
$r_{cut}^{BCG}$), keeping its position fixed at the center of the
light distribution.

\subsection{Lens modelling definitions}
\label{subsec:lenstool}

The strong lensing modelling is performed using the public software \emph{lenstool} \citep{1996ApJ...471..643K, 2007NJPh....9..447J}.
Once the model mass components are defined, the best-fitting model parameters are found by minimizing the distance between the observed and model-predicted positions of the multiple images, and the parameter covariance is quantified using a Bayesian Markov chain Monte Carlo (MCMC) technique.

In detail, to find the best-fitting model, we define the lens plane $\chi^2$ function as follows:
\begin{equation}
\label{eq:chi2_lens}
    \chi^2(\vec{\Pi})  :=  \sum_{j=1}^{N_{\rm fam}} \sum_{i=1}^{N^{j}_{\rm im}}\left( \frac{\left| \vec{\theta}^{\rm obs}_{i,j} -
              \vec{\theta}_{i,j}^{\rm pred}\left( \vec{\Pi}\right) \right|}{\sigma^{\rm obs}_{i,j}} \right)^2,
\end{equation}
where $N_{\rm fam}$ and $N^{j}_{\rm im}$ are the number of families
and the number of multiple images belonging to the family $j$,
respectively.  $\vec{\theta}^{\rm obs}$ and $\vec{\theta}^{\rm pred}$
are the observed and the model-predicted positions of the multiple images,
and $\sigma^{\rm obs}$ is the uncertainty in the observed position.
The model-predicted position of an image is a function of the both lens parameters and the
cosmological parameters, all represented by the vector $\vec{\Pi}$. In this work we adopt flat priors on all parameters, thus the
set of parameters $\vec{\Pi}$ that provides the minimum value of the $\chi^2$ function ($\chi^2_{min}$) is called the best-fitting model, while the predicted
positions of this model are referred to as $\vec{\theta}^{\rm bf}$.
Note that for some multiple image families we do not have measured spectroscopic redshifts.  In
these cases, the family redshift is also a free parameter optimized in the calculation of the $\chi^2_{min}$ with a flat prior.

Besides the value of the $\chi^2_{min}$, we can quantify the goodness of the fit with the root-mean-square between the observed and reconstructed positions of the multiple images:
\begin{equation}
\Delta_{rms} = \sqrt{ \frac{1}{N}\sum_{i=1}^{N} \left| \vec{\theta}^{\rm obs}_i - \vec{\theta}_i^{\rm bf} \right|^2},
\label{eq:rms_def}
\end{equation}
where $N$ is the total number of multiple images. This quantity does not depend on the value of the observed uncertainties $\sigma^{\rm obs}$, making it suitable when comparing results of different works.
Finally, we also define the displacement of a single multiple image $i$ as
\begin{equation}
\Delta_i = \left| \vec{\theta}^{\rm obs}_i - \vec{\theta}_i^{\rm bf} \right|.
\label{eq:rms_singe}
\end{equation}

The posterior probability distribution function of the free parameters is given by the product of the likelihood function and the prior
\begin{equation}
P\left(\vec{\theta}^{\rm obs}|\vec{\Pi}\right) \propto \exp{ \left(-\frac{\chi^2(\vec{\Pi})}{2}\right) }P(\vec{\Pi}),
\label{eq:likelihood}
\end{equation}
where P(\vec{\Pi}) is the prior, which in this work we consider to be flat for all free parameters.
In order to properly sample the parameter space and obtain the posterior distribution of the parameter values, we use a MCMC with at least $10^5$ points and a convergence speed rate of 0.1 (a parameter of the \emph{BayeSys}\footnote{\url{http://www.inference.phy.cam.ac.uk/bayesys/}} algorithm used by the \emph{lenstool} software). We have checked that these values ensure the convergence of the chains.
We remark that all computations are performed estimating the value of the $\chi^2$ function on the image plane, which is formally more accurate than working on the source plane \citep[e.g.,][]{2001astro.ph..2340K}.

\subsection{Cosmological parameters}
\label{subsec:cosmological_parameters}

The availability of a large number of multiple images, with spectroscopic
redshifts spanning a wide range, in a relatively regular mass
distribution, makes RXC~J2248 a suitable cluster lens to
test the possibility of constraining cosmological parameters.
Strong lensing is sensitive to the underlying geometry of the Universe
via the angular diameter distances from the observer to the lens
($D_{OL}$) and source ($D_{OS}$), and from the lens to the source
($D_{LS}$).  For one source, the lens equation can be written as
\begin{equation}
\vec{\theta} = \vec{\beta} + \frac{D_{LS}}{D_{OS}}\hat{\vec{\alpha}}\left(\vec{\theta}\right)
\label{eq:lens_eq_cosmo}
\end{equation}
where $\vec{\theta}$ and $\vec{\beta}$ are the angular positions on the lens and source planes, respectively, $\hat{\vec{\alpha}}$ is the deflection angle and the cosmological dependence is embedded into the angular diameter distances.
In general, the ratio between the cosmological distances can be absorbed by the parameters of the lens mass distribution (i.e. the factor that multiplies the mass distribution), $\sigma_v^2$ in the PIEMD case.
However, when a significant number of multiply lensed sources at different redshifts is present, this degeneracy can be broken and a leverage on cosmological parameters can be obtained via the so called family ratio:
\begin{equation}
\Xi_{S1,S2}(\vec{\pi}) = \frac{D(\vec{\pi})_{LS,1} D(\vec{\pi})_{OS,2}}{D(\vec{\pi})_{LS,2} D(\vec{\pi})_{OS,1}}, 
\label{eq:cosmo_prober}
\end{equation}
where $\vec{\pi}$ is the set of cosmological parameters and 1 and 2 are two different sources at redshifts $z_{s1}$ and $z_{s2}$.  This
technique has been applied in \citet{2004A&A...417L..33S} and
\citet{2010Sci...329..924J} for the galaxy clusters Abell~2218 and
Abell~1689, respectively.  In this work, we use the $\Lambda$CDM
cosmological model, which includes as free parameters the energy
density of the total matter of the universe (ordinary and dark matter)
$\Omega_m$, the dark energy density $\Omega_{\Lambda}$
and the equation of state parameter of this last component, $w=P/\rho$.
All the results from our lens models are described in Section
\ref{sec:results_sl}.

\subsection{Multiple image selection}
\label{subsec:multple_image_selection}

In the previous strong lensing studies of RXC~J2248, 19 candidates of
multiple image families were identified \citep{2014MNRAS.438.1417M,
  2014ApJ...797...48J, 2014MNRAS.444..268R}, however some of them are
not secure candidates. The selection of secure multiple image systems,
i.e. systems with spectroscopic confirmation or multiple images with
correct parity and/or consistent colors, is essential to avoid
systematics in lensing models. This criterion leaves us with 16
families, whose properties are summarized in Table \ref{tab:families}.
Based on this strict criterion, we do not include the counter image 16c in our
models, since the corresponding model-predicted
position from our best-fitting models (see Section
\ref{sec:results_sl}) is close to two objects with similar colors,
leaving the identification of this counter image uncertain.

Given our relatively simple models to parametrize the cluster mass
distribution and the total mass-luminosity scaling relation of the cluster members, we
also avoid multiple images in the vicinity of the members.
Their truncated PIEMD mass with a constant total mass-to-light ratio might not be able to
accurately reproduce the positions of the multiple images close to the core of the members, introducing a bias in the best fits.  Quantitatively, we
do not include multiple images closer than $3\rm\, kpc$ ($\approx
0\arcsec.6$) to a cluster member, which is approximately the
Einstein's radius of a galaxy with $\sigma_{v}=160{\rm\, km/s}$ for a
source at $z=3$.  As a result, the multiple images which are not
included in our fiducial lens model are 3b, 8a, 8b and 17b.  In the
case of family 8, we are left with only one multiple image which does
add any constraint to the models, we therefore exclude the entire
family.

Finally, in all different models we analyse, family 11 presents a
much larger offset between observed and reconstructed images ($\Delta_i\approx1\arcsec$), when compared with the
other families ($\approx 0\arcsec.3$).  We conduct several tests in
the effort to improve the fit of this family: 1) we freely vary the mass parameter of the nearby cluster members when optimizing the model;
2) we introduce a dark
halo in the vicinity of the images; 3) we consider an external shear
component represented by a PIEMD in the south-west region of the cluster with
free mass. The latter is suggested by the apparent discontinuity in the X-ray
emission from the {\it Chandra} data, $\sim\! 30\arcsec$ SW of the BCG.
We verify that it is difficult to reduce the average value of $\Delta_i$ below $1\arcsec$ for the images of family 11 by only adding an extra mass component.

We also consider the effect of the large-scale structure along the
line of sight, which we can sample with our redshift
survey. Specifically, in order to investigate whether this family
could be lensed by a galaxy behind the cluster, we map the positions
of the three multiple images onto a plane at the redshift of each
background source in the Tables \ref{tab:families} and
\ref{tab:high_mag_src}.  In this analysis, we find that the background
galaxy ID B7 at $z=1.270$ (see Table \ref{tab:high_mag_src}) could
significantly perturb the positions of the multiple images of family
11, since its distance is $\approx 25 \rm \,kpc$, or
$\approx3\arcsec$, from the positions of the multiple images $11a$ and
$11b$ on the $z=1.27$ plane.
In this
configuration, assuming a velocity dispersion value of $150 \rm
\,km/s$ for B7, the deflection angle induced in the multiple images of
family 11 is $\approx 0''.4$.  We therefore argue that the effect of
this background galaxy can partially explain the large $\Delta_i$
value of this family, although more complex multi-plane ray-tracing
procedures should be employed to fully account for such a deviation.
A detailed modelling of the background effects on the strong lensing
analyses of RXC~J2248 is out of the scope of this work, however in
Section \ref{subsec:line_of_sign_structure} we will return to this
non-negligible issue by estimating the statistical effect on the image
positions due to the line of sight mass structure and show that it can
have an important impact on high-precision lensing modelling.

To summarise, in the effort to minimize possible sources of systematic
uncertainties, we decide not to include the multiple images 3b, 16c
and 17b, and the families 8 and 11 in our strong lensing analysis.  In
the end, we consider a total of 38 multiple images of which 19 are
spectroscopically confirmed, belonging to 14 families at different
redshifts.  We leave to future work the detailed study of individual
sources, including a further refinement of the mass distribution
modeling which takes into account the influence of mass structures
along the line of sight, as the spectroscopic work particularly with
VLT/MUSE continues.

\section{Results}
\label{sec:results_sl}
\begin{table*}[!ht]
\centering
\caption{Summary of the best fit models}
\begin{tabular}{l c c c c l} \hline \hline
Model ID & $DOF$ & $N_{images}$ & $\Delta_{rms} ['']$ & $\chi^2_{min, ref}$ & Description \\
\hline
F1                    & 16 & 20 & 0.33 & 8.5  & fixed cosmology, only spec families\\
F2                    & 31 & 38 & 0.31 & 14.8 & fixed cosmology, all families ({\sl reference model})\\
F1a                   & 26 & 27 & 0.82 & 72.2 & fixed cosmology, all spec families (including families 8, 11 and image 3b)\\
F1-5th                & 18 & 21 & 0.34 & 9.6  & fixed cosmology, all spec families including the 5th image of family 14\\
N1                    & 16 & 20 & 1.15 & 106.0& fixed cosmology, only spec families and NFW instead of PIEMD\\
N2                    & 31 & 38 & 1.20 & 217.4& fixed cosmology, all families and NFW instead of PIEMD\\
W1                    & 14 & 20 & 0.29 & 6.7  & free $\Omega_m$ and $w$ in a flat universe, only spec families\\
W2                    & 29 & 38 & 0.30 & 13.3 & free $\Omega_m$ and $w$ in a flat universe, all families\\
W3                    & 23 & 34 & 0.29 & 11.1 & free $\Omega_m$ and $w$ in a flat universe, all families except family 14 ($z=6.111$)\\
L1                    & 14 & 20 & 0.29 & 6.7  & free $\Omega_m$ and $\Omega_{\Lambda}$, only spec families\\
L2                    & 29 & 38 & 0.30 & 13.8 & free $\Omega_m$ and $\Omega_{\Lambda}$, all families\\
L3                    & 23 & 34 & 0.29 & 11.2 & free $\Omega_m$ and $\Omega_{\Lambda}$, all families except family 14\\
WL1                   & 13 & 20 & 0.29 & 6.7  & free $\Omega_m$, $\Omega_{\Lambda}$ and $w$, only spec families\\
WL2                   & 28 & 38 & 0.30 & 13.3 & free $\Omega_m$, $\Omega_{\Lambda}$ and $w$, all families\\
FZ1                   & 9  & 20 & 0.25 & 4.9  & fixed cosmology, only spec families but free redshift\\
FZ2                   & 24 & 38 & 0.28 & 11.9 & fixed cosmology, all families but free redshift\\
WZ1                   & 7  & 20 & 0.25 & 4.9  & free $\Omega_m$ and $w$ in a flat universe, only spec families but free redshift\\
WZ2                   & 22 & 38 & 0.28 & 11.9 & free $\Omega_m$ and $w$ in a flat universe, all families but free redshift\\
NW1                   & 14 & 20 & 0.63 & 32.0 & free $\Omega_m$ and $w$ in a flat universe, only spec families and NFW instead of PIEMD\\
NW2                   & 29 & 38 & 0.62 & 57.9 & free $\Omega_m$ and $w$ in a flat universe, all families and NFW instead of PIEMD\\
Wa1                   & 13 & 20 & 0.29 & 6.6  & free $\Omega_m$, $w$ and $w_{a}$ in a flat universe, only spec families\\
Wa2                   & 28 & 38 & 0.29 & 12.7 & free $\Omega_m$, $w$ and $w_{a}$ in a flat universe, all families\\
\hline \hline
\end{tabular}
\label{tab:summary_bf}
\tablefoot{Summary of the considered strong lensing models and their global results. Columns show the model IDs, the number of degrees of freedom ($DOF$), the number of input images used, the  best fit positional $\Delta_{rms}$ (see Equation \ref{eq:rms_def}), the value of the reference $\chi^2_{min}$ (computed considering an image positional error of $0''.5$, see Equation \ref{eq:chi2_lens}) and a short description of each model.}
\end{table*}
\subsection{A collection of lens models}
\label{subsec:collection_lens_models}

We explore a number of strong lensing models based on different
samples of secure multiple-image systems, as described above, and
different model parameters.

First we define two samples of multiple-image families: ``family
sample 1'' includes families with spectroscopic confirmation, namely
families with IDs 2, 3, 4, 6, 7, 14 and 18, totalling 20 multiple
images in 7 families; ``family sample 2'' contains all the secure
families, including also multiple images with no spectroscopic
confirmation, but with correct colours and parity as expected from
gravitational lensing theory. This extended sample includes family IDs
9, 13, 15, 16, 17, 20 and 21 in addition to the 7 spectroscopic
families, totalling 38 multiple images in 14 families.  Note that the
redshift of the compact multiple images of family 20 is conservatively
considered a free parameter here, since the multiple images are not
necessarily associated with the extended Ly$\alpha$ emission (see
Section \ref{subsec:muse_redshift}).  We find however, 
that the best fit redshift of family 20 obtained from the strong
lensing model is in very good agreement with that derived from the LAB
emission, confirming a posteriori that the compact sources of family
20 are associated with the extended Ly$\alpha$ emission (see
Figure \ref{fig:family_20_spec}). Since this extra information does
not improve the lens models significantly, we did not recompute the
MCMC analyses for all the reference models.

To optimize the models, we adopt an image positional error of
$0\arcsec.5$ in the positions of the multiple images, which is in agreement with predictions of the effects of matter density fluctuations along the line of sight on the positions of multiple images \citep{2012MNRAS.420L..18H}.
In all cases this choice leads to a $\chi^2_{min,ref}$ lower
than the number of degrees of freedom ($DOF$, defined as the
difference between the number of constraints and the number of free
parameters of a model).  Positional errors ($\sigma^{\rm obs}$ in
Equation \ref{eq:chi2_lens}) are then rescaled to yield a
$\chi^2_{min}$ value equal to the $DOF$ when probing the space
parameter using the MCMC technique.  The values of the rescaled
$\sigma^{\rm obs}$ are approximately $0\arcsec.33$ for all the models
under study.  This can effectively account for, e.g., line of sight
mass structures and the scatter in the adopted total mass-to-light relation
of the cluster members.

We exploit different lensing models to assess possible systematic
effects stemming from our assumptions on the cluster total mass
distribution, multiple image systems and adopted free parameters. A
list of all models is given in Table \ref{tab:summary_bf}, including
their main parameters and a brief description for each model.

The IDs of the models are composed by letters indicating the model
assumption.  ``F'' indicates a fixed cosmology (flat $\rm \Lambda CDM$
cosmology with $\Omega_m = 0.3$ and $H_0 = 70\rm\, km/s/Mpc$), ``N''
indicates that we use a PNFW mass profile to represent the smooth dark matter mass distribution instead of a PIEMD.
``W'' indicates that we vary the parameters $\Omega_m$ and $w$ (the dark energy equation of state, in a flat universe) while ``L'' indicates free $\Omega_m$ and $\Omega_\Lambda$ and fixed $w=-1$ (i.e. we vary the curvature of the Universe).
``WL'' indicates we are varying all the three cosmological parameters
at the same time. Finally, the ID ``Wa'' stands for a model where we
consider a variation of $w$ with redshift parametrized by $w(z) =
w_0 + w_a\, z/(1+z)$.

The numbers in the IDs indicate three different multiple image inputs.
Number ``1'' indicates that we consider only the family sample 1,
while ``2'' refers to all the secure families (family sample 2).
Moreover, for two models we also explore the effect on the best
fitting parameters (indicated by ``3'') of removing the highest
redshift source (family 14).  The letter ``Z'' indicates the models
in which we do not use any information on the spectroscopic redshifts,
i.e. the redshifts of all families are free parameters in the
optimization.  
For completeness, we also quote results for the model F1a, in which all spectroscopic families are included, i.e. the model F1 with the addition of the families 11 and 8, and image 3b.
Although the model F1a has a worst overall fit due to the systematics introduced by the non bona-fide multiple images, it is more accurate to compute lensing quantities, such as the magnification, of these specific multiple images.
Finally, after the identification of the  fifth image, belonging to family 14 and close to the BCG, we include it into an additional model.
The model F1-5th considers the family sample 1 plus this extra image.
We therefore present best fit models for 22 different cases.

For a subset of models in Table \ref{tab:summary_bf}, we compute 
parameter uncertainties  by performing an MCMC analysis.  Since this
can be very time consuming, we do not consider models N1, N2, NW1 and NW2
because they do not describe accurately the properties of the lens
mass distribution (see Section
\ref{subsec:mass_distribution_parameters}).  We also exclude the
models Wa1 and Wa2 because we find that the multiple image positions are not very sensitive to the variation of the $w_a$ parameter.
In Table \ref{tab:cl_all}, we show the best-fit
parameters and their errors (68\% confidence level) for the 12 models for which
the MCMC analysis was performed (the model IDs refer to Table
\ref{tab:summary_bf}).
Note that for the models FZ1 and FZ2 we do not show the estimated
redshift of the family sample 1 for better
visualisation.  In the next sections, we discuss the results from the best
fit models on the mass distribution of RXC~J2248, the cosmological
parameters, and the degree of degeneracy among the different model
parameters.

\begin{table*}[!ht]
\centering
\small
\caption{Results of the MCMC statistical analysis for the strong lensing models of Table \ref{tab:summary_bf}.}
\begin{tabular}{|c|cc|cc|cc|cc|cc|cc|} \hline \hline
~ & \multicolumn{2}{c|}{ F1} & \multicolumn{2}{c|}{ F2 } & \multicolumn{2}{c|}{ W1 }& \multicolumn{2}{c|}{ W2}& \multicolumn{2}{c|}{ L1}& \multicolumn{2}{c|}{ L2}\\
    ~ & Median & $68\%$ CL & Median & $68\%$ CL & Median & $68\%$ CL & Median & $68\%$ CL& Median & $68\%$ CL& Median & $68\%$ CL \\ \hline
$x (\arcsec)$               &$-0.52$  &$_{-0.21}^{+0.22}$   &$-0.42$  &$_{-0.16}^{+0.17}$  &$-0.59$  &$_{-0.22}^{+0.21}$   &$-0.42$  &$_{-0.14}^{+0.15}$  &$-0.57$  &$_{-0.20}^{+0.20}$   &$-0.42$  &$_{-0.15}^{+0.15}$ \\
$y (\arcsec)$               &$0.54$   &$_{-0.14}^{+0.13}$   &$0.56$   &$_{-0.11}^{+0.10}$  &$0.56$   &$_{-0.14}^{+0.13}$   &$0.57$   &$_{-0.10}^{+0.11}$  &$0.54$   &$_{-0.13}^{+0.13}$   &$0.58$   &$_{-0.11}^{+0.11}$\\
$\varepsilon$          &$0.61$   &$_{-0.03}^{+0.03}$   &$0.59$   &$_{-0.01}^{+0.02}$  &$0.61$   &$_{-0.03}^{+0.03}$   &$0.58$   &$_{-0.02}^{+0.02}$  &$0.61$   &$_{-0.03}^{+0.03}$   &$0.58$   &$_{-0.02}^{+0.02}$\\
$\theta(\rm{deg})$          &$-37.43$ &$_{-0.22}^{+0.21}$   &$-37.29$ &$_{-0.12}^{+0.12}$  &$-37.45$ &$_{-0.20}^{+0.20}$   &$-37.29$ &$_{-0.12}^{+0.12}$  &$-37.44$ &$_{-0.20}^{+0.20}$   &$-37.29$ &$_{-0.12}^{+0.12}$\\
$r_{core}(\arcsec)$         &$19.95$  &$_{-1.26}^{+1.66}$   &$19.04$  &$_{-0.59}^{+0.69}$  &$21.33$  &$_{-1.85}^{+2.26}$   &$19.44$  &$_{-0.78}^{+0.89}$  &$21.35$  &$_{-1.76}^{+2.01}$   &$19.56$  &$_{-0.80}^{+0.91}$\\
$\sigma_v(\rm{km/s})$       &$1535$&$_{-16}^{+14}       $   &$1532$&$_{-13   }^{+9   }$ &$1540$&$_{-34   }^{+23   }$ &$1528$&$_{-32   }^{+21   }$&$1590$&$_{-54   }^{+34   } $&$1580$&$_{-53   }^{+29   }$\\
$\sigma_v^{BCG}(\rm{km/s})$ &$270  $ &$_{-126   }^{+99   }$&$166  $ &$_{-92   }^{+71   }$&$318  $ &$_{-114   }^{+97   }$&$181  $ &$_{-94  }^{+74  }$&$325  $ &$_{-113   }^{+97   }$&$191   $ &$_{-93   }^{+73   }$\\
$r_{cut}^{BCG}(\arcsec)$    &$86   $  &$_{-62   }^{+76   }$ &$95   $  &$_{-69   }^{+71   }$&$83   $  &$_{-65   }^{+80   }$ &$88   $  &$_{-72   }^{+76   }$&$83   $  &$_{-66   }^{+79   }$ &$93   $  &$_{-72   }^{+74   }$\\
$r_{cut}^{gals}(\arcsec)$   &$14.7 $  &$_{-7.4 }^{+16.3 }$  &$16.7 $  &$_{-6.1 }^{+11.2 }$ &$12.0 $  &$_{-5.6 }^{+10.3 }$  &$15.3 $  &$_{-5.3 }^{+8.4 }$  &$11.4 $  &$_{-5.3 }^{+8.7 }$   &$14.9 $  &$_{-5.3 }^{+8.3 }$\\
$\sigma_v^{gals}(\rm{km/s})$&$125  $ &$_{-21   }^{+24   }$ &$128  $ &$_{-14   }^{+14   }$&$132  $ &$_{-20   }^{+24   }$ &$130  $ &$_{-13   }^{+15   }$&$139  $ &$_{-20   }^{+26   }$ &$136  $ &$_{-14   }^{+16   }$\\
$\Omega_m$             &---      &---                  &---      &---                 &$0.25$   &$_{-0.14}^{+0.11}$   &$0.25$   &$_{-0.16}^{+0.13}$  &$0.31$   &$_{-0.13}^{+0.12}$   &$0.35$   &$_{-0.14}^{+0.11}$\\
$\Omega_\Lambda$       &---      &---                  &---      &---                 &---      &---                  &---      &---                 &$0.38$   &$_{-0.27}^{+0.38}$   &$0.36$   &$_{-0.26}^{+0.40}$\\
$w$                    &---      &---                  &---      &---                 &$-1.20$  &$_{-0.47}^{+0.24}$   &$-1.07$  &$_{-0.42}^{+0.16}$  &---      &---                  &---      &---\\
$z_{9}$                &---      &---                  &$2.48$   &$_{-0.05}^{+0.05}$  &---      &---                  &$2.48$   &$_{-0.06}^{+0.06}$  &---      &---                  &$2.47$   &$_{-0.05}^{+0.06}$\\
$z_{13}$               &---      &---                  &$1.27$   &$_{-0.03}^{+0.03}$  &---      &---                  &$1.26$   &$_{-0.03}^{+0.03}$  &---      &---                  &$1.26$   &$_{-0.03}^{+0.03}$\\
$z_{15}$               &---      &---                  &$3.14$   &$_{-0.09}^{+0.10}$  &---      &---                  &$3.14$   &$_{-0.10}^{+0.11}$  &---      &---                  &$3.12$   &$_{-0.09}^{+0.10}$\\
$z_{16}$               &---      &---                  &$1.43$   &$_{-0.02}^{+0.02}$  &---      &---                  &$1.43$   &$_{-0.02}^{+0.02}$  &---      &---                  &$1.43$   &$_{-0.02}^{+0.02}$\\
$z_{17}$               &---      &---                  &$2.39$   &$_{-0.06}^{+0.05}$  &---      &---                  &$2.41$   &$_{-0.07}^{+0.07}$  &---      &---                  &$2.40$   &$_{-0.06}^{+0.06}$\\
$z_{20}$               &---      &---                  &$3.11$   &$_{-0.10}^{+0.11}$  &---      &---                  &$3.11$   &$_{-0.11}^{+0.11}$  &---      &---                  &$3.09$   &$_{-0.10}^{+0.11}$\\
$z_{21}$               &---      &---                  &$3.49$   &$_{-0.12}^{+0.13}$  &---      &---                  &$3.51$   &$_{-0.13}^{+0.14}$  &---      &---                  &$3.49$   &$_{-0.12}^{+0.14}$\\
\hline\hline
~ & \multicolumn{2}{c|}{ FZ1} & \multicolumn{2}{c|}{ FZ2 } & \multicolumn{2}{c|}{ W3 }& \multicolumn{2}{c|}{ L3}& \multicolumn{2}{c|}{ WL1}& \multicolumn{2}{c|}{ WL2}\\
    ~ & Median & $68\%$ CL & Median & $68\%$ CL & Median & $68\%$ CL & Median & $68\%$ CL& Median & $68\%$ CL& Median & $68\%$ CL \\ \hline
$x (\arcsec)$               &$-0.62$  &$_{-0.55}^{+0.49}$   &$-0.31$  &$_{-0.30}^{+0.31}$  &$-0.49$  &$_{-0.20}^{+0.19}$   &$-0.46$  &$_{-0.19}^{+0.19}$  &$-0.56$  &$_{-0.23}^{+0.23}$   &$-0.42$  &$_{-0.15}^{+0.15}$ \\
$y (\arcsec)$               &$0.56$   &$_{-0.37}^{+0.40}$   &$0.42$   &$_{-0.24}^{+0.24}$  &$0.59$   &$_{-0.13}^{+0.14}$   &$0.59$   &$_{-0.13}^{+0.13}$  &$0.54$   &$_{-0.14}^{+0.14}$   &$0.57$   &$_{-0.10}^{+0.10}$\\
$\varepsilon$          &$0.69$   &$_{-0.05}^{+0.05}$   &$0.61$   &$_{-0.02}^{+0.03}$  &$0.59$   &$_{-0.02}^{+0.03}$   &$0.58$   &$_{-0.02}^{+0.03}$  &$0.61$   &$_{-0.03}^{+0.03}$   &$0.58$   &$_{-0.01}^{+0.02}$\\
$\theta(\rm{deg})$          &$-37.43$ &$_{-0.33}^{+0.32}$   &$-37.23$ &$_{-0.14}^{+0.14}$  &$-37.19$ &$_{-0.14}^{+0.14}$   &$-37.21$ &$_{-0.14}^{+0.14}$  &$-37.44$ &$_{-0.21}^{+0.21}$   &$-37.29$ &$_{-0.12}^{+0.12}$\\
$r_{core}(\arcsec)$         &$21.07$  &$_{-2.38}^{+2.89}$   &$19.50$  &$_{-1.11}^{+1.20}$  &$19.89$  &$_{-1.02}^{+1.23}$   &$19.80$  &$_{-0.91}^{+1.11}$  &$21.46$  &$_{-1.89}^{+2.15}$   &$19.56$  &$_{-0.77}^{+0.89}$\\
$\sigma_v(\rm{km/s})$       &$1487$&$_{-35   }^{+29   }$ &$1495$&$_{-20   }^{+15   }$&$1565$&$_{-34   }^{+24   }$ &$1575$&$_{-53   }^{+33   }$&$1600$&$_{-42   }^{+29   }$ &$1590$&$_{-36   }^{+23   }$\\
$\sigma_v^{BCG}(\rm{km/s})$ &$383  $ &$_{-96   }^{+78   }$ &$243  $ &$_{-72   }^{+60   }$&$244  $ &$_{-134   }^{+106   }$&$219   $ &$_{-131}^{+111}    $&$334  $ &$_{-114   }^{+92   }$&$199  $ &$_{-89   }^{+81   }$\\
$r_{cut}^{BCG}(\arcsec)$    &$101   $ &$_{-67   }^{+66   }$ &$94   $  &$_{-71   }^{+71   }$&$96   $  &$_{-68   }^{+71   }$ &$97   $  &$_{-73   }^{+71   }$&$90   $  &$_{-68   }^{+75   }$ &$90   $  &$_{-76   }^{+76   }$\\
$r_{cut}^{gals}(\arcsec)$   &$28.1 $  &$_{-16.2 }^{+72.8 }$ &$16.1 $  &$_{-6.1 }^{+16.1 }$ &$11.7 $  &$_{-4.3 }^{+6.8 }$   &$12.2 $  &$_{-4.4 }^{+7.9 }$  &$12.3 $  &$_{-5.9 }^{+12.1 }$  &$14.7 $  &$_{-4.8 }^{+7.1 }$\\
$\sigma_v^{gals}(\rm{km/s})$&$116  $ &$_{-30   }^{+31   }$ &$134  $ &$_{-21   }^{+18   }$&$151  $ &$_{-20   }^{+23   }$ &$147  $ &$_{-20   }^{+22   }$&$137  $ &$_{-23   }^{+27   }$ &$137  $ &$_{-13   }^{+15   }$\\
$\Omega_m$             &---      &---                  &---      &---                 &$0.49$   &$_{-0.26}^{+0.26}$   &$0.41$   &$_{-0.18}^{+0.20}$  &$0.32$   &$_{-0.16}^{+0.12}$   &$0.33$   &$_{-0.19}^{+0.12}$\\
$\Omega_\Lambda$       &---      &---                  &---      &---                 &---      &---                  &$0.40$   &$_{-0.29}^{+0.39}$  &$0.29$   &$_{-0.20}^{+0.32}$   &$0.32$   &$_{-0.23}^{+0.30}$\\
$w$                    &---      &---                  &---      &---                 &$-1.07$  &$_{-0.57}^{+0.42}$   &---      &---                 &$-0.97$  &$_{-0.67}^{+0.61}$   &$-0.83$  &$_{-0.56}^{+0.41}$\\
$z_{9}$                &---      &---                  &$3.30$   &$_{-0.19}^{+0.21}$  &$2.39$   &$_{-0.07}^{+0.09}$   &$2.43$   &$_{-0.08}^{+0.09}$  &---      &---                  &$2.46$   &$_{-0.06}^{+0.06}$\\
$z_{13}$               &---      &---                  &$1.42$   &$_{-0.05}^{+0.06}$  &$1.26$   &$_{-0.03}^{+0.03}$   &$1.26$   &$_{-0.03}^{+0.03}$  &---      &---                  &$1.26$   &$_{-0.03}^{+0.03}$\\
$z_{15}$               &---      &---                  &$4.75$   &$_{-0.37}^{+0.38}$  &$2.92$   &$_{-0.12}^{+0.16}$   &$3.00$   &$_{-0.14}^{+0.19}$  &---      &---                  &$3.10$   &$_{-0.09}^{+0.10}$\\
$z_{16}$               &---      &---                  &$1.60$   &$_{-0.07}^{+0.07}$  &$1.43$   &$_{-0.02}^{+0.02}$   &$1.43$   &$_{-0.02}^{+0.02}$  &---      &---                  &$1.43$   &$_{-0.02}^{+0.02}$\\
$z_{17}$               &---      &---                  &$3.03$   &$_{-0.21}^{+0.23}$  &$2.31$   &$_{-0.08}^{+0.10}$   &$2.36$   &$_{-0.08}^{+0.09}$  &---      &---                  &$2.39$   &$_{-0.06}^{+0.06}$\\
$z_{20}$               &---      &---                  &$4.68$   &$_{-0.38}^{+0.40}$  &$2.93$   &$_{-0.13}^{+0.16}$   &$3.00$   &$_{-0.14}^{+0.18}$  &---      &---                  &$3.08$   &$_{-0.10}^{+0.11}$\\
$z_{21}$               &---      &---                  &$5.62$   &$_{-0.53}^{+0.59}$  &$3.25$   &$_{-0.16}^{+0.21}$   &$3.36$   &$_{-0.18}^{+0.25}$  &---      &---                  &$3.47$   &$_{-0.12}^{+0.14}$\\
\hline\hline
    \end{tabular}
\label{tab:cl_all}
\tablefoot{IDs correspond to the models in the Table \ref{tab:summary_bf}. For the models FZ1 and FZ2 the best-fit redshift values of the families 2, 3, 4, 6, 7, 14 and 18 are omitted to improve visualisation. The values of all velocity dispersions ($\sigma_v$) are corrected by the factor $\sqrt{2/3}$ as described in the \emph{lenstool} manual (see \url{http://projets.lam.fr/projects/lenstool/wiki/PIEMD}).}
\end{table*}
\subsection{Mass distribution parameters}
\label{subsec:mass_distribution_parameters}

Firstly, we notice that the PNFW models provide a significantly worse
fit than the PIEMD ones (compare models F1 and F2 with N1 and N2 in
Table \ref{tab:summary_bf}). The final positional $\Delta_{rms}$ of N1
and N2 is a factor of 3.5 and 3.9 higher than that of F1 and F2,
respectively.  The main reason for this difference is that RXC~J2248 is
characerized by a relatively shallow inner mass density distribution,
as pointed out by previous works
\citep{2014ApJ...797...48J,2014MNRAS.438.1417M}.  Moreover, when we
let the values of the cosmological parameters, $\Omega_m$ and $w$, to
vary in the model optimization, the $\Delta_{rms}$ is reduced by a
factor of 2 for the NW1 and NW2 models.  This indicates that a different
set of cosmological parameters partially compensates the effects
of the presence of a core in the total mass density profile of the
cluster.  Specifically, we obtain best fit cosmological parameters
$\Omega_m \approx 0.0$ and $w<-2.0$, which are completely non physical
and in disagreement with other cosmological probes.  Moreover, the
large $\Delta_{rms}$ value in this case
indicates that this mass distribution
profile is such a bad representation of the real one that the lensing
models are unable to probe parameters related to the background cosmology
and halo substructures.

By comparing the models with fixed cosmology using the redshift
information of the family sample 1, i.e. models F1 and F2 in the
Tables \ref{tab:summary_bf} and \ref{tab:cl_all}, we notice that the extra information of the family sample 2 does not significantly change the strong lens modelling.
This is indicated by the
fact that the $\Delta_{rms}$ values of these two models are very
similar (see Table \ref{tab:summary_bf}) and the values of all
parameters are also consistent within their $1\sigma$ confidence
levels in Table \ref{tab:cl_all}.  A larger deviation is obtained
for the BCG parameters $\sigma_v^{BCG}$ and $r_{cut}^{BCG}$, which are
not estimated very precisely due to the degeneracies with the other
parameters and the lack of multiple images close to the BCG.
This behaviour is present in all the 12 different models
we analysed.
We remark that the inclusion of image 14e allows us to obtain a more precise estimate of the value of the effective velocity dispersion of the BCG, but not of its truncation radius.
In detail, we find that the median values and the 68\% confidence levels for these two parameters from the model F1-5th are $\sigma_v^{BCG}=363^{+25}_{-26}$ $\rm km/s$ and $r_{cut}^{BCG} = 75_{-52}^{+78}$$\arcsec$.

Interestingly, the addition of extra families of the sample 2 allows
us to reduce the errors on the best-fitting parameters
and to place significant constraints on the redshifts of these
extra multiple image families.
We consider the model F2 as our reference model, since we use here the maximum possible and secure information of the clean sample of multiple images.
In Figure \ref{fig:src_lens_planes}, we show $4\arcsec$ wide cutouts of the multiple images used in this model.
The red circles have $0\arcsec.5$ radius and locate the observed input positions listed in Table \ref{tab:families}. The yellow crosses are the predicted positions of the lens model F2.
We notice that all multiple images are very well reproduced by the model and there is no systematic offset in the predicted positions.
\begin{figure}[!t]
  \includegraphics[scale=0.5]{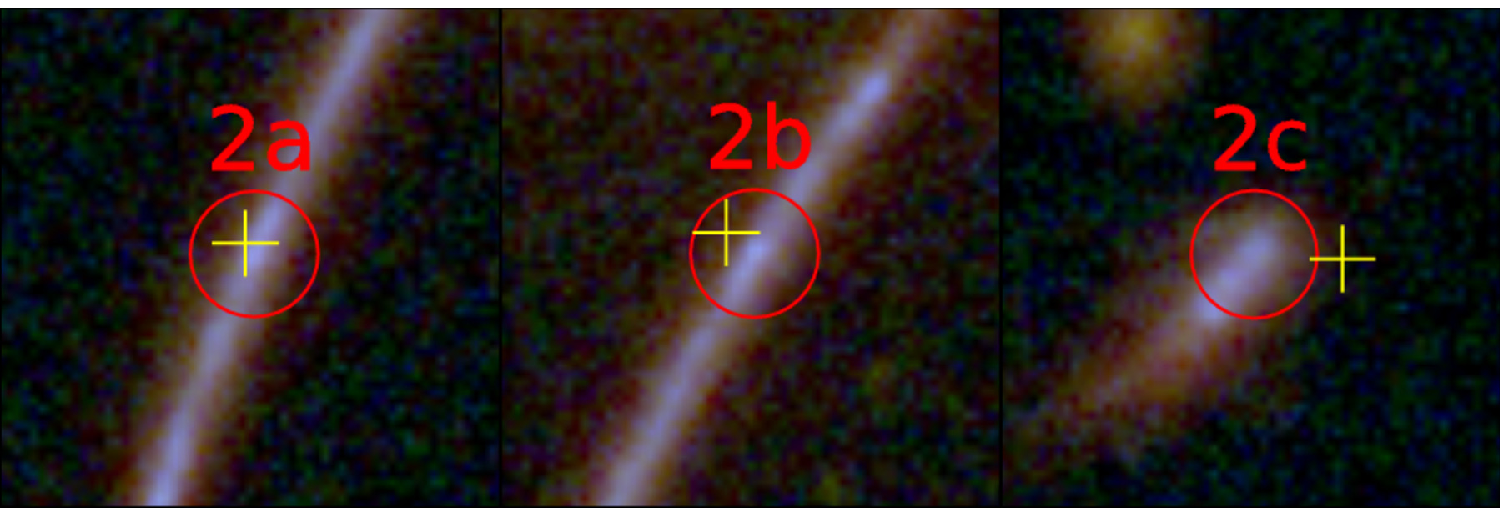}
  \includegraphics[scale=0.5]{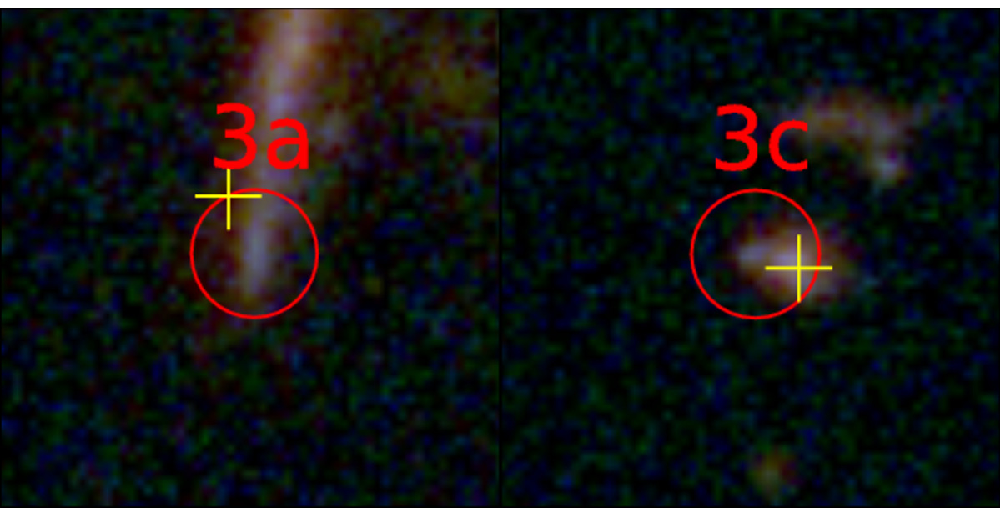}
  \includegraphics[scale=0.5]{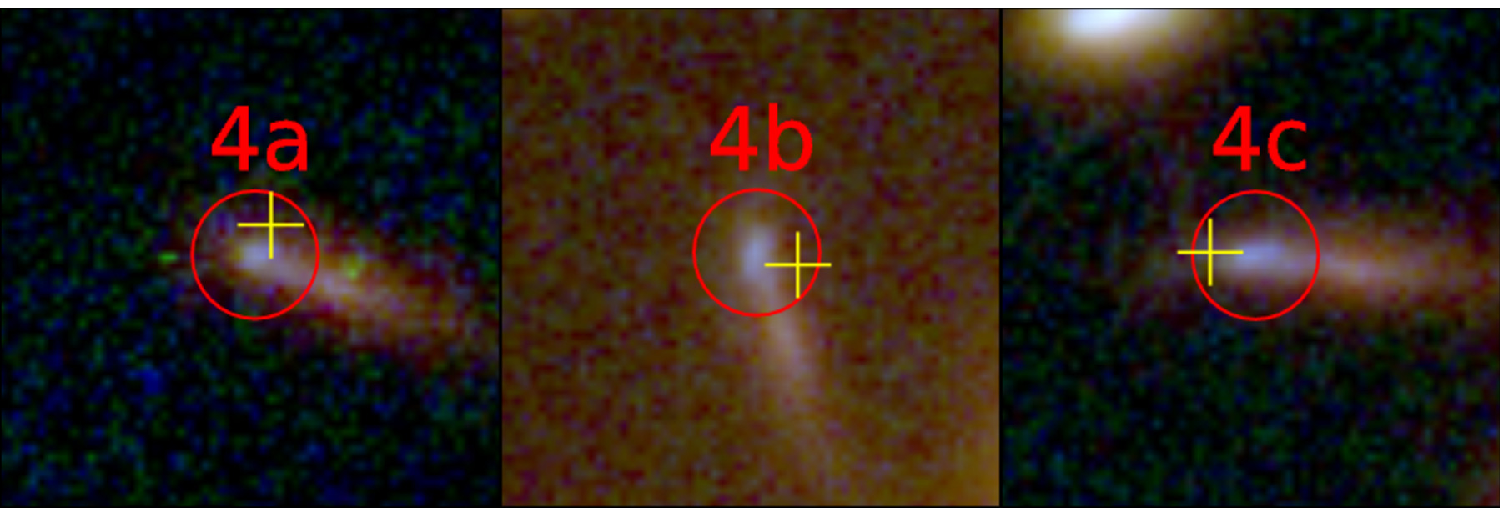}
  \includegraphics[scale=0.5]{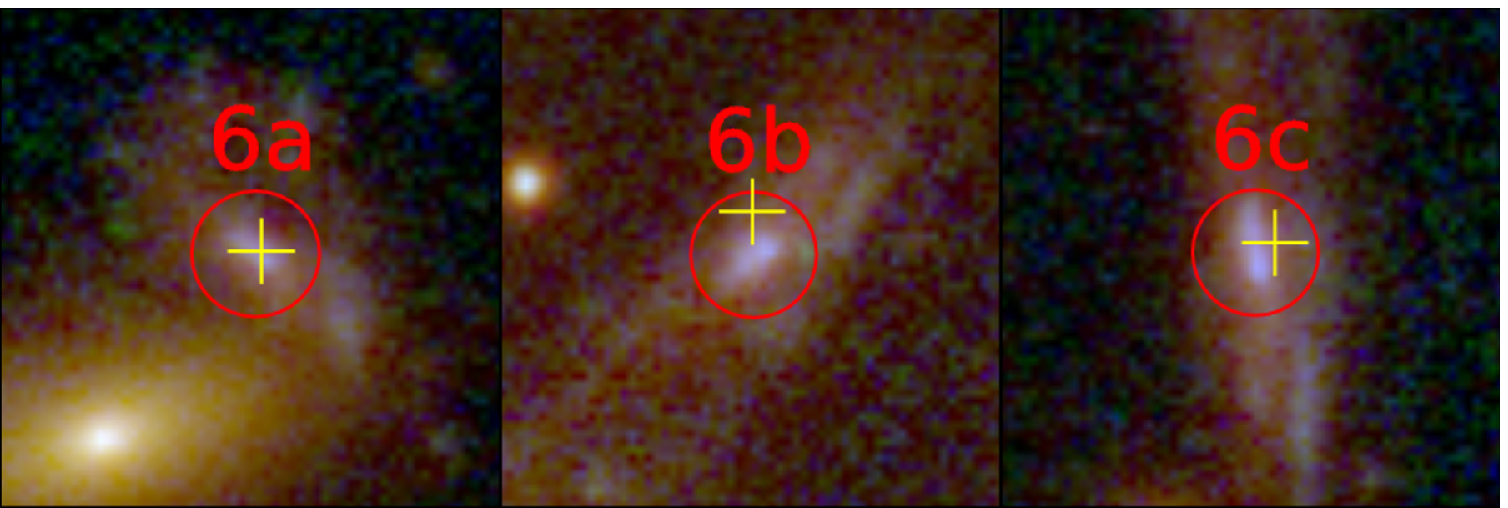}
  \includegraphics[scale=0.5]{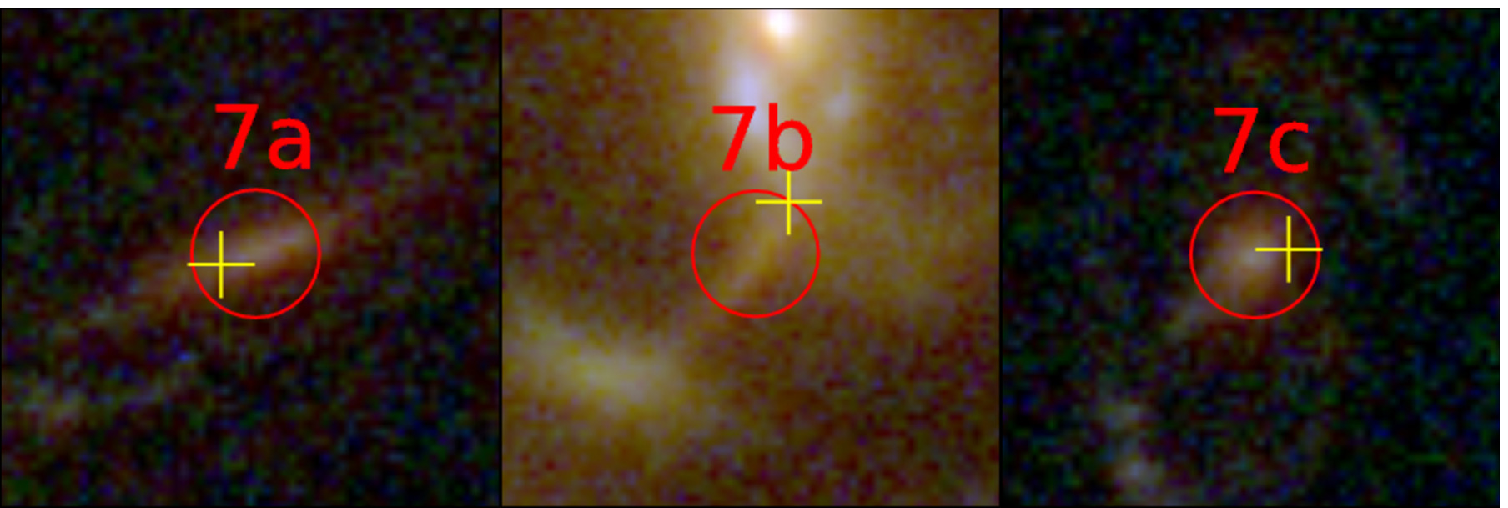}
  \includegraphics[scale=0.5]{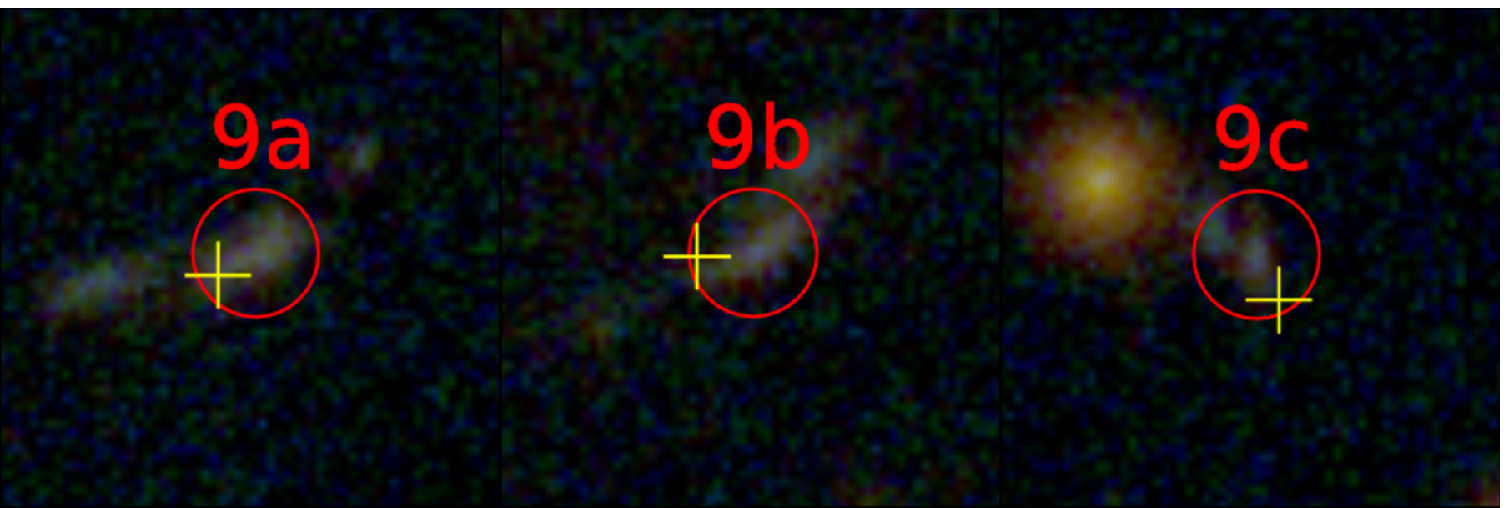}
  \includegraphics[scale=0.5]{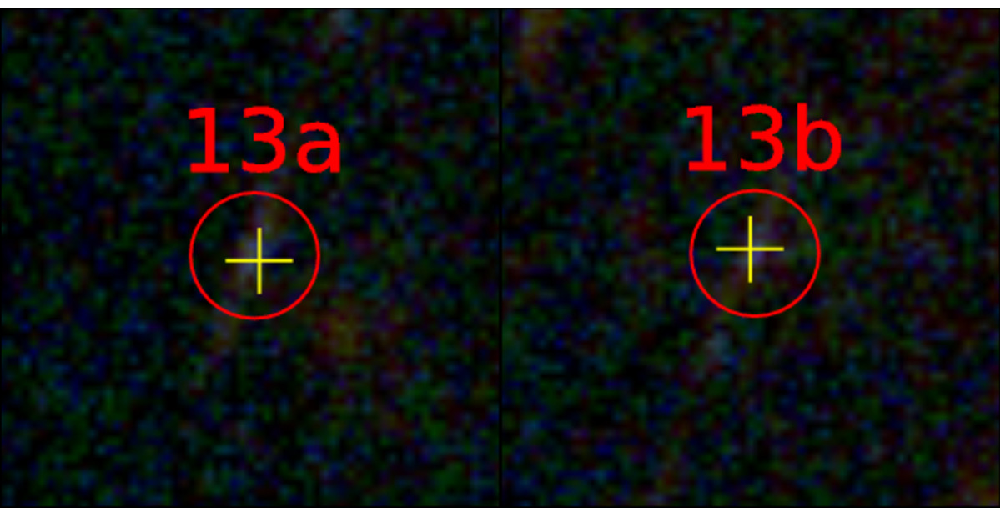}
  \caption{CLASH/\emph{HST} color cutouts ($4\arcsec$ wide) of all multiple images
    used in the reference model F2. Red circles ($0\arcsec.5$ radius)
    indicate the observed positions while the yellow crosses the model-predicted positions. The multiple image ID 14e is shown in the HFF cutout and the best fitting position (blue cross) is given by the model F1-5th.}
  \label{fig:src_lens_planes}
\end{figure}
\begin{figure}[!t]
  \ContinuedFloat
  \includegraphics[scale=0.5]{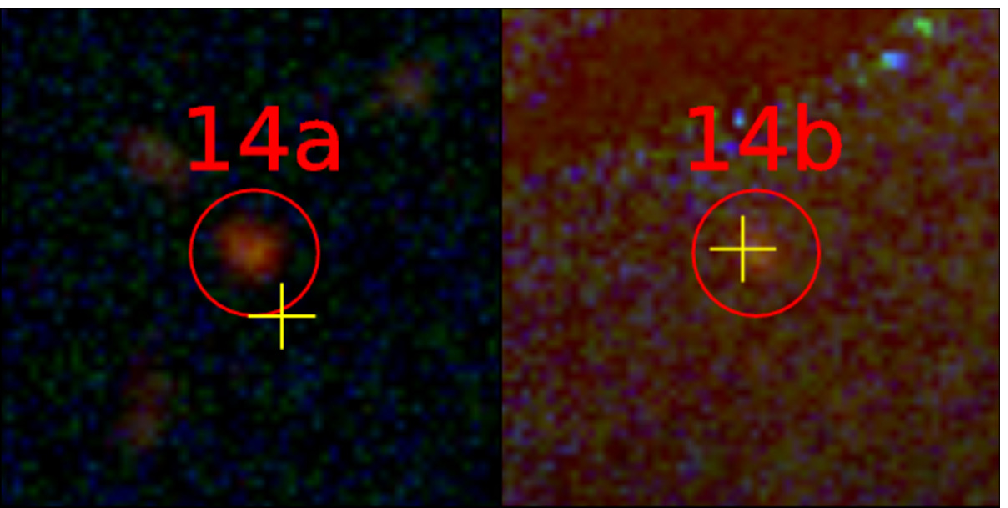}
  \includegraphics[width = 0.848\columnwidth]{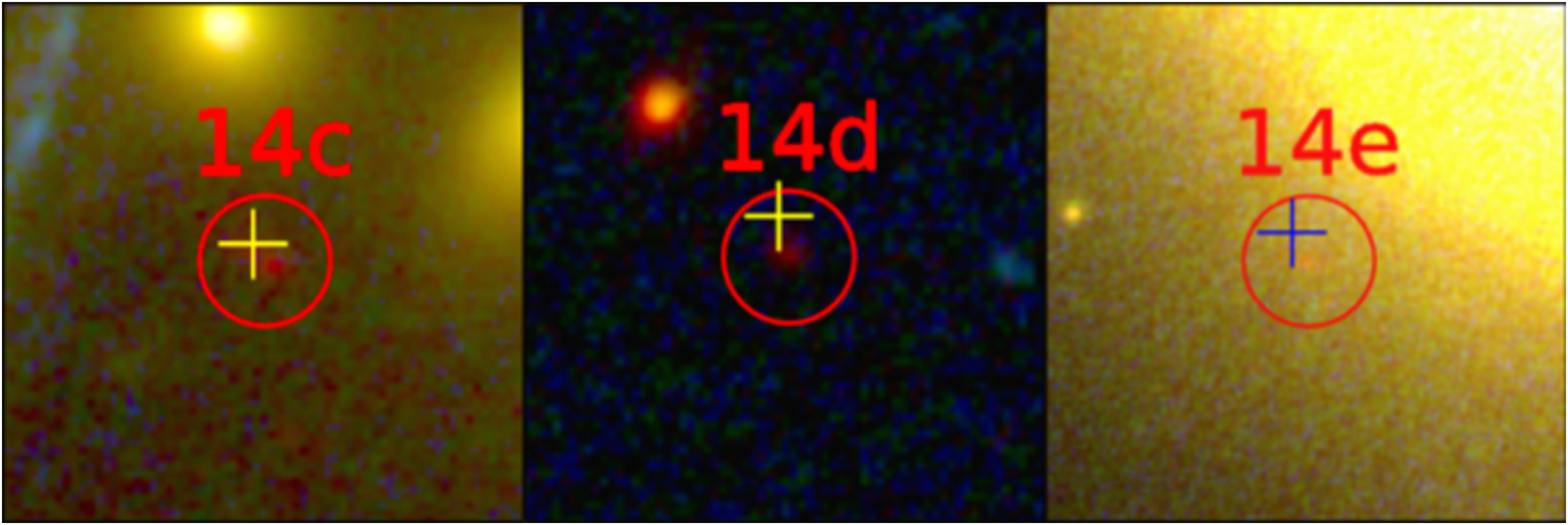}
  \includegraphics[scale=0.5]{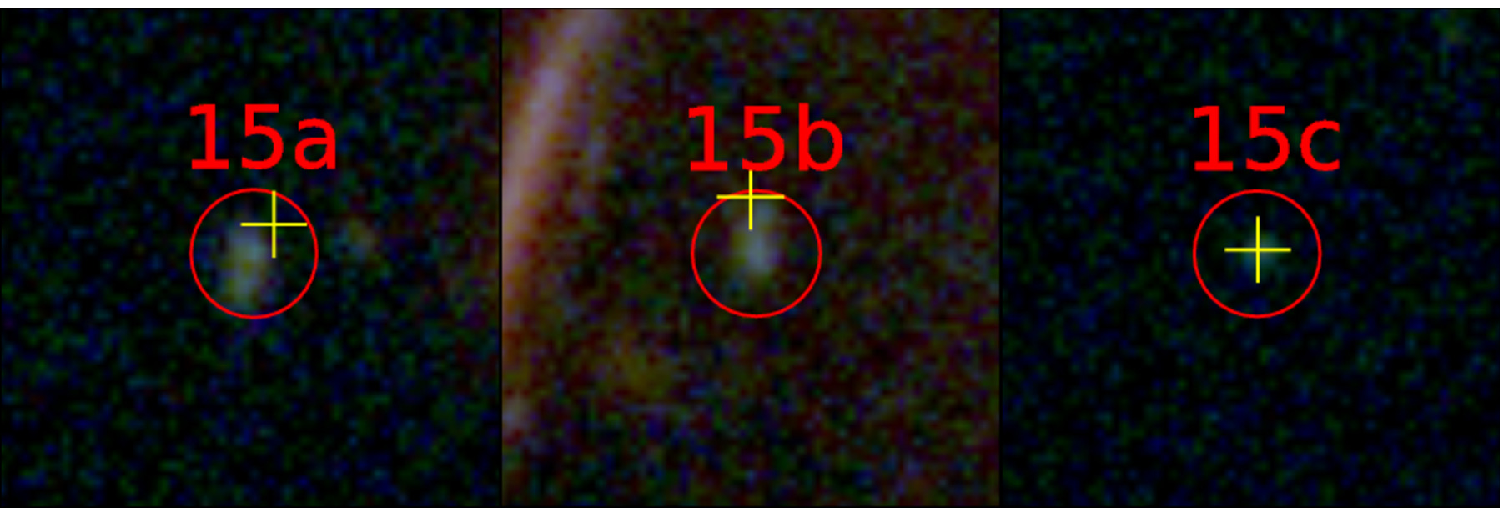}
  \includegraphics[scale=0.5]{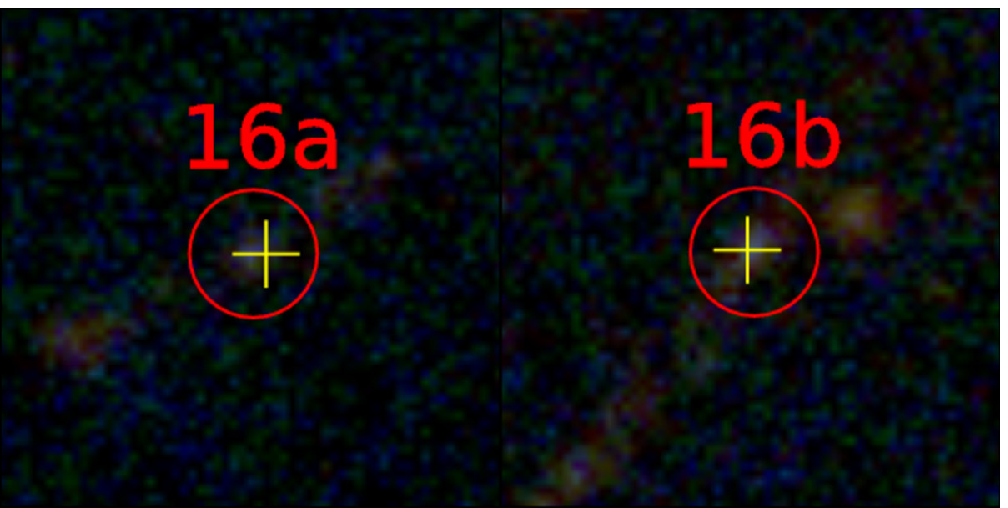}
  \includegraphics[scale=0.5]{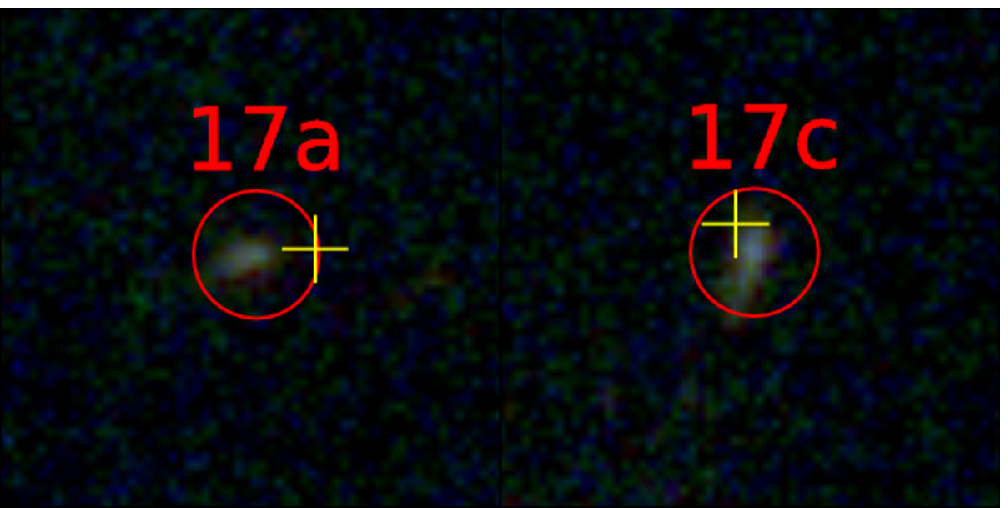}
  \includegraphics[scale=0.5]{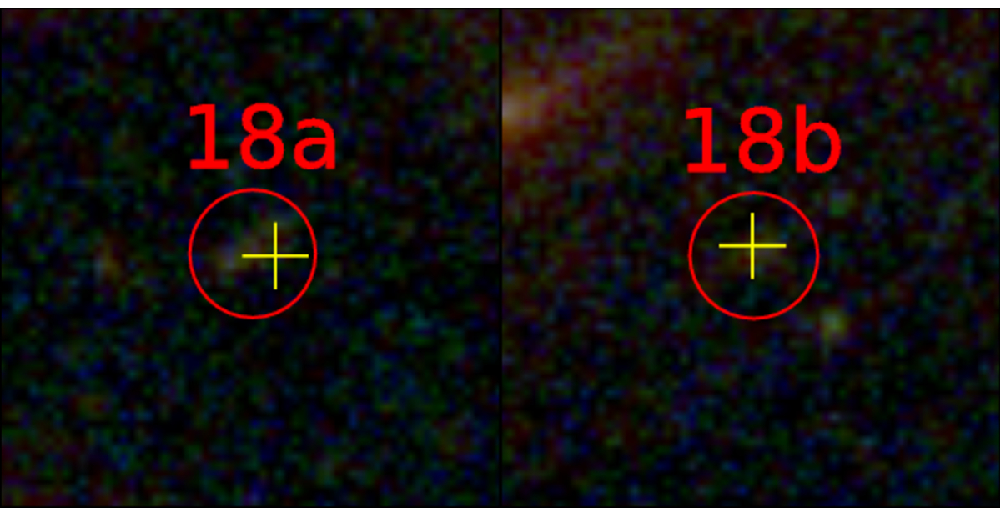}
  \includegraphics[scale=0.5]{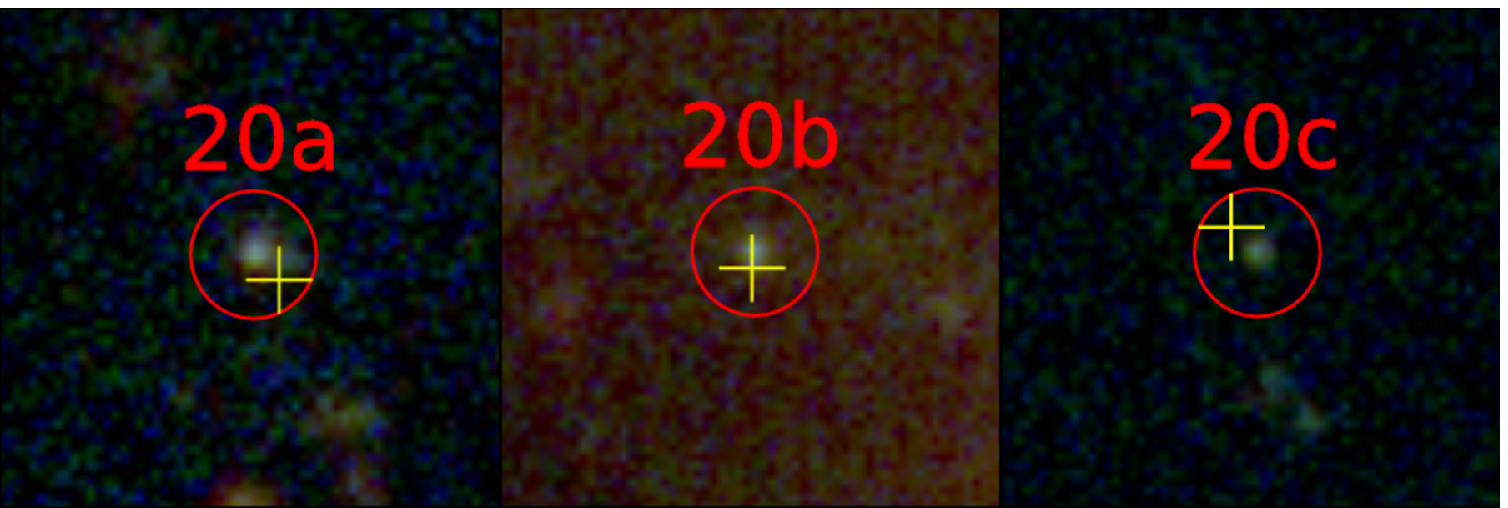}
  \includegraphics[scale=0.5]{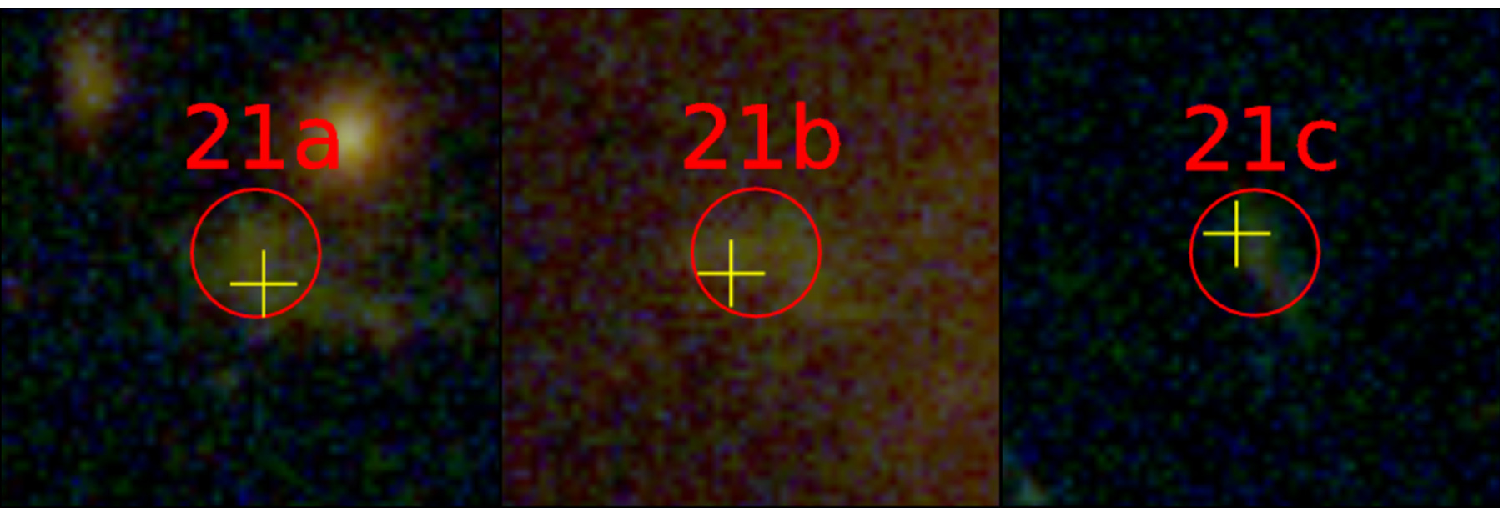}
  \caption{(Continued)}
  \label{fig:src_lens_planes}
\end{figure}

\begin{figure}[!ht]
  \centering
  \includegraphics[width = .5\textwidth]{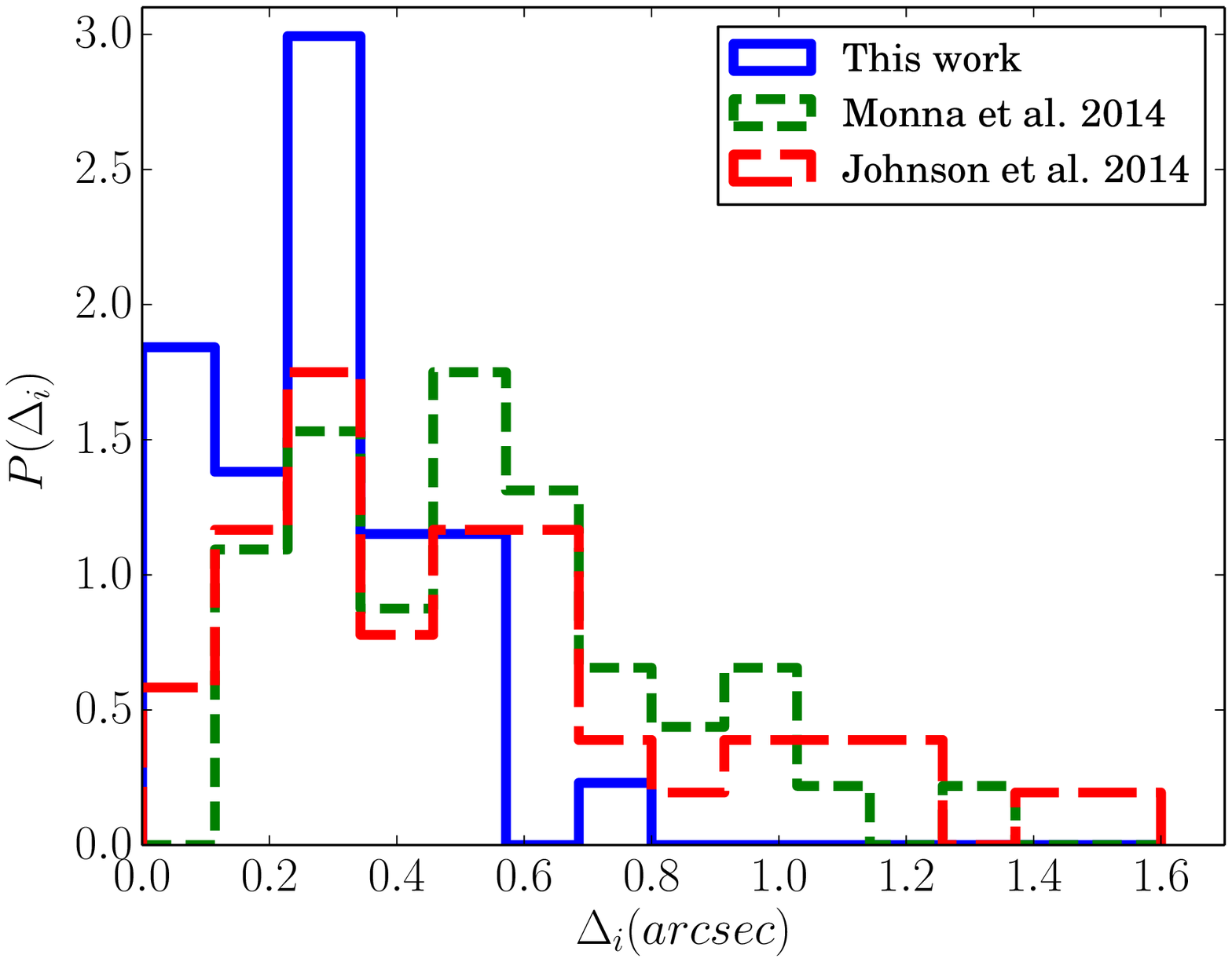}
  \caption{Distribution of the displacement values of the multiple images (absolute values of the observed minus the reconstructed positions, see Equation \ref{eq:rms_singe}) obtained from our reference model F2 (solid blue line) and in previous works by \citet{2014MNRAS.438.1417M} (green dashed) and \citet{2014ApJ...797...48J} (red long dashed), for the cluster RXC~J2248.}
  \label{fig:rms_f2}
\end{figure}
In Figure \ref{fig:rms_f2}, we compare the distribution of the
displacements of each multiple image ($\Delta_i$, see Equation
\ref{eq:rms_singe}) of the reference model F2 with those relative to
the multiple images considered in \citet{2014MNRAS.438.1417M} and
\citet{2014ApJ...797...48J} (the only studies that made this
information available).  This figure shows that our model reproduces
the multiple image positions with better accuracy compared to previous
works.  Specifically, the final $\Delta_{rms}$ (Equation
\ref{eq:rms_def}) of our model is $0\arcsec.31$, while is
$0\arcsec.61$ in \citet{2014MNRAS.438.1417M} and $0\arcsec.64$ in
\citet{2014ApJ...797...48J}.  Note that these models use different
assumptions, such as extra dark matter halos and different cluster
member selections, slightly distinct multiple image families and
different redshift information.

We can compare the projected total mass values within an aperture of
$250\, {\rm kpc}$ from these studies.  Using the $1\sigma$ confidence
level of our F2 model, we find
$2.90_{-0.02}^{+0.02}\times10^{14}M_{\odot}$ (the errors are given by the 68\% confidence level), somewhat higher than the
values of $2.68^{+0.03}_{-0.05}\times10^{14}M_{\odot}$ and
$2.67^{+0.08}_{-0.08}\times10^{14}M_{\odot}$ presented in
\citet{2014ApJ...797...48J} and \citet{2014MNRAS.438.1417M},
respectively.  Although these measurements are not consistent within
the estimated errors, the mean values do not differ more than 10\%,
and are likely due to different assumptions in these studies, as well as
aforementiond systematics arising from a non bona-fide set of multiple
images.

Since strong lensing modelling in galaxy clusters is often not supported by
extensive spectroscopy of lensed background sources, we
examine the impact of not using spectroscopic information in the
lens modelling.  We initially compute the best fit model assuming all
families's redshifts as free parameters for a fixed cosmology (models FZ1 and FZ2), and varying $\Omega_m$ and $w$ (models WZ1 and WZ2, 
see Table \ref{tab:summary_bf}).  Comparing models F1 and F2 with
FZ1 and FZ2, we note the value of $\sigma_v$ decreases by $\approx
3\%$ for the models F1 and FZ1, and $\approx 2\%$ for F2 and FZ2.
Even within the $1\sigma$ confidence level, this difference is more
likely to be caused by systematics introduced by the missing redshift
information than statistical fluctuations.
\begin{figure}[!ht]
  \centering
  \includegraphics[width = .5\textwidth]{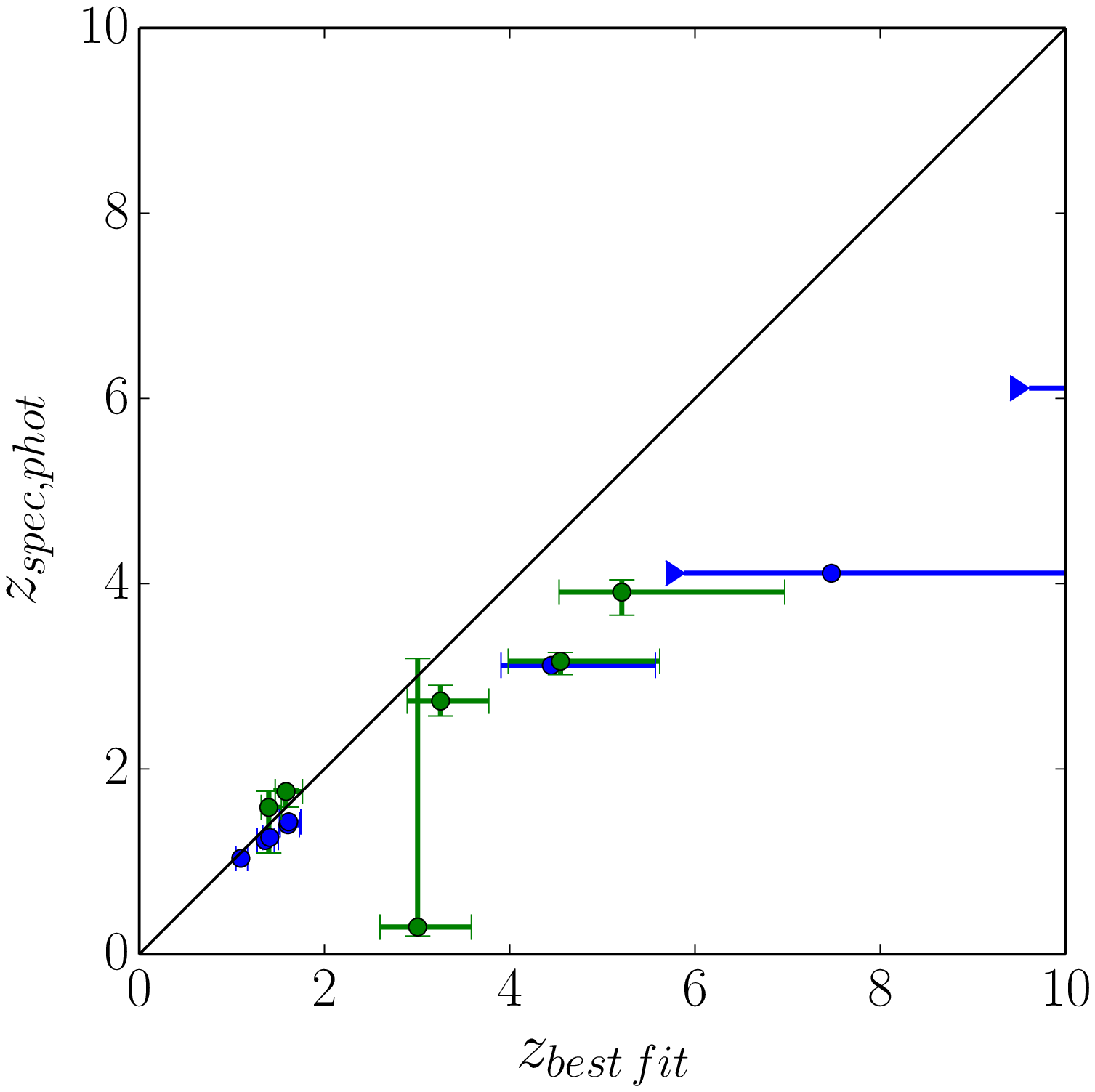}
  \caption{Best fit redshift values of the multiple image families compared
    with the spectroscopic, in blue, and photometric, in green, redshift values. The arrows indicate the unconstrained redshifts and the
    black line the relation $z_{spec,phot} = z_{best\,fit}$. For
    family 20 we use the spectroscopic redshift value measured from the
    Ly$\alpha$ blob.}
  \label{fig:zlens_zspec}
\end{figure}

There is a well-known degeneracy between the mass of a lens
(parametrised by $\sigma_v$) and the distance of a lensed
source. Simplifying, as the source distance increases the lens mass has to
decrease in order to match the same multiple image positions. From
Table \ref{tab:summary_bf}, the best-fitting redshift values of the
model FZ2 are all systematically larger than those of the model F2.
In Figure \ref{fig:zlens_zspec}, we compare the model-predicted
redshifts ($z_{best\,fit}$) of all the 14 multiple image families with
the spectroscopic (blue marks) and photometric (green) estimates from
\citet{2014A&A...562A..86J}, with $95\%$ confidence level error
bars. To associate only one photometric redshift value with each
family, we choose the multiple image with the highest value of the
\emph{odds} parameter from the photo-\emph{z} algorithm \citep[see
Section 3.3 of ][]{2014A&A...562A..86J}.  As expected, for the low
redshift families ($z_{spec,phot} < 2$) the agreement between the
model predictions and the measurements is very good.  However, for
families at higher redshifts the difference increases significantly
and progressively, always leading to overestimate the redshift value.
For families with $z_{spec,phot} > 4$, redshifts are basically
unconstrained, indicating that spectroscopic measurements for these
sources become critical to avoid significant systematic uncertainties
on the mass (and cosmological) parameters.
\begin{figure}[!ht]
  \centering
  \includegraphics[width=.475\textwidth]{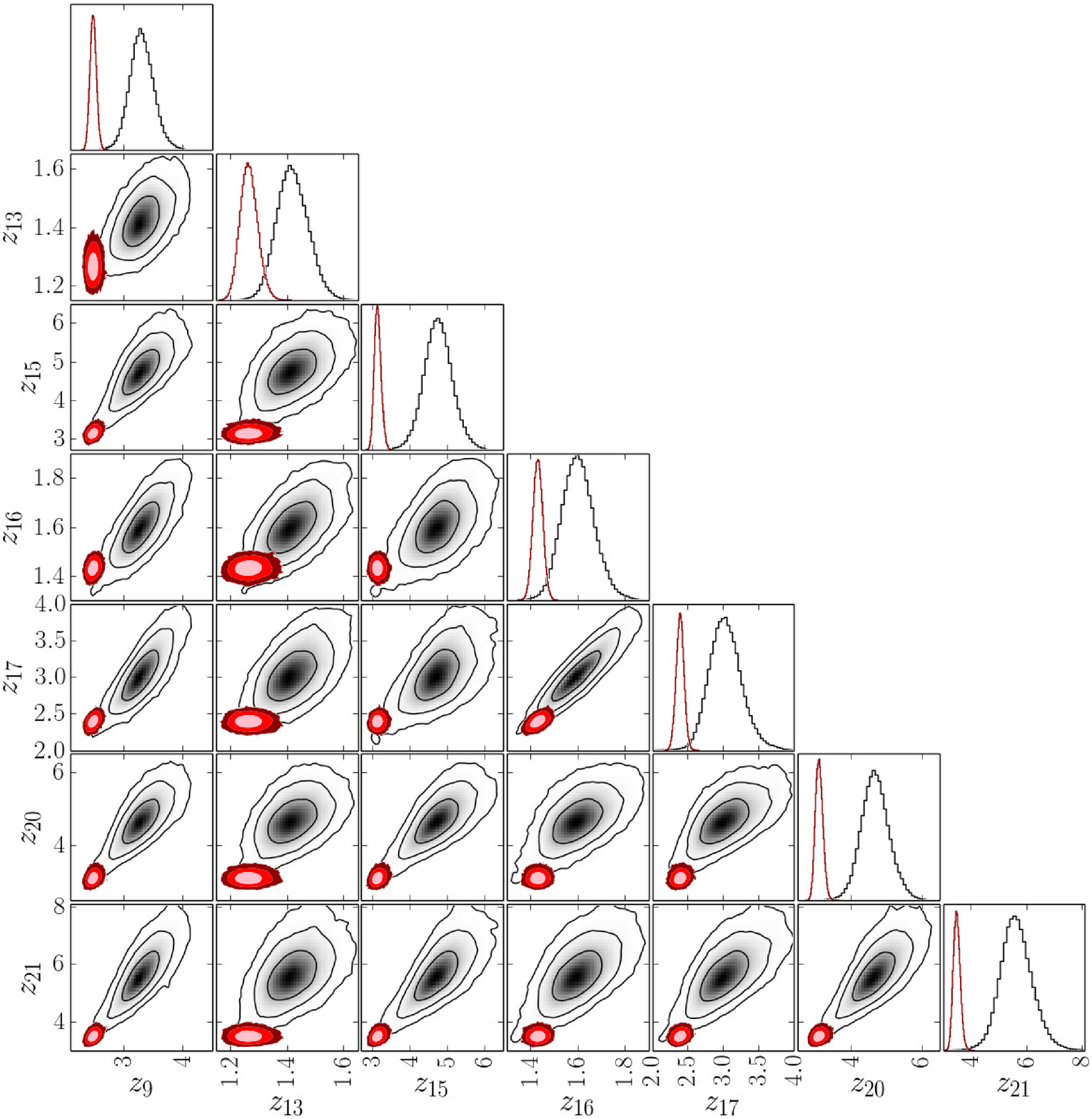}
  \caption{Confidence regions for the redshifts values of the
    families with no spectroscopic confirmation. Red regions: the
    redshift values of the spectroscopically confirmed families are
    fixed. Grey regions: the redshift values of all families are left
    free. These correspond to models F2 and FZ2, respectively, in Table \ref{tab:summary_bf}. The contours represent the $68\%$,
    $95.4\%$ and $99.7\%$ confidence levels.}
  \label{fig:comp_z}
\end{figure}

In Figure \ref{fig:comp_z}, we show the confidence regions, estimated from the MCMC analysis, of the best-fit redshifts for the model FZ2 (in grey) and F2 (in red).
We note that in the model FZ2 (all redshifts left free) the redshift
values are all strongly correlated.  This effect becomes larger, in
absolute values, for the sources at higher redshifts.
For the model F2, the confidence regions are much smaller and the
correlation much less pronounced.  Moreover, the overlap of the
confidence regions for the two models occur only at low redshifts, and
only in the $3\sigma$ area of the model FZ2, indicating again the bias
introduced by the lack of spectroscopic information.
We notice that the absence of information about the source redshifts results in a best-fitting model with a lower total mass for the cluster that is compensated by higher values for the source redshifts.
The degeneracy between the total cluster mass and source redshifts explains
the difference of $\approx 3\%$ in the value of the effective velocity
dispersion ($\sigma_v$), linked to the total cluster mass, of the
models F1, F2 and FZ1, FZ2.  For the model FZ2, we find a total mass
projected within $250 \,{\rm kpc}$ of
$2.78_{-0.02}^{+0.02}\times10^{14}M_{\odot}$, a difference of
approximately 4\% when compared with to F2.  This shows that the
measurements of the projected total mass are similar, despite the
large redshift bias.
On the contrary, since the best-fit redshift values are biased, we
expect that quantities that depend directly on cosmological distances,
such as $\Xi$ in Equation \ref{eq:cosmo_prober}, will also be biased
if spectroscopic redshifts are not available.

By leaving the redshift values of all families free, we increase the
number of free parameters by 7 and 14 for family sample 1 and 2,
respectively. Clearly, the larger number of free parameters reduces
the value of the final $\Delta_{rms}$ (and consequently
$\chi^2_{min,ref}$), but biases the recovered parameters, principally
the cosmological ones.  For the models WZ1 and WZ2, the best-fit
cosmological parameters are $\Omega_m=$ 1.0 and 0.6, and $w=$ $-1.4$
and $-1.3$, respectively.  These values are in disagreement with other
established cosmological probes, showing that missing information on
the background source redshifts makes cosmological constraints
unreliable.
In Section \ref{sec:cosmological_constraints}, we however show that if
one starts with a large sample of spectroscopically confirmed multiple
image families, the addition of more secure families with no redshift
information does not bias the estimates of the cosmological
parameters.

\subsection{Cosmological parameters}
\label{sec:cosmological_constraints}

We focus here on the ability of the lensing model to constrain the
cosmological parameters, by considering three different $\rm \Lambda
CDM$ models: 1) a flat cosmological model with free $\Omega_m$ and $w$
(ID~W); 2) a cosmological model with fixed $w=-1$, but allowing
different curvature values, i.e. free $\Omega_m$ and $\Omega_\Lambda$
(ID L); 3) a cosmological model with the three parameters free,
$\Omega_m$, $\Omega_{\Lambda}$, and $w$, (ID WL).

From Table \ref{tab:summary_bf}, we notice that the models with fixed
cosmological parameters, F1 and F2, have larger $\Delta_{rms}$ values
than those allowing some freedom in the background cosmological model,
showing the leverage of the cosmological parameters on the multiple
image positions.
For instance, the reduced $\chi^2$ ($\chi^2_{min,ref}/DOF$) decreases by $\approx 13\%$ when we compare the model F1, including the spectroscopic confirmed families with fixed cosmology, with the models W1 and L1, where the value of the cosmological parameters are left free.

\begin{figure}[!ht]
  \centering
  \includegraphics[width = .242\textwidth]{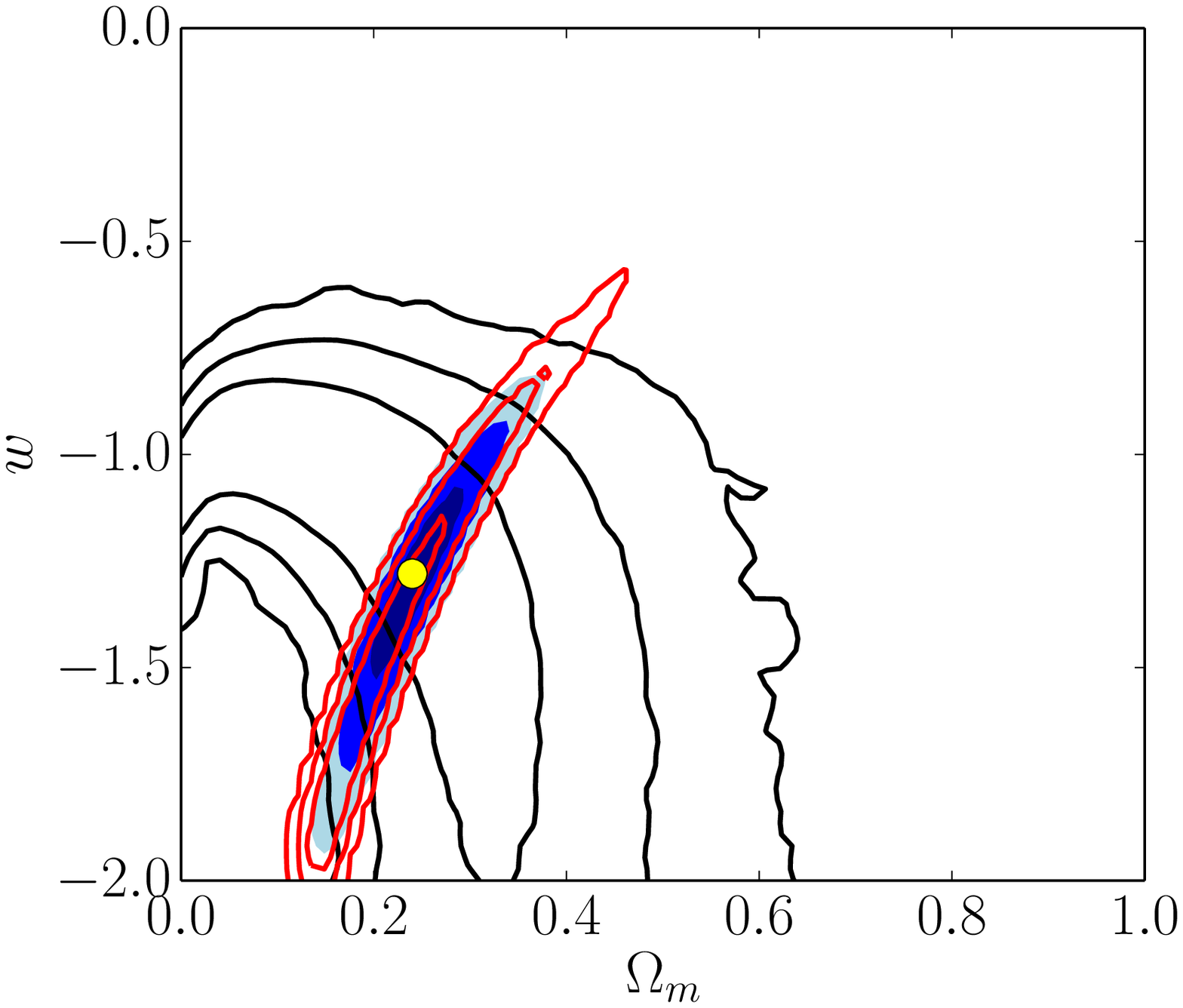}
  \includegraphics[width = .242\textwidth]{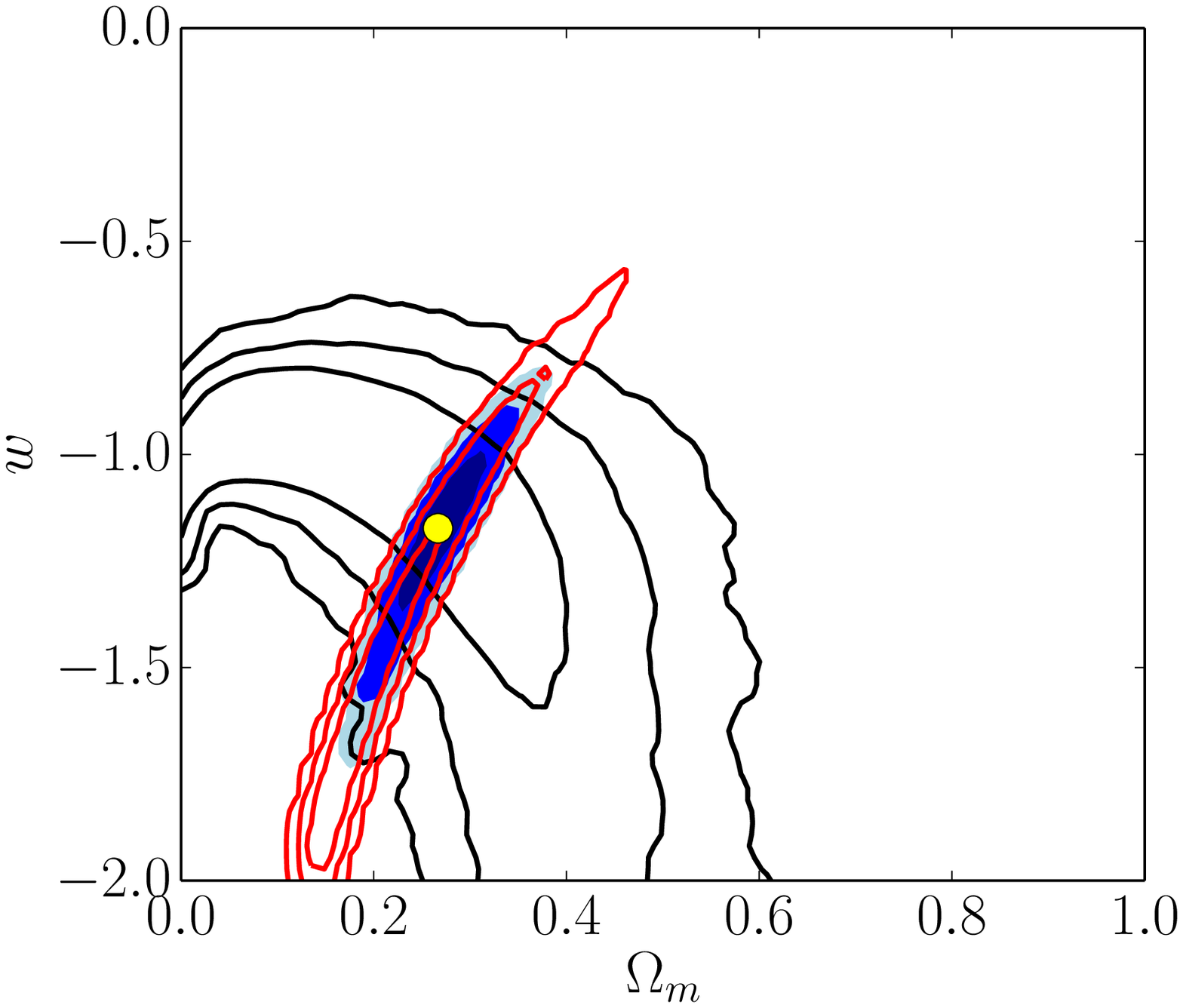}

  \includegraphics[width = .242\textwidth]{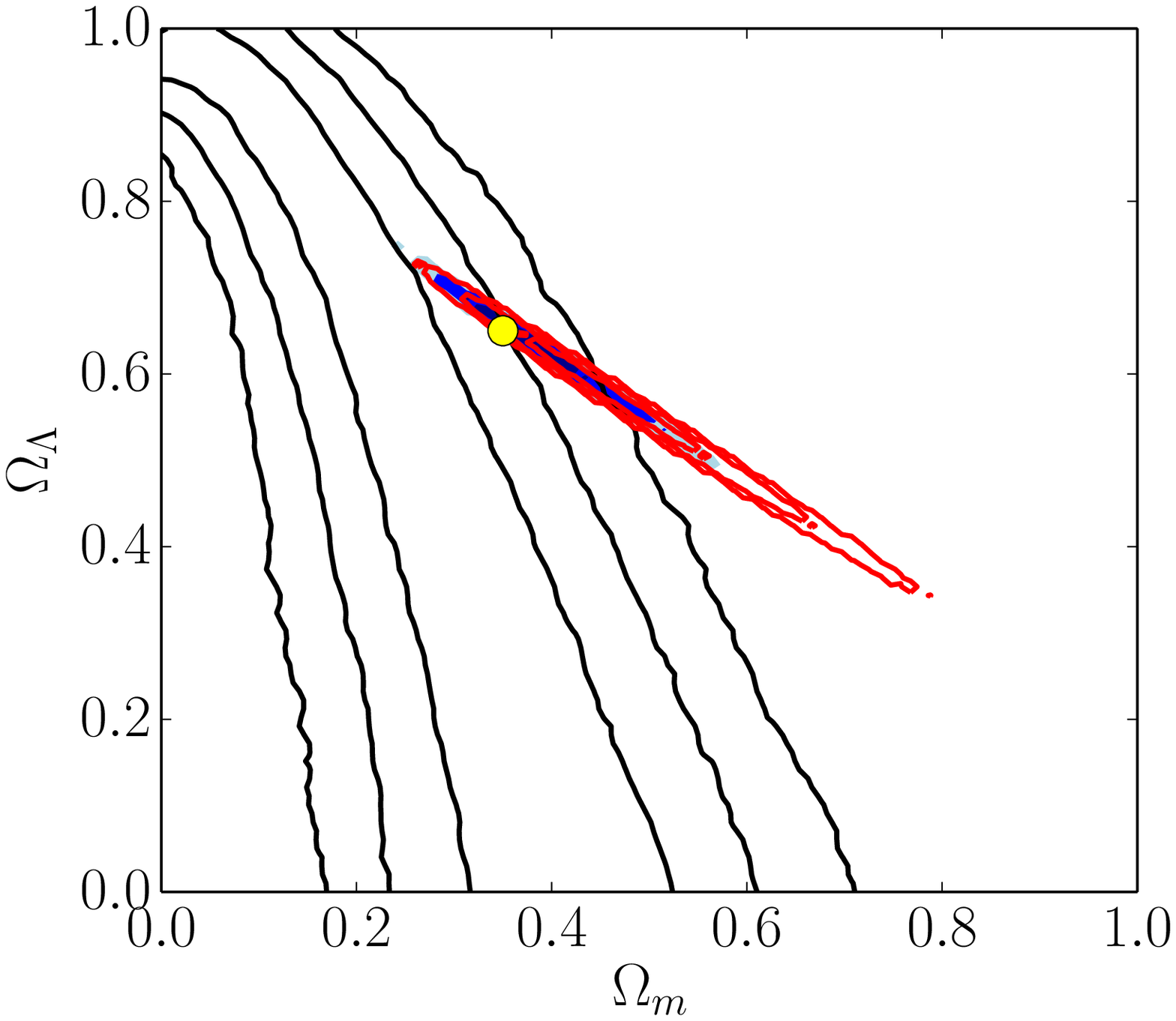}
  \includegraphics[width = .242\textwidth]{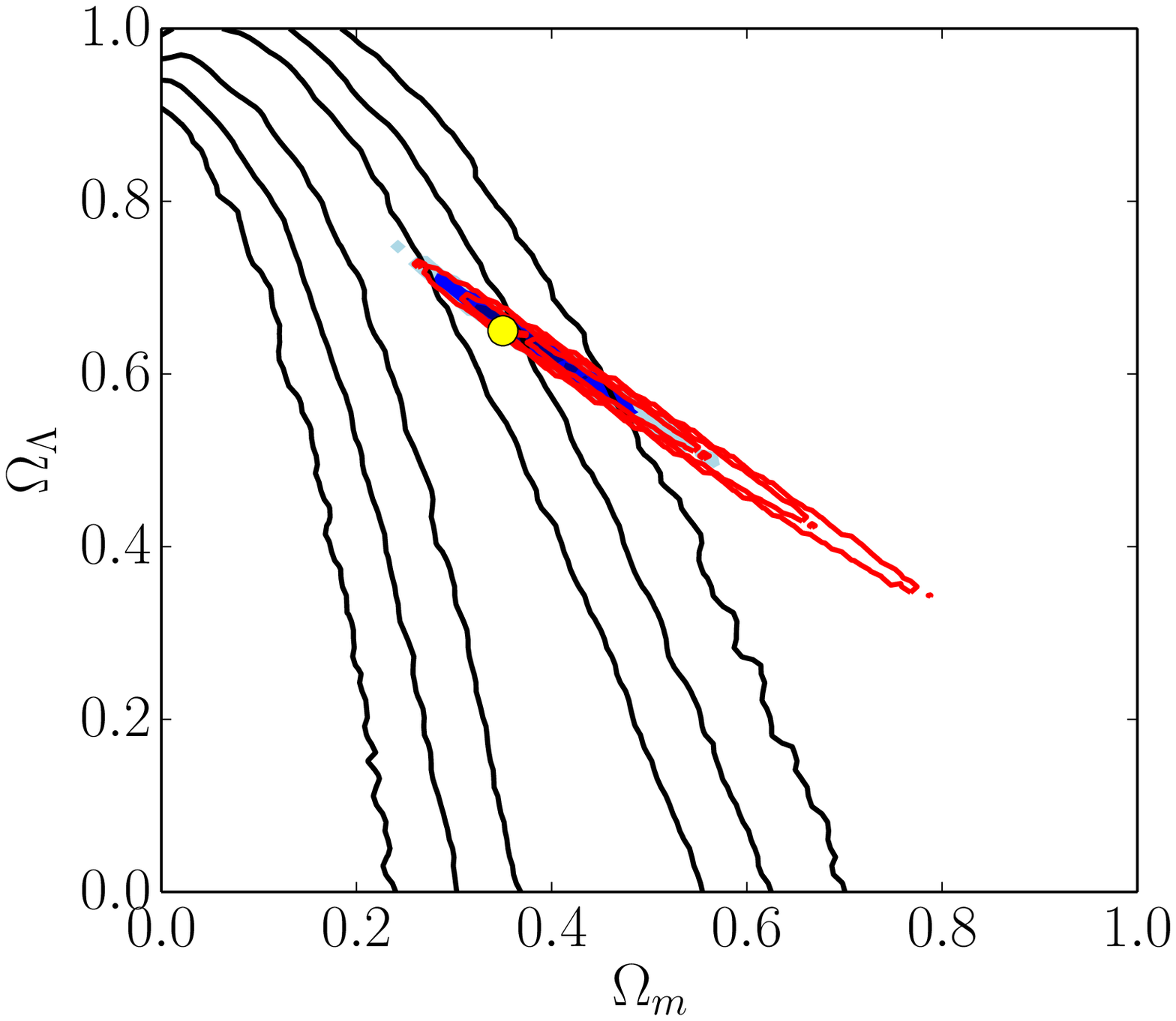}
  \caption{Confidence levels (black lines) for the cosmological parameters of  models W1 and W2 ($\Omega_m+\Omega_\Lambda=1$, top panels) and L1 and L2 ($w=-1$, bottom panels). Left panels refer to strong lensing models using only spectroscopic families (L1, W1), models in the right panels include all families (L2, W2). Red lines: contours from Planck Data Release 2 data. Blue regions: combined constraints. The yellow circles indicate the maximum likelihood peak in this projection.}
  \label{fig:cosmo_contour}
\end{figure}

In flat cosmological models, the 68\% confidence levels for each
parameter yield: $\Omega_m = 0.25_{-0.14}^{+0.11}$,
$w=-1.20_{-0.47}^{+0.25}$ and $\Omega_m = 0.25_{-0.16}^{+0.13}$,
$w=-1.07_{-0.42}^{+0.16}$, for the models W1 and W2, respectively.  By
including family sample 2 (secure multiple images with unknown
redshift), the statistical uncertainties on $w$ is $\approx 20\%$
smaller, but that on $\Omega_m$ increases by $\approx 14\%$.  This is
caused by a tilt in the orientation of the degeneracy between these
two parameters.
It appears that the extra information included in the additional
multiple image families leads to an improvement of the overall model,
i.e. to smaller errors on the values of the lens mass distribution
parameters and, consequently, of the cosmological parameters.
The 68\%, 95.4\% and 99.7\% confidence regions
on the cosmological parameter plane are shown in the top panels of Figure
\ref{fig:cosmo_contour}, for the models W1 and W2, respectively.  The
red contours indicate the confidence regions from the Planck satellite
Data Release 2 \citep{2015arXiv150201589P} and the blue regions the
combination with the likelihood from our strong lensing models.  The
agreement with the results from the CMB data, $\Omega_m = 0.3089\pm0.0062$ and
$w=-1.019_{-0.080}^{+0.075}$, is very good \citep[see Tables 4 and 5
in][]{2015arXiv150201589P} and it is interesting to note the complementarity of the two different cosmological probes, making their combination in principle powerful.

In the bottom panels of Figure \ref{fig:cosmo_contour}, we show the
confidence regions of the cosmological parameters for the models L1
and L2 ($\Omega_m$ and $\Omega_\Lambda$ free to vary and $w=-1$).
Here, we find a clear degeneracy between the values of $\Omega_m$ and
$\Omega_{\Lambda}$, with the value of $\Omega_m$ smaller than 0.7 at
$99.7\%$ confidence level and that of $\Omega_{\Lambda}$ essentially
unconstrained.
Indeed, the results of the simulations performed by \citet{2009MNRAS.396..354G}
showed that the values of the family ratios of Equation (\ref{eq:cosmo_prober}), predicted by strong lensing models, are not very sensitive to changes in the value of the dark energy density parameter.
For the models L1 and L2, we obtain
$\Omega_m = 0.31_{-0.13}^{+0.14}$,
$\Omega_\Lambda=0.38_{-0.27}^{+0.38}$ and $\Omega_m =
0.35_{-0.14}^{+0.11}$, $\Omega_\Lambda=0.36_{-0.26}^{+0.40}$ (68\%
confidence level), respectively.

\begin{figure*}[!ht]
  \centering
  \includegraphics[width = .99\textwidth]{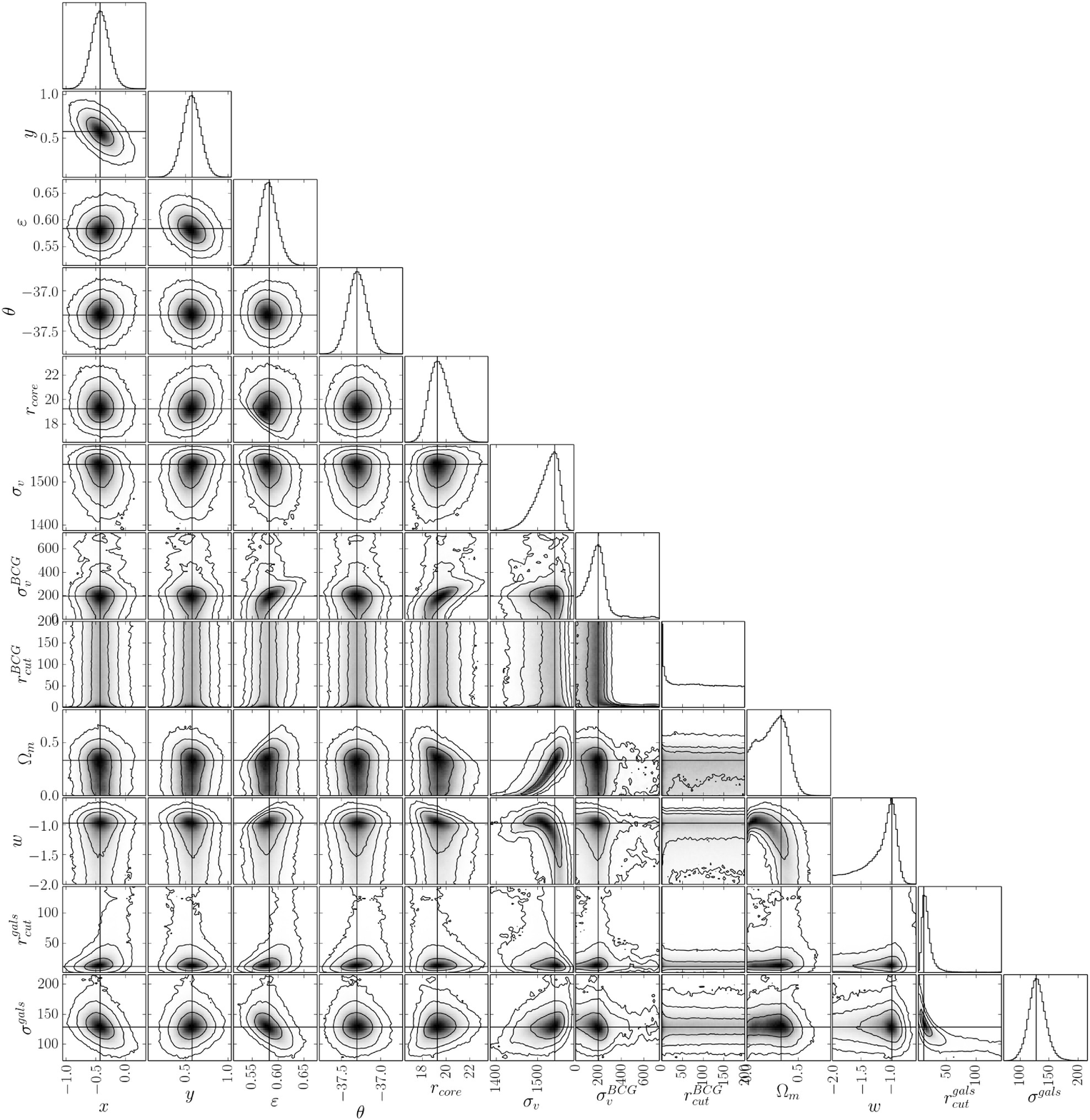}
  \caption{Confidence regions of the free parameters in the model
    considering all multiple image families and varying $\Omega_m$ and
    $w$ in a flat universe (model W2). The contours represent the
    $68\%$, $95.4\%$ and $99.7\%$ confidence levels. The lines indicate the maximum
    likelihood in the projection on each single parameter.  Contours
    associated to constrained redshifts are omitted for clarity.}
  \label{fig:cont_all_Omw}
\end{figure*}

In Figure \ref{fig:cont_all_Omw}, we show for the model ID W2 the correlation between the parameters describing the total mass distribution of the lens and those related to the cosmological model.
The histograms represent the probability density distributions of each free parameter, marginalized over all the other parameters.
For visualization clarity, we do not show in this figure the redshifts of families 9, 13, 15, 16, 17, 20 and 21, although they are also free parameters (see the model ID W2 in Table \ref{tab:cl_all}).

Figure \ref{fig:cont_all_Omw} shows that the cosmological parameters are mainly degenerate with the $\sigma_v$ parameter, associated to the mass of the cluster dark-matter halo: $\Omega_m$ and $\sigma_v$ are positively correlated, while $w$ and $\sigma_v$ are strongly anticorrelated for, respectively, low ($w<-1$) and high ($> 1500$ km/s) values.
This result suggests that independent information on the total
mass of a cluster, for example from galaxy dynamics \citep[e.g.][]{2013A&A...558A...1B} or weak lensing \citep[e.g.][]{2014ApJ...795..163U} could further reduce the statistical uncertainties on recovered cosmological parameters. It remains
important however to consider the impact of a number of systematics
inherent in different methods of mass measurements.

In a previous work, \citet{2010Sci...329..924J} studied the same
cosmological model using the galaxy cluster Abell 1689, a merging cluster
located at $z=0.184$.  In that work, starting from a sample of
102 secure-spectroscopic multiple images, they considered a subsample of 28 multiple images from 8 different families distributed in redshift between 1.50 and 3.05.
Figure 2 of \citet{2010Sci...329..924J} shows the confidence
regions in the $\Omega_m$-$w$ plane, as obtained from their strong lensing
analysis only and in combination with the results from CMB observations.
Those results are qualitatively similar to our findings in Figure \ref{fig:cosmo_contour}.
Small differences in the confidence regions of the two studies can be
ascribed to the different parametrization of the total mass of the
clusters (including a careful selection of the cluster members in our
model), the configuration of the adopted multiple images, the source and cluster redshifts, and the
treatment of the positional errors of multiple images.

\begin{figure}[!ht]
  \centering
  \includegraphics[width = .242\textwidth]{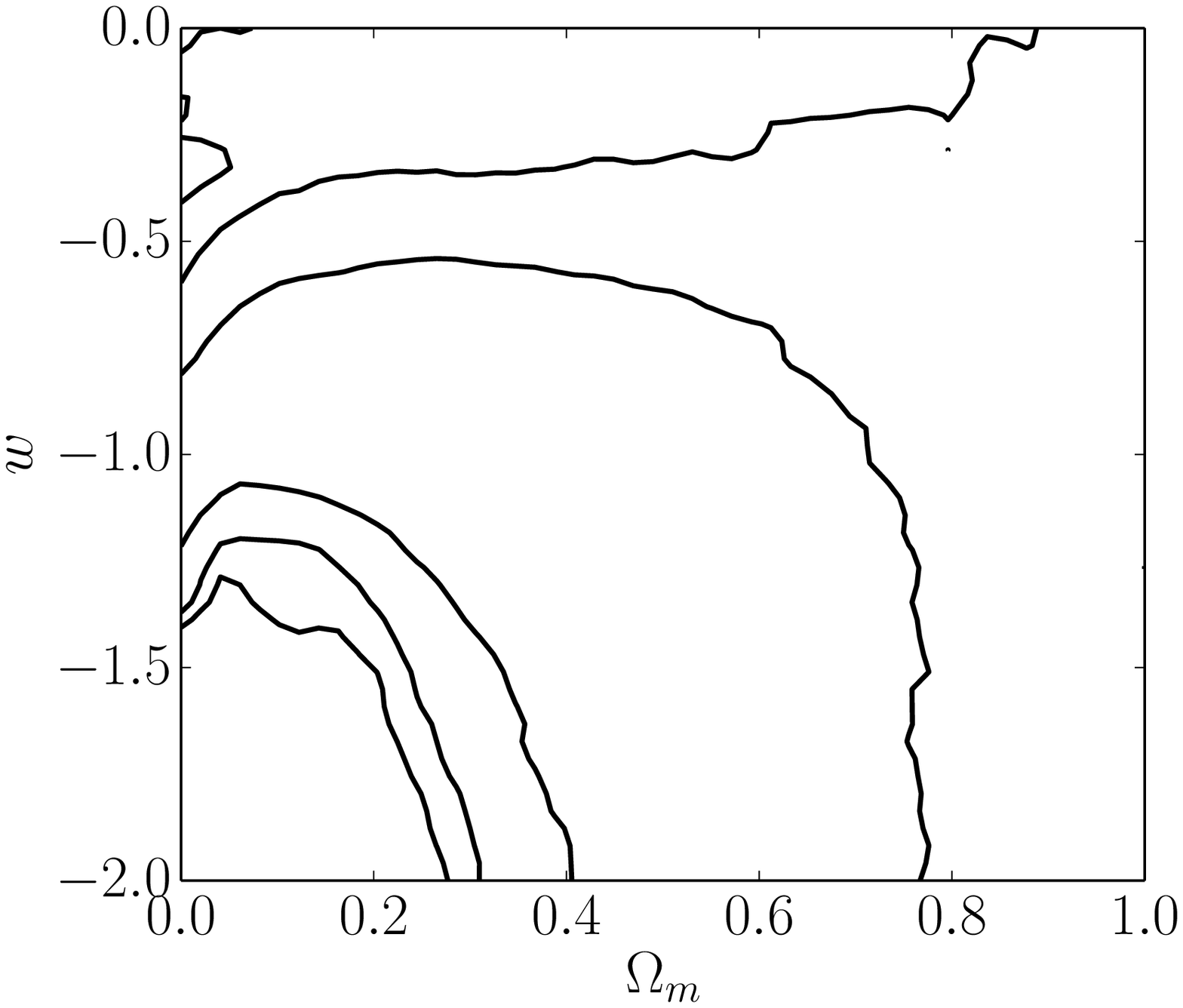}
  \includegraphics[width = .242\textwidth]{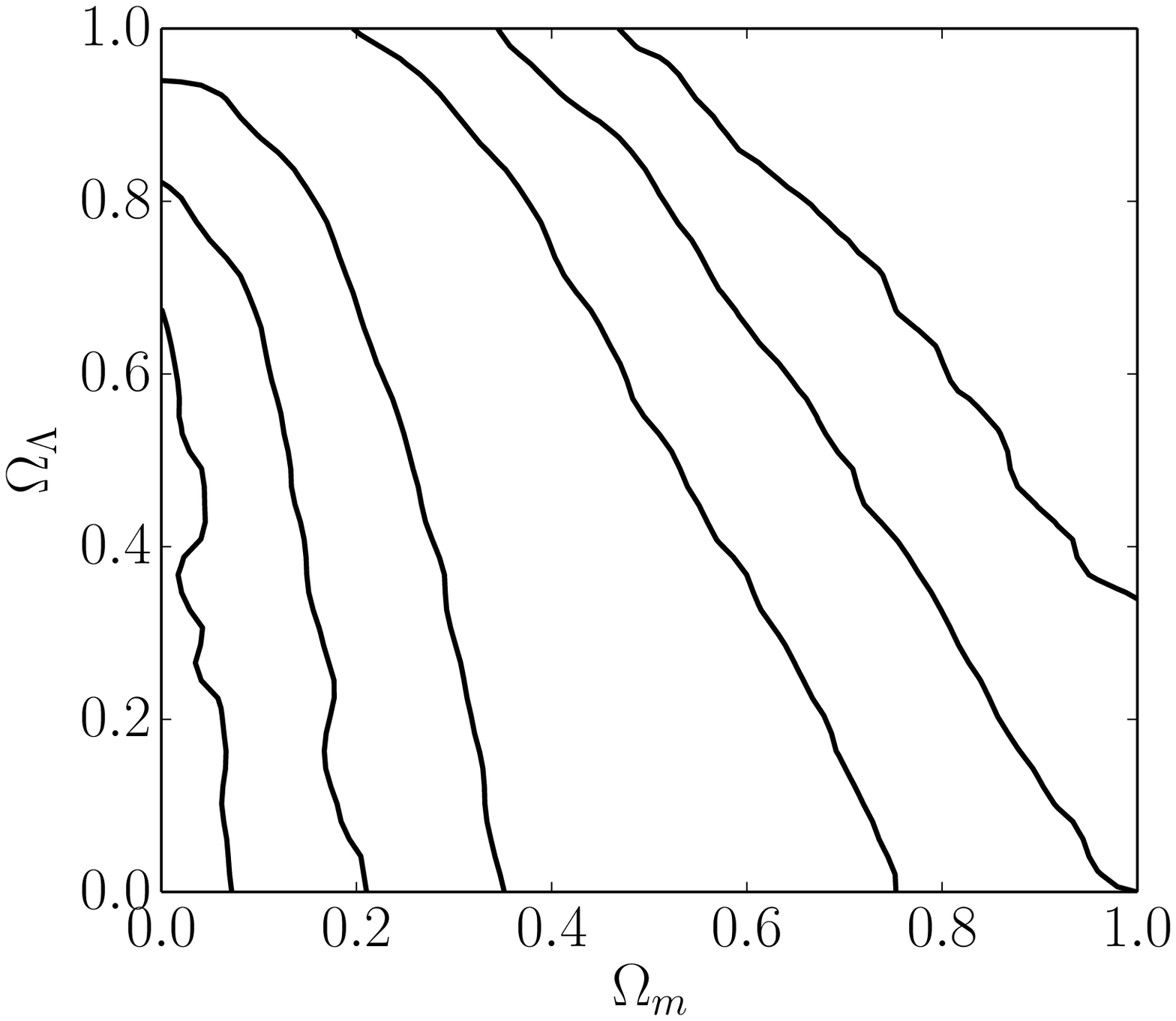}
  \caption{Confidence regions on the cosmological parameter planes when all multiple image systems
    are used with the exception of the highest redshift ($z=6.111$)
    family (left: model W3, right: model L3).}
  \label{fig:cosmo_no_highz}
\end{figure}
In order to highlight the importance of having multiply lensed sources
over a wide range of redshift when trying to constrain cosmological parameters, we also study specific models (W3 and L3 in Tables \ref{tab:summary_bf} and \ref{tab:cl_all}), in which we exclude the family at the highest redshift, z = 6.111 (ID 14).
In these models the
final positional $\Delta_{rms}$ remains basically unchanged
when compared to models W2 and L2, 
however the constraints on the cosmological parameters become much
weaker, as shown in Figure \ref{fig:cosmo_no_highz} (model W3/L3:
left/right panel).  Although the confidence regions of the lens mass
distribution parameters increase by less than $\sim\! 10\%$, this
high-redshift system has a significant leverage on the estimate of
cosmological parameter.  In this case, we find that the same confidence
regions of $\Omega_m$ and $w$ increase by $\approx50\%$ from model W2
to W3, with $\Omega_m$ becoming now largely unconstrained. Such a
deterioration is even more evident for the model L3, when compared to
L2. This test highlights the importance of probing the widest possible
redshift range, with spectroscopic multiply lensed systems, when
exploring cosmography with strong lensing techniques.
Similar results from cluster-scale strong lensing simulations were presented by \citet{2002A&A...387..788G}, confirming the essential role played by spectroscopically confirmed systems over a large redshift range for accurate measurements of the values of the cosmological parameters.
Finally, we
mention that the models Wa1 and Wa2, which include a variation with
redshift in the dark energy equation of state, can reproduce only
slightly better the observed multiple image positions.  This indicates
very little sensitivity on the $w_a$ parameter in our current
strong lensing models of RXC~J2248.

\section{Line of sight mass structure}
\label{subsec:line_of_sign_structure}

To estimate the perturbing lensing effect of mass structures along the
line of sight not included in the single-plane lens modelling of RXC~J2248,
we perform the following simplified tests with the {\sc Glee}
software, developed by A. Halkola and S. H. Suyu
\citep{2010A&A...524A..94S, 2012ApJ...750...10S}.  We mimic the strong lensing geometry observed in RXC~J2248 as close
as possible, both in
terms of angular positions and redshifts of the multiple images.  In
detail, we consider 8 different sources lensed into 24 multiple
images, distributed within a circle of $\approx 1\arcmin$ in
radius from the cluster centre and covering a redshift range between
1.0 and 6.1, thus following the observed configuration (see Table
\ref{tab:families} and Figure \ref{fig:multiple_image_systems}).  The
starting \emph{unperturbed} positions of the 24 images are perfectly
fitted, i.e. with a null $rms$ offset, by only one PIEMD mass profile,
with parameter values very close to those shown in Table
\ref{tab:summary_bf} and within a fixed cosmological model with
$\Omega_{m}=0.3$, $\Omega_{\Lambda}=0.7$, $w=-1.0$, and $H_0 = 70$ $\rm km/s/Mpc$.
Then, we simulate plausible lens galaxies along the line of sight, modeled as
dPIE mass distributions with vanishing ellipticity and core radius, we
introduce their mass components in the lensing model and calculate the
\emph{perturbed} multiple image positions.  We use these new
positions as observables to optimize the parameters of the PIEMD
mass profile, neglecting the contribution of the secondary line of
sight deflectors.  This should represent the typical lensing modelling
situation in which the parameters describing the total mass
distribution of a galaxy cluster, acting as primary deflector on
background sources, are measured by fitting the positions of a set of
multiple images, in the single-plane lens approximation, i.e. ignoring
the effect of possible mass structures along the line of sight.

Initially, we add a single dPIE perturber to the PIEMD mass component.
We fix the values of its distance from the cluster centre and
effective velocity dispersion to 60 arcsec and 200 km/s,
respectively.  We then vary the redshift value of the dPIE component
from 0.05 to 0.65, with a constant step of 0.1.  To obtain non
negligible perturbing lensing effects, we purposely simulate such a
massive galaxy, close in projection to the Einstein radius of the
cluster for the source at the highest redshift.  The optimized PIEMD
mass models (without varying the cosmological parameters) can
reproduce the perturbed multiple-image positions with $\Delta_{rms}$ values
that range from 0.3 to 0.1 arcsec, decreasing systematically with the
redshift of the perturber.  This simple test confirms the results of
previous studies \citep[e.g.,][]{2014MNRAS.443.3631M} which have shown
that statistically, at fixed total mass values, mass concentrations in
the foreground of a main deflector affect the lensed positions of the
multiple images more significantly than mass concentrations in the
background.  We notice that it rarely happens that massive foreground
or member galaxies are not included in the lensing model of a galaxy
cluster because these galaxies are usually very luminous and easily
identified as important lensing components \citep[e.g., see Section 3.4
in][]{2015ApJ...800...38G}.  Therefore, rms values of 0.3 arcsec or
larger associated to only one massive and neglected line of sight
structure are not very likely in detailed strong lensing models.

\begin{figure*}[!ht]
  \centering
  \includegraphics[width = .32\textwidth]{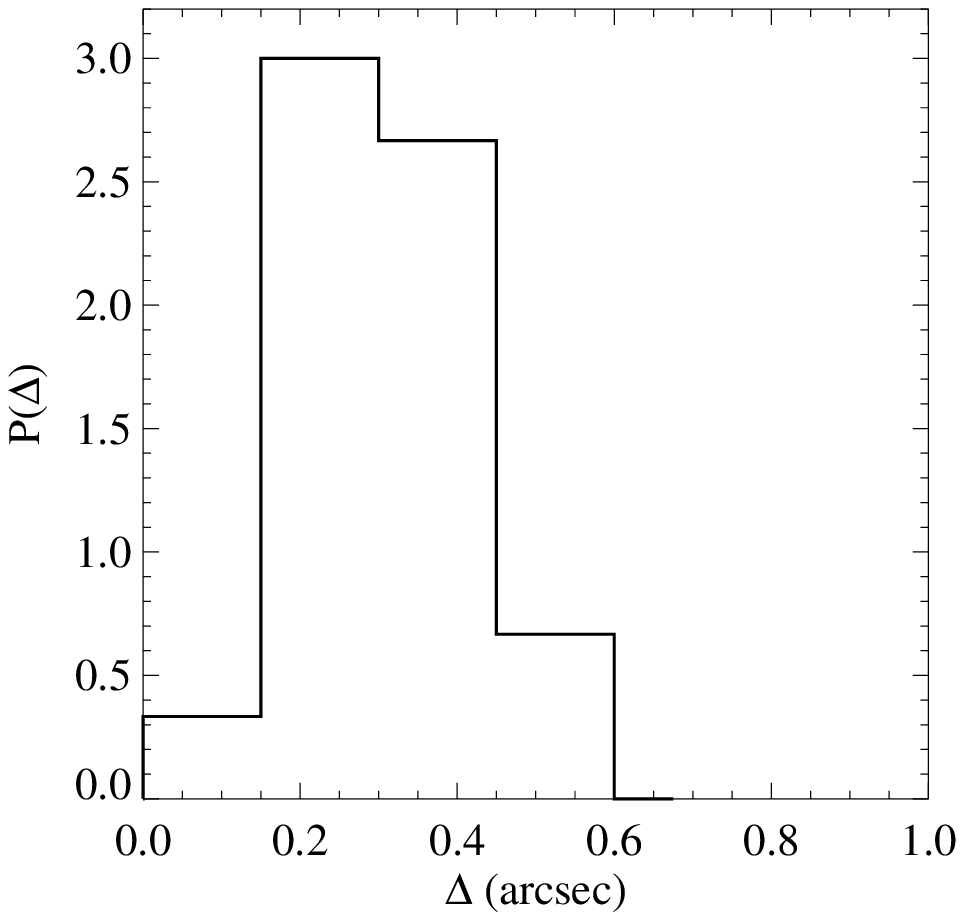}
  \includegraphics[width = .32\textwidth]{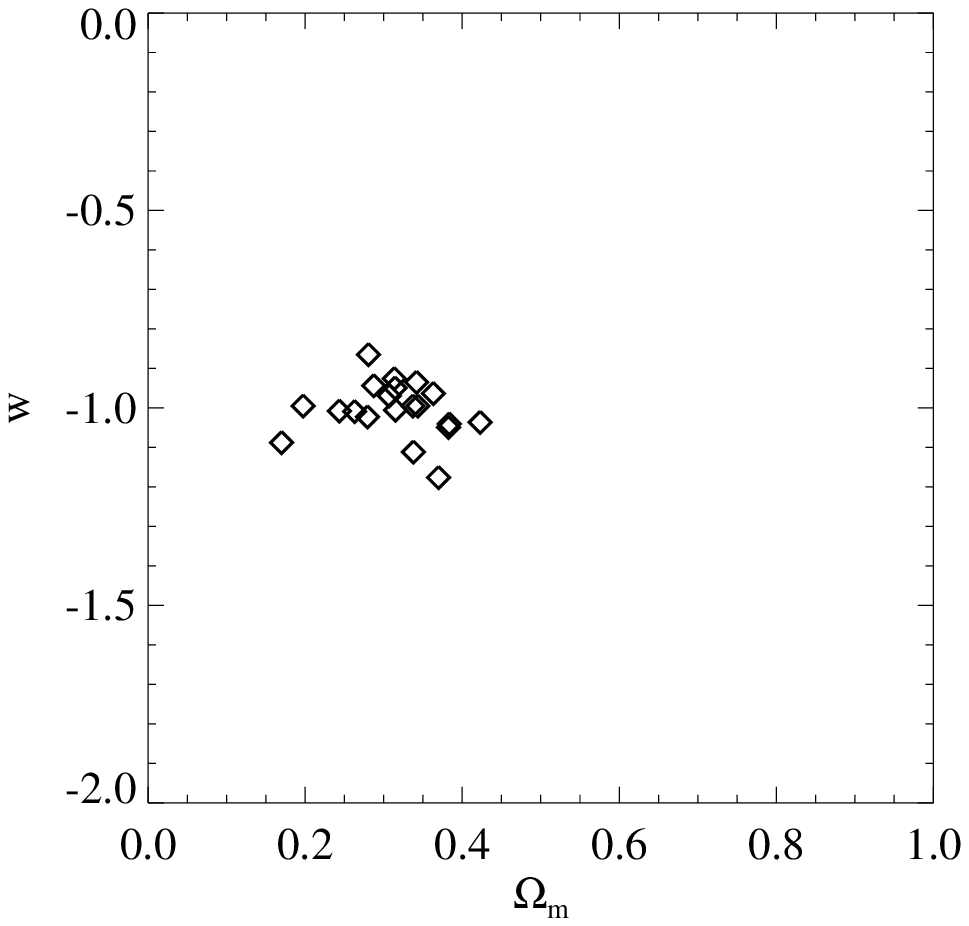}
  \includegraphics[width = .32\textwidth]{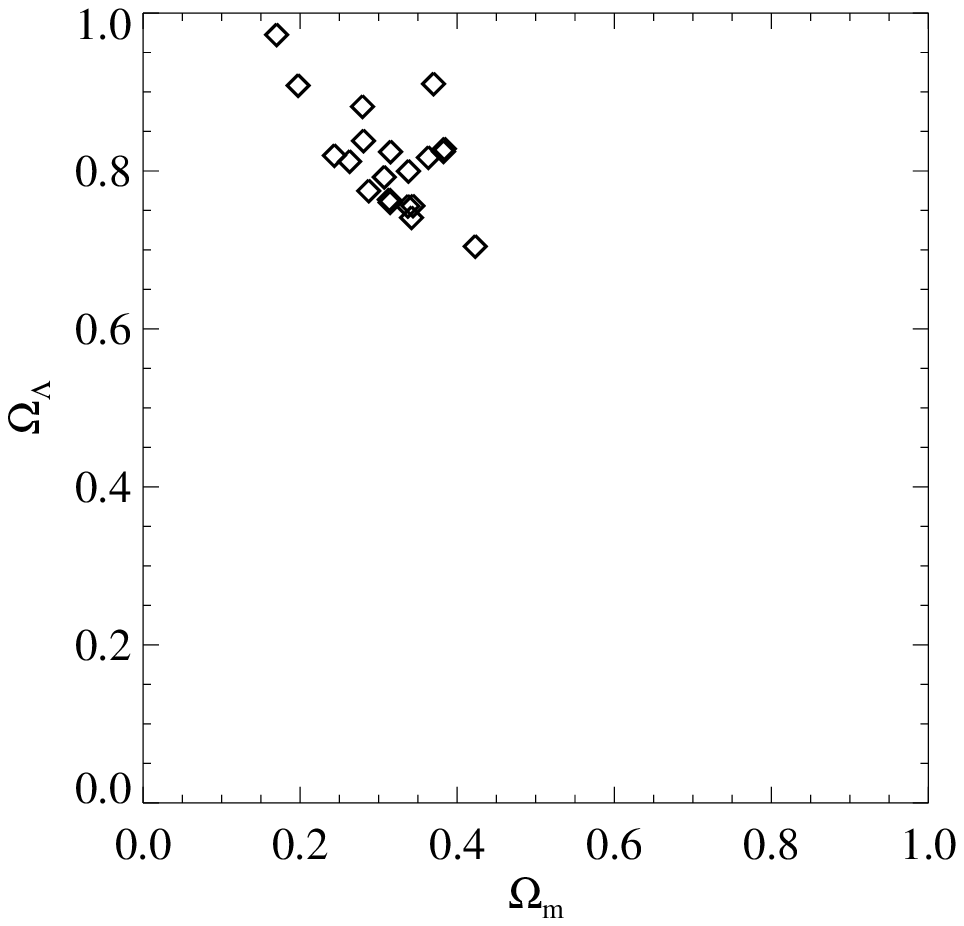}
  \caption{Left panel: Probability distribution function, obtained from the set of twenty simulations, of the $rms$ offset between the perturbed and best-fitting model-predicted positions of the 24 multiple images. The latter are determined by optimizing the parameters of the total mass distribution of the main lens and not including in the modelling the mass perturbers along the line of sight. Middle and Right panels: Best-fitting values of the parameters $\Omega_{m}$, $\Omega_\Lambda$ and $w$ from the set of twenty simulations. These values are determined by optimizing the parameters of the total mass distribution of the main lens and of the cosmological model and not including in the modelling the mass perturbers along the line of sight.}
  \label{fig:bf_los}
\end{figure*}

Next, we  consider a set of twenty more realistic simulations,
each of which containing ten different dPIE mass components.  The
position and mass parameters of these components are extracted from
uniform distributions in the following ranges: mass centers, in
angular coordinates $x$ and $y$ from the cluster center, between $-60$
and $60$ arcsec, redshift values between 0 and 0.8, and effective
velocity dispersion values between 25 and 175 km/s.
As above, for each simulation we optimize the PIEMD mass parameters,
not modelling the perturbers and not changing the cosmological
parameters, and estimate the $rms$ offset $\Delta$ between the
perturbed and model-predicted positions of the multiple images.  The
results are summarized in the first panel of Figure \ref{fig:bf_los}.
We remark that the probability distribution function of $\Delta$ has
mean and standard deviation values of 0.3 and 0.1 arcsec,
respectively.  Interestingly, an offset of approximately 0.3 arcsec in
the reproduction of the observed multiple-image positions has been
found in our best-fitting strong lensing models of RXC~J2248 (see
Section \ref{sec:results_sl}), MACS J0416 \citep[e.g., see
Section 3.5.1 in][]{2015ApJ...800...38G} and MACS J1149 \citep[e.g., see Section 3.4 in ][]{2015arXiv151104093G}, which have been obtained in
the single-plane lens approximation, as in these simulations.
Moreover, we investigate the systematic uncertainty on the
values of the cosmological parameters introduced by neglecting the
mass structure along the line of sight.  To do so, we add the
values of $\Omega_{m}$, $\Omega_{\Lambda}$, and $w$ to the PIEMD mass
parameters in the modelling optimization performed on the same sample
of twenty sets of perturbed multiple-image positions.  Allowing three
additional (cosmological) parameters to vary, leads to average
$rms$ values that are approximately $10\%$ smaller than the previous
ones.  We show in the second and third panels of Figure
\ref{fig:bf_los} the best-fitting values of $\Omega_{m}$,
$\Omega_{\Lambda}$ and $w$ and estimate, respectively, median with
standard deviation values of $0.3 \pm 0.1$, $0.8 \pm 0.1$ and $-1.0
\pm 0.1$.  The comparison of Figure \ref{fig:cosmo_contour} and Figure
\ref{fig:bf_los} suggests that the total (statistical+systematic)
degeneracy between the values of $\Omega_{m}$ and $w$ is likely not driven by the systematic effect of unmodeled line of sight mass
structure.  The results of Figure \ref{fig:cont_all_Omw} (in
particular the $\sigma_{v}$-$w$ panel) and Figure \ref{fig:bf_los}
indicate that additional information on the total mass is needed from different diagnostics in order to reduce the uncertainties on the values of $\Omega_m$ and $w$.
 
We caution that the previous results have been obtained through
simplified total mass models of a galaxy cluster that does not contain
cluster members.  We have purposely chosen to do so in order to reduce
the degeneracies among the parameters describing the relative mass
contributions of the cluster and cluster members and thus facilitate
the interpretation of the test outcome.  We postpone to a future work
a more thorough analysis including the cluster members.  We remark
that the spectroscopic CLASH-VLT program and additional VLT/MUSE
follow-up campaigns
(e.g., in RXC~2248, see \citealt{2015A&A...574A..11K}, and in MACS~J1149, see \citealt{2015arXiv151005750T} and \citealt{2015arXiv151104093G})
have identified the
mass structures along the line of sight that should be incorporated in
the ultimate strong lensing models of galaxy clusters.  Unfortunately,
at the time when this analysis was performed none of the lensing
codes available could fully model line of sight mass
structures and carefully quantify the impact of this effect on the
reconstructed values of the cluster mass and cosmological parameters.
The {\sc Glee} software has recently been updated to include multiple
plane lensing, which will be presented in future works.

\section{Conclusions}
\label{sec:discussions}
In this paper, we perform a comprehensive strong lensing analysis of
the galaxy cluster RXC~J2248 based on \emph{HST} imaging and new extensive VLT
spectroscopy with the VIMOS and MUSE instruments.
Interestingly, we also find an extended Ly$\alpha$ emitter at redshift
3.118, which is one of the first cases of a multiply lensed ``Ly$\alpha$
blob'' identified.  We consider 22 lensing models with different mass model
parametrization, samples of multiple images and assumptions on the
free parameters.  We show that RXC~J2248 is a massive cluster, which is
particularly suitable for constraining the background geometry of the
Universe with strong lensing modelling, due to its unique combination
of regular shape, a large number of multiple images spanning a wide
redshift range, and a relatively modest presence of intervening
large-scale structure, as revealed by our spectroscopic survey. We
show that the accuracy with which we reproduce the observed positions
of the multiple images ($\Delta_{rms}\simeq 0\arcsec.3$) is such that
the perturbing effect of mass structures along the line of sight needs
to be taken properly into account for further improvements.  Future
work will also need to focus on reducing systematics in the total mass-light
scaling relation of the subhalo population and this can be achieved by
using measured velocity dispersions of the BCG and other bright
cluster galaxies \citep[e.g. ][]{2015MNRAS.447.1224M}.
The main results of this study can be summarised as follows:

\begin{enumerate}
\item We reconstruct the observed positions of 38 multiple images from
  14 different sources in the redshift range $1.035-6.111$, with an
  accuracy of $0.\arcsec31$ in our reference model F2 (see Table
  \ref{tab:summary_bf}).
\item By testing different lensing models we show that the total mass
  density distribution in the center ($R\lesssim 300$ kpc) of RXC~J2248  is better represented by a PIEMD
  profile rather than a NFW.
  This is basically due to the
  existence of a significant core in the inner regions.
\item Owing to the wide redshift range of secure multiply lensed
  sources and the regular mass distribution of RXC~J2248, we are able
  to significantly alleviate degeneracies when fitting simultaneously
  the background geometry of the Universe and the total mass
  distribution of the lens in our strong lensing analyses. We thus
  find in the strong lensing analyses only that $\Omega_m=0.25^{+0.13}_{-0.16}$ and $w=-1.07^{+0.16}_{-0.42}$
  for a flat $\Lambda$CDM model and $\Omega_m=0.31^{+0.12}_{-0.13}$
  and $\Omega_\Lambda=0.38^{+0.38}_{-0.27}$ for a Universe with free
  curvature but $w=-1$.
\item We show that spectroscopic information is key for a
  high-precision strong lensing model. The lack of spectroscopic
  measurements of the multiply lensed sources or the use of photometric
  redshifts can bias the results on the values of the cosmological
  parameters, although the impact on the estimate of the total mass of
  the lens is not very significant. Moreover, a wide redshift range of
  multiply lensed sources is also critical to increase the leverage on
  cosmology.
\item Simple simulations, aimed at estimating the impact of line of
  sight perturbers on the lens modelling, show that this effect can
  introduce a scatter of $(0.3\pm 0.1) \arcsec$ in the multiple image
  positions, which is of the same order of the statistical errors
  achieved by our models.
\end{enumerate}
 
  We anticipate that repeating this experiment
  on other CLASH-VLT clusters, with similar high-quality samples of
  multiple images, leads instead to very loose contraints on
  cosmological parameters in cases where the spectroscopic campaign
  reveals significant large-scale structure along the line of
  sight. This suggests that a more sophisticated treatment of the
  oberved line of sight effects is needed in the lensing models to
  overcome this fundamental limit of lensing techniques. 
  This will be the subject of future papers.

\begin{acknowledgements}
  We thank the ESO User Support group for the excellent support on the
  implementation of the Large Programme 186.A-0798.  The CLASH
  Multi-Cycle Treasury Program is based on observations made with the
  NASA/ESA {\it Hubble Space Telescope}.
  This work made use of MUSE data taken under programme ID 60.A-9345(A), during the science verification period.
  The authors thank the referee Marceau Limousin for useful comments on this paper.
  G.B.C. is supported by the CAPES-ICRANET program through the grant BEX 13946/13-7. C.G. and E.M. acknowledge support by VILLUM FONDEN Young Investigator Programme through grant no. 10123.  This work
  made use of the CHE cluster, managed and funded by ICRA/CBPF/MCTI,
  with financial support from FINEP (grant 01.07.0515.00 from CT-INFRA
  - 01/2006) and FAPERJ (grants E-26/171.206/2006 and
  E-26/110.516/2012). We acknowledge support from PRIN-INAF 2014
  1.05.01.94.02 (PI M. Nonino). PR acknowledges the hospitality and
  support of the visitor program of the DFG cluster of excellence
  ``Origin and Structure of the Universe''. A.Z. is supported by NASA through Hubble Fellowship grant \#HST-HF2-51334.001-A awarded by STScI.
\end{acknowledgements}

\bibliographystyle{aa}

\bibliography{references.bib}

\begin{thebibliography}{70}
\expandafter\ifx\csname natexlab\endcsname\relax\def\natexlab#1{#1}\fi

\bibitem[{{Abell} {et~al.}(1989){Abell}, {Corwin}, \&
  {Olowin}}]{1989ApJS...70....1A}
{Abell}, G.~O., {Corwin}, Jr., H.~G., \& {Olowin}, R.~P. 1989, \apjs, 70, 1

\bibitem[{{Bacon} {et~al.}(2010){Bacon}, {Accardo}, {Adjali}, {Anwand},
  {Bauer}, {Biswas}, {Blaizot}, {Boudon}, {Brau-Nogue}, {Brinchmann},
  {Caillier}, {Capoani}, {Carollo}, {Contini}, {Couderc}, {Daguis{\'e}},
  {Deiries}, {Delabre}, {Dreizler}, {Dubois}, {Dupieux}, {Dupuy}, {Emsellem},
  {Fechner}, {Fleischmann}, {Fran{\c c}ois}, {Gallou}, {Gharsa}, {Glindemann},
  {Gojak}, {Guiderdoni}, {Hansali}, {Hahn}, {Jarno}, {Kelz}, {Koehler},
  {Kosmalski}, {Laurent}, {Le Floch}, {Lilly}, {Lizon}, {Loupias}, {Manescau},
  {Monstein}, {Nicklas}, {Olaya}, {Pares}, {Pasquini}, {P{\'e}contal-Rousset},
  {Pell{\'o}}, {Petit}, {Popow}, {Reiss}, {Remillieux}, {Renault}, {Roth},
  {Rupprecht}, {Serre}, {Schaye}, {Soucail}, {Steinmetz}, {Streicher}, {Stuik},
  {Valentin}, {Vernet}, {Weilbacher}, {Wisotzki}, \&
  {Yerle}}]{2010SPIE.7735E..08B}
{Bacon}, R., {Accardo}, M., {Adjali}, L., {et~al.} 2010, in Society of
  Photo-Optical Instrumentation Engineers (SPIE) Conference Series, Vol. 7735,
  Society of Photo-Optical Instrumentation Engineers (SPIE) Conference Series,
  8

\bibitem[{{Balestra} {et~al.}(2013){Balestra}, {Vanzella}, {Rosati}, {Monna},
  {Grillo}, {Nonino}, {Mercurio}, {Biviano}, {Bradley}, {Coe}, {Fritz},
  {Postman}, {Seitz}, {Scodeggio}, {Tozzi}, {Zheng}, {Ziegler}, {Zitrin},
  {Annunziatella}, {Bartelmann}, {Benitez}, {Broadhurst}, {Bouwens}, {Czoske},
  {Donahue}, {Ford}, {Girardi}, {Infante}, {Jouvel}, {Kelson}, {Koekemoer},
  {Kuchner}, {Lemze}, {Lombardi}, {Maier}, {Medezinski}, {Melchior},
  {Meneghetti}, {Merten}, {Molino}, {Moustakas}, {Presotto}, {Smit}, \&
  {Umetsu}}]{2013A&A...559L...9B}
{Balestra}, I., {Vanzella}, E., {Rosati}, P., {et~al.} 2013, \aap, 559, L9

\bibitem[{{Bertin} \& {Arnouts}(1996)}]{1996A&AS..117..393B}
{Bertin}, E. \& {Arnouts}, S. 1996, \aaps, 117, 393

\bibitem[{{Biviano} {et~al.}(2013){Biviano}, {Rosati}, {Balestra}, {Mercurio},
  {Girardi}, {Nonino}, {Grillo}, {Scodeggio}, {Lemze}, {Kelson}, {Umetsu},
  {Postman}, {Zitrin}, {Czoske}, {Ettori}, {Fritz}, {Lombardi}, {Maier},
  {Medezinski}, {Mei}, {Presotto}, {Strazzullo}, {Tozzi}, {Ziegler},
  {Annunziatella}, {Bartelmann}, {Benitez}, {Bradley}, {Brescia}, {Broadhurst},
  {Coe}, {Demarco}, {Donahue}, {Ford}, {Gobat}, {Graves}, {Koekemoer},
  {Kuchner}, {Melchior}, {Meneghetti}, {Merten}, {Moustakas}, {Munari}, {Reg{\H
  o}s}, {Sartoris}, {Seitz}, \& {Zheng}}]{2013A&A...558A...1B}
{Biviano}, A., {Rosati}, P., {Balestra}, I., {et~al.} 2013, \aap, 558, A1

\bibitem[{{Boone} {et~al.}(2013){Boone}, {Cl{\'e}ment}, {Richard}, {Schaerer},
  {Lutz}, {Wei{\ss}}, {Zemcov}, {Egami}, {Rawle}, {Walth}, {Kneib}, {Combes},
  {Smail}, {Swinbank}, {Altieri}, {Blain}, {Chapman}, {Dessauges-Zavadsky},
  {Ivison}, {Knudsen}, {Omont}, {Pell{\'o}}, {P{\'e}rez-Gonz{\'a}lez},
  {Valtchanov}, {van der Werf}, \& {Zamojski}}]{2013A&A...559L...1B}
{Boone}, F., {Cl{\'e}ment}, B., {Richard}, J., {et~al.} 2013, \aap, 559, L1

\bibitem[{{Bouwens} {et~al.}(2011){Bouwens}, {Illingworth}, {Oesch},
  {Labb{\'e}}, {Trenti}, {van Dokkum}, {Franx}, {Stiavelli}, {Carollo},
  {Magee}, \& {Gonzalez}}]{2011ApJ...737...90B}
{Bouwens}, R.~J., {Illingworth}, G.~D., {Oesch}, P.~A., {et~al.} 2011, \apj,
  737, 90

\bibitem[{{Collett} \& {Auger}(2014)}]{2014MNRAS.443..969C}
{Collett}, T.~E. \& {Auger}, M.~W. 2014, \mnras, 443, 969

\bibitem[{{Efstathiou} {et~al.}(2002){Efstathiou}, {Moody}, {Peacock},
  {Percival}, {Baugh}, {Bland-Hawthorn}, {Bridges}, {Cannon}, {Cole},
  {Colless}, {Collins}, {Couch}, {Dalton}, {de Propris}, {Driver}, {Ellis},
  {Frenk}, {Glazebrook}, {Jackson}, {Lahav}, {Lewis}, {Lumsden}, {Maddox},
  {Norberg}, {Peterson}, {Sutherland}, \& {Taylor}}]{2002MNRAS.330L..29E}
{Efstathiou}, G., {Moody}, S., {Peacock}, J.~A., {et~al.} 2002, \mnras, 330,
  L29

\bibitem[{{Eisenstein} {et~al.}(2005){Eisenstein}, {Zehavi}, {Hogg},
  {Scoccimarro}, {Blanton}, {Nichol}, {Scranton}, {Seo}, {Tegmark}, {Zheng},
  {Anderson}, {Annis}, {Bahcall}, {Brinkmann}, {Burles}, {Castander},
  {Connolly}, {Csabai}, {Doi}, {Fukugita}, {Frieman}, {Glazebrook}, {Gunn},
  {Hendry}, {Hennessy}, {Ivezi{\'c}}, {Kent}, {Knapp}, {Lin}, {Loh}, {Lupton},
  {Margon}, {McKay}, {Meiksin}, {Munn}, {Pope}, {Richmond}, {Schlegel},
  {Schneider}, {Shimasaku}, {Stoughton}, {Strauss}, {SubbaRao}, {Szalay},
  {Szapudi}, {Tucker}, {Yanny}, \& {York}}]{2005ApJ...633..560E}
{Eisenstein}, D.~J., {Zehavi}, I., {Hogg}, D.~W., {et~al.} 2005, \apj, 633, 560

\bibitem[{{El{\'{\i}}asd{\'o}ttir} {et~al.}(2007){El{\'{\i}}asd{\'o}ttir},
  {Limousin}, {Richard}, {Hjorth}, {Kneib}, {Natarajan}, {Pedersen}, {Jullo},
  \& {Paraficz}}]{2007arXiv0710.5636E}
{El{\'{\i}}asd{\'o}ttir}, {\'A}., {Limousin}, M., {Richard}, J., {et~al.} 2007,
  ArXiv e-prints [\eprint[arXiv]{0710.5636}]

\bibitem[{{Francis} {et~al.}(2001){Francis}, {Williger}, {Collins}, {Palunas},
  {Malumuth}, {Woodgate}, {Teplitz}, {Smette}, {Sutherland}, {Danks}, {Hill},
  {Lindler}, {Kimble}, {Heap}, \& {Hutchings}}]{2001ApJ...554.1001F}
{Francis}, P.~J., {Williger}, G.~M., {Collins}, N.~R., {et~al.} 2001, \apj,
  554, 1001

\bibitem[{{Fynbo} {et~al.}(1999){Fynbo}, {M{\o}ller}, \&
  {Warren}}]{1999MNRAS.305..849F}
{Fynbo}, J.~U., {M{\o}ller}, P., \& {Warren}, S.~J. 1999, \mnras, 305, 849

\bibitem[{{Gilmore} \& {Natarajan}(2009)}]{2009MNRAS.396..354G}
{Gilmore}, J. \& {Natarajan}, P. 2009, \mnras, 396, 354

\bibitem[{{Gilmour} {et~al.}(2009){Gilmour}, {Best}, \&
  {Almaini}}]{2009MNRAS.392.1509G}
{Gilmour}, R., {Best}, P., \& {Almaini}, O. 2009, \mnras, 392, 1509

\bibitem[{{Golse} \& {Kneib}(2002)}]{2002A&A...390..821G}
{Golse}, G. \& {Kneib}, J.-P. 2002, \aap, 390, 821

\bibitem[{{Golse} {et~al.}(2002){Golse}, {Kneib}, \&
  {Soucail}}]{2002A&A...387..788G}
{Golse}, G., {Kneib}, J.-P., \& {Soucail}, G. 2002, \aap, 387, 788

\bibitem[{{G{\'o}mez} {et~al.}(2012){G{\'o}mez}, {Valkonen}, {Romer},
  {Lloyd-Davies}, {Verdugo}, {Cantalupo}, {Daub}, {Goldstein}, {Kuo}, {Lange},
  {Lueker}, {Holzapfel}, {Peterson}, {Ruhl}, {Runyan}, {Reichardt}, \&
  {Sabirli}}]{2012AJ....144...79G}
{G{\'o}mez}, P.~L., {Valkonen}, L.~E., {Romer}, A.~K., {et~al.} 2012, \aj, 144,
  79

\bibitem[{{Grillo} {et~al.}(2015{\natexlab{a}}){Grillo}, {Karman}, {Suyu},
  {Rosati}, {Balestra}, {Mercurio}, {Lombardi}, {Treu}, {Caminha}, {Halkola},
  {Rodney}, {Gavazzi}, \& {Caputi}}]{2015arXiv151104093G}
{Grillo}, C., {Karman}, W., {Suyu}, S.~H., {et~al.} 2015{\natexlab{a}}, ArXiv
  e-prints [\eprint[arXiv]{1511.04093}]

\bibitem[{{Grillo} {et~al.}(2008){Grillo}, {Lombardi}, \&
  {Bertin}}]{2008A&A...477..397G}
{Grillo}, C., {Lombardi}, M., \& {Bertin}, G. 2008, \aap, 477, 397

\bibitem[{{Grillo} {et~al.}(2015{\natexlab{b}}){Grillo}, {Suyu}, {Rosati},
  {Mercurio}, {Balestra}, {Munari}, {Nonino}, {Caminha}, {Lombardi}, {De
  Lucia}, {Borgani}, {Gobat}, {Biviano}, {Girardi}, {Umetsu}, {Coe},
  {Koekemoer}, {Postman}, {Zitrin}, {Halkola}, {Broadhurst}, {Sartoris},
  {Presotto}, {Annunziatella}, {Maier}, {Fritz}, {Vanzella}, \&
  {Frye}}]{2015ApJ...800...38G}
{Grillo}, C., {Suyu}, S.~H., {Rosati}, P., {et~al.} 2015{\natexlab{b}}, \apj,
  800, 38

\bibitem[{{Gruen} {et~al.}(2013){Gruen}, {Brimioulle}, {Seitz}, {Lee}, {Young},
  {Koppenhoefer}, {Eichner}, {Riffeser}, {Vikram}, {Weidinger}, \&
  {Zenteno}}]{2013MNRAS.432.1455G}
{Gruen}, D., {Brimioulle}, F., {Seitz}, S., {et~al.} 2013, \mnras, 432, 1455

\bibitem[{{Halkola} {et~al.}(2008){Halkola}, {Hildebrandt}, {Schrabback},
  {Lombardi}, {Brada{\v c}}, {Erben}, {Schneider}, \&
  {Wuttke}}]{2008A&A...481...65H}
{Halkola}, A., {Hildebrandt}, H., {Schrabback}, T., {et~al.} 2008, \aap, 481,
  65

\bibitem[{{Halkola} {et~al.}(2006){Halkola}, {Seitz}, \&
  {Pannella}}]{2006MNRAS.372.1425H}
{Halkola}, A., {Seitz}, S., \& {Pannella}, M. 2006, \mnras, 372, 1425

\bibitem[{{Host}(2012)}]{2012MNRAS.420L..18H}
{Host}, O. 2012, \mnras, 420, L18

\bibitem[{{Johnson} {et~al.}(2014){Johnson}, {Sharon}, {Bayliss}, {Gladders},
  {Coe}, \& {Ebeling}}]{2014ApJ...797...48J}
{Johnson}, T.~L., {Sharon}, K., {Bayliss}, M.~B., {et~al.} 2014, \apj, 797, 48

\bibitem[{{Jouvel} {et~al.}(2014){Jouvel}, {Host}, {Lahav}, {Seitz}, {Molino},
  {Coe}, {Postman}, {Moustakas}, {Ben{\`i}tez}, {Rosati}, {Balestra}, {Grillo},
  {Bradley}, {Fritz}, {Kelson}, {Koekemoer}, {Lemze}, {Medezinski}, {Mercurio},
  {Moustakas}, {Nonino}, {Scodeggio}, {Zheng}, {Zitrin}, {Bartelmann},
  {Bouwens}, {Broadhurst}, {Donahue}, {Ford}, {Graves}, {Infante},
  {Jimenez-Teja}, {Lazkoz}, {Melchior}, {Meneghetti}, {Merten}, {Ogaz}, \&
  {Umetsu}}]{2014A&A...562A..86J}
{Jouvel}, S., {Host}, O., {Lahav}, O., {et~al.} 2014, \aap, 562, A86

\bibitem[{{Jullo} {et~al.}(2007){Jullo}, {Kneib}, {Limousin},
  {El{\'{\i}}asd{\'o}ttir}, {Marshall}, \& {Verdugo}}]{2007NJPh....9..447J}
{Jullo}, E., {Kneib}, J.-P., {Limousin}, M., {et~al.} 2007, New Journal of
  Physics, 9, 447

\bibitem[{{Jullo} {et~al.}(2010){Jullo}, {Natarajan}, {Kneib}, {D'Aloisio},
  {Limousin}, {Richard}, \& {Schimd}}]{2010Sci...329..924J}
{Jullo}, E., {Natarajan}, P., {Kneib}, J.-P., {et~al.} 2010, Science, 329, 924

\bibitem[{{Karman} {et~al.}(2015){Karman}, {Caputi}, {Grillo}, {Balestra},
  {Rosati}, {Vanzella}, {Coe}, {Christensen}, {Koekemoer}, {Kr{\"u}hler},
  {Lombardi}, {Mercurio}, {Nonino}, \& {van der Wel}}]{2015A&A...574A..11K}
{Karman}, W., {Caputi}, K.~I., {Grillo}, C., {et~al.} 2015, \aap, 574, A11

\bibitem[{{Kassiola} \& {Kovner}(1993)}]{1993ApJ...417..450K}
{Kassiola}, A. \& {Kovner}, I. 1993, \apj, 417, 450

\bibitem[{{Keeton}(2001)}]{2001astro.ph..2340K}
{Keeton}, C.~R. 2001, ArXiv Astrophysics e-prints [\eprint{astro-ph/0102340}]

\bibitem[{{Kneib}(2002)}]{2002sgdh.conf...50K}
{Kneib}, J.-P. 2002, in The Shapes of Galaxies and their Dark Halos, ed.
  P.~{Natarajan}, 50--57

\bibitem[{{Kneib} {et~al.}(1996){Kneib}, {Ellis}, {Smail}, {Couch}, \&
  {Sharples}}]{1996ApJ...471..643K}
{Kneib}, J.-P., {Ellis}, R.~S., {Smail}, I., {Couch}, W.~J., \& {Sharples},
  R.~M. 1996, \apj, 471, 643

\bibitem[{{Koekemoer} {et~al.}(2011){Koekemoer}, {Faber}, {Ferguson}, {Grogin},
  {Kocevski}, {Koo}, {Lai}, {Lotz}, {Lucas}, {McGrath}, {Ogaz}, {Rajan},
  {Riess}, {Rodney}, {Strolger}, {Casertano}, {Castellano}, {Dahlen},
  {Dickinson}, {Dolch}, {Fontana}, {Giavalisco}, {Grazian}, {Guo}, {Hathi},
  {Huang}, {van der Wel}, {Yan}, {Acquaviva}, {Alexander}, {Almaini}, {Ashby},
  {Barden}, {Bell}, {Bournaud}, {Brown}, {Caputi}, {Cassata}, {Challis},
  {Chary}, {Cheung}, {Cirasuolo}, {Conselice}, {Roshan Cooray}, {Croton},
  {Daddi}, {Dav{\'e}}, {de Mello}, {de Ravel}, {Dekel}, {Donley}, {Dunlop},
  {Dutton}, {Elbaz}, {Fazio}, {Filippenko}, {Finkelstein}, {Frazer}, {Gardner},
  {Garnavich}, {Gawiser}, {Gruetzbauch}, {Hartley}, {H{\"a}ussler},
  {Herrington}, {Hopkins}, {Huang}, {Jha}, {Johnson}, {Kartaltepe},
  {Khostovan}, {Kirshner}, {Lani}, {Lee}, {Li}, {Madau}, {McCarthy},
  {McIntosh}, {McLure}, {McPartland}, {Mobasher}, {Moreira}, {Mortlock},
  {Moustakas}, {Mozena}, {Nandra}, {Newman}, {Nielsen}, {Niemi}, {Noeske},
  {Papovich}, {Pentericci}, {Pope}, {Primack}, {Ravindranath}, {Reddy},
  {Renzini}, {Rix}, {Robaina}, {Rosario}, {Rosati}, {Salimbeni}, {Scarlata},
  {Siana}, {Simard}, {Smidt}, {Snyder}, {Somerville}, {Spinrad}, {Straughn},
  {Telford}, {Teplitz}, {Trump}, {Vargas}, {Villforth}, {Wagner}, {Wandro},
  {Wechsler}, {Weiner}, {Wiklind}, {Wild}, {Wilson}, {Wuyts}, \&
  {Yun}}]{2011ApJS..197...36K}
{Koekemoer}, A.~M., {Faber}, S.~M., {Ferguson}, H.~C., {et~al.} 2011, \apjs,
  197, 36

\bibitem[{{Komatsu} {et~al.}(2011){Komatsu}, {Smith}, {Dunkley}, {Bennett},
  {Gold}, {Hinshaw}, {Jarosik}, {Larson}, {Nolta}, {Page}, {Spergel},
  {Halpern}, {Hill}, {Kogut}, {Limon}, {Meyer}, {Odegard}, {Tucker}, {Weiland},
  {Wollack}, \& {Wright}}]{2011ApJS..192...18K}
{Komatsu}, E., {Smith}, K.~M., {Dunkley}, J., {et~al.} 2011, \apjs, 192, 18

\bibitem[{{Laporte} \& {White}(2015)}]{2015MNRAS.451.1177L}
{Laporte}, C.~F.~P. \& {White}, S.~D.~M. 2015, \mnras, 451, 1177

\bibitem[{{Le F{\`e}vre} {et~al.}(2003){Le F{\`e}vre}, {Saisse}, {Mancini},
  {Brau-Nogue}, {Caputi}, {Castinel}, {D'Odorico}, {Garilli}, {Kissler-Patig},
  {Lucuix}, {Mancini}, {Pauget}, {Sciarretta}, {Scodeggio}, {Tresse}, \&
  {Vettolani}}]{2003SPIE.4841.1670L}
{Le F{\`e}vre}, O., {Saisse}, M., {Mancini}, D., {et~al.} 2003, in Society of
  Photo-Optical Instrumentation Engineers (SPIE) Conference Series, Vol. 4841,
  Instrument Design and Performance for Optical/Infrared Ground-based
  Telescopes, ed. M.~{Iye} \& A.~F.~M. {Moorwood}, 1670--1681

\bibitem[{{Maga{\~n}a} {et~al.}(2015){Maga{\~n}a}, {Motta}, {C{\'a}rdenas},
  {Verdugo}, \& {Jullo}}]{2015ApJ...813...69M}
{Maga{\~n}a}, J., {Motta}, V., {C{\'a}rdenas}, V.~H., {Verdugo}, T., \&
  {Jullo}, E. 2015, \apj, 813, 69

\bibitem[{{Maughan} {et~al.}(2008){Maughan}, {Jones}, {Forman}, \& {Van
  Speybroeck}}]{2008ApJS..174..117M}
{Maughan}, B.~J., {Jones}, C., {Forman}, W., \& {Van Speybroeck}, L. 2008,
  \apjs, 174, 117

\bibitem[{{McCully} {et~al.}(2014){McCully}, {Keeton}, {Wong}, \&
  {Zabludoff}}]{2014MNRAS.443.3631M}
{McCully}, C., {Keeton}, C.~R., {Wong}, K.~C., \& {Zabludoff}, A.~I. 2014,
  \mnras, 443, 3631

\bibitem[{{Melchior} {et~al.}(2015){Melchior}, {Suchyta}, {Huff}, {Hirsch},
  {Kacprzak}, {Rykoff}, {Gruen}, {Armstrong}, {Bacon}, {Bechtol}, {Bernstein},
  {Bridle}, {Clampitt}, {Honscheid}, {Jain}, {Jouvel}, {Krause}, {Lin},
  {MacCrann}, {Patton}, {Plazas}, {Rowe}, {Vikram}, {Wilcox}, {Young}, {Zuntz},
  {Abbott}, {Abdalla}, {Allam}, {Banerji}, {Bernstein}, {Bernstein}, {Bertin},
  {Buckley-Geer}, {Burke}, {Castander}, {da Costa}, {Cunha}, {Depoy}, {Desai},
  {Diehl}, {Doel}, {Estrada}, {Evrard}, {Neto}, {Fernandez}, {Finley},
  {Flaugher}, {Frieman}, {Gaztanaga}, {Gerdes}, {Gruendl}, {Gutierrez},
  {Jarvis}, {Karliner}, {Kent}, {Kuehn}, {Kuropatkin}, {Lahav}, {Maia},
  {Makler}, {Marriner}, {Marshall}, {Merritt}, {Miller}, {Miquel}, {Mohr},
  {Neilsen}, {Nichol}, {Nord}, {Reil}, {Roe}, {Roodman}, {Sako}, {Sanchez},
  {Santiago}, {Schindler}, {Schubnell}, {Sevilla-Noarbe}, {Sheldon}, {Smith},
  {Soares-Santos}, {Swanson}, {Sypniewski}, {Tarle}, {Thaler}, {Thomas},
  {Tucker}, {Walker}, {Wechsler}, {Weller}, \& {Wester}}]{2015MNRAS.449.2219M}
{Melchior}, P., {Suchyta}, E., {Huff}, E., {et~al.} 2015, \mnras, 449, 2219

\bibitem[{{Merten} {et~al.}(2015){Merten}, {Meneghetti}, {Postman}, {Umetsu},
  {Zitrin}, {Medezinski}, {Nonino}, {Koekemoer}, {Melchior}, {Gruen},
  {Moustakas}, {Bartelmann}, {Host}, {Donahue}, {Coe}, {Molino}, {Jouvel},
  {Monna}, {Seitz}, {Czakon}, {Lemze}, {Sayers}, {Balestra}, {Rosati},
  {Ben{\'{\i}}tez}, {Biviano}, {Bouwens}, {Bradley}, {Broadhurst}, {Carrasco},
  {Ford}, {Grillo}, {Infante}, {Kelson}, {Lahav}, {Massey}, {Moustakas},
  {Rasia}, {Rhodes}, {Vega}, \& {Zheng}}]{2015ApJ...806....4M}
{Merten}, J., {Meneghetti}, M., {Postman}, M., {et~al.} 2015, \apj, 806, 4

\bibitem[{{Monna} {et~al.}(2014){Monna}, {Seitz}, {Greisel}, {Eichner},
  {Drory}, {Postman}, {Zitrin}, {Coe}, {Halkola}, {Suyu}, {Grillo}, {Rosati},
  {Lemze}, {Balestra}, {Snigula}, {Bradley}, {Umetsu}, {Koekemoer}, {Kuchner},
  {Moustakas}, {Bartelmann}, {Ben{\'{\i}}tez}, {Bouwens}, {Broadhurst},
  {Donahue}, {Ford}, {Host}, {Infante}, {Jimenez-Teja}, {Jouvel}, {Kelson},
  {Lahav}, {Medezinski}, {Melchior}, {Meneghetti}, {Merten}, {Molino},
  {Moustakas}, {Nonino}, \& {Zheng}}]{2014MNRAS.438.1417M}
{Monna}, A., {Seitz}, S., {Greisel}, N., {et~al.} 2014, \mnras, 438, 1417

\bibitem[{{Monna} {et~al.}(2015){Monna}, {Seitz}, {Zitrin}, {Geller}, {Grillo},
  {Mercurio}, {Greisel}, {Halkola}, {Suyu}, {Postman}, {Rosati}, {Balestra},
  {Biviano}, {Coe}, {Fabricant}, {Hwang}, \& {Koekemoer}}]{2015MNRAS.447.1224M}
{Monna}, A., {Seitz}, S., {Zitrin}, A., {et~al.} 2015, \mnras, 447, 1224

\bibitem[{{Navarro} {et~al.}(1996){Navarro}, {Frenk}, \&
  {White}}]{1996ApJ...462..563N}
{Navarro}, J.~F., {Frenk}, C.~S., \& {White}, S.~D.~M. 1996, \apj, 462, 563

\bibitem[{{Navarro} {et~al.}(1997){Navarro}, {Frenk}, \&
  {White}}]{1997ApJ...490..493N}
{Navarro}, J.~F., {Frenk}, C.~S., \& {White}, S.~D.~M. 1997, \apj, 490, 493

\bibitem[{{Nilsson} {et~al.}(2006){Nilsson}, {Fynbo}, {M{\o}ller},
  {Sommer-Larsen}, \& {Ledoux}}]{2006A&A...452L..23N}
{Nilsson}, K.~K., {Fynbo}, J.~P.~U., {M{\o}ller}, P., {Sommer-Larsen}, J., \&
  {Ledoux}, C. 2006, \aap, 452, L23

\bibitem[{{Perlmutter} {et~al.}(1999){Perlmutter}, {Aldering}, {Goldhaber},
  {Knop}, {Nugent}, {Castro}, {Deustua}, {Fabbro}, {Goobar}, {Groom}, {Hook},
  {Kim}, {Kim}, {Lee}, {Nunes}, {Pain}, {Pennypacker}, {Quimby}, {Lidman},
  {Ellis}, {Irwin}, {McMahon}, {Ruiz-Lapuente}, {Walton}, {Schaefer}, {Boyle},
  {Filippenko}, {Matheson}, {Fruchter}, {Panagia}, {Newberg}, {Couch}, \&
  {Project}}]{1999ApJ...517..565P}
{Perlmutter}, S., {Aldering}, G., {Goldhaber}, G., {et~al.} 1999, \apj, 517,
  565

\bibitem[{{Planck Collaboration} {et~al.}(2014){Planck Collaboration}, {Ade},
  {Aghanim}, {Armitage-Caplan}, {Arnaud}, {Ashdown}, {Atrio-Barandela},
  {Aumont}, {Baccigalupi}, {Banday}, \& et~al.}]{2014A&A...571A..16P}
{Planck Collaboration}, {Ade}, P.~A.~R., {Aghanim}, N., {et~al.} 2014, \aap,
  571, A16

\bibitem[{{Planck Collaboration} {et~al.}(2015){Planck Collaboration}, {Ade},
  {Aghanim}, {Arnaud}, {Ashdown}, {Aumont}, {Baccigalupi}, {Banday},
  {Barreiro}, {Bartlett}, \& et~al.}]{2015arXiv150201589P}
{Planck Collaboration}, {Ade}, P.~A.~R., {Aghanim}, N., {et~al.} 2015, ArXiv
  e-prints [\eprint[arXiv]{1502.01589}]

\bibitem[{{Postman} {et~al.}(2012{\natexlab{a}}){Postman}, {Coe},
  {Ben{\'{\i}}tez}, {Bradley}, {Broadhurst}, {Donahue}, {Ford}, {Graur},
  {Graves}, {Jouvel}, {Koekemoer}, {Lemze}, {Medezinski}, {Molino},
  {Moustakas}, {Ogaz}, {Riess}, {Rodney}, {Rosati}, {Umetsu}, {Zheng},
  {Zitrin}, {Bartelmann}, {Bouwens}, {Czakon}, {Golwala}, {Host}, {Infante},
  {Jha}, {Jimenez-Teja}, {Kelson}, {Lahav}, {Lazkoz}, {Maoz}, {McCully},
  {Melchior}, {Meneghetti}, {Merten}, {Moustakas}, {Nonino}, {Patel},
  {Reg{\"o}s}, {Sayers}, {Seitz}, \& {Van der Wel}}]{2012ApJS..199...25P}
{Postman}, M., {Coe}, D., {Ben{\'{\i}}tez}, N., {et~al.} 2012{\natexlab{a}},
  \apjs, 199, 25

\bibitem[{{Postman} {et~al.}(2012{\natexlab{b}}){Postman}, {Lauer}, {Donahue},
  {Graves}, {Coe}, {Moustakas}, {Koekemoer}, {Bradley}, {Ford}, {Grillo},
  {Zitrin}, {Lemze}, {Broadhurst}, {Moustakas}, {Ascaso}, {Medezinski}, \&
  {Kelson}}]{2012ApJ...756..159P}
{Postman}, M., {Lauer}, T.~R., {Donahue}, M., {et~al.} 2012{\natexlab{b}},
  \apj, 756, 159

\bibitem[{{Richard} {et~al.}(2014){Richard}, {Jauzac}, {Limousin}, {Jullo},
  {Cl{\'e}ment}, {Ebeling}, {Kneib}, {Atek}, {Natarajan}, {Egami}, {Livermore},
  \& {Bower}}]{2014MNRAS.444..268R}
{Richard}, J., {Jauzac}, M., {Limousin}, M., {et~al.} 2014, \mnras, 444, 268

\bibitem[{{Riess} {et~al.}(1998){Riess}, {Filippenko}, {Challis},
  {Clocchiatti}, {Diercks}, {Garnavich}, {Gilliland}, {Hogan}, {Jha},
  {Kirshner}, {Leibundgut}, {Phillips}, {Reiss}, {Schmidt}, {Schommer},
  {Smith}, {Spyromilio}, {Stubbs}, {Suntzeff}, \&
  {Tonry}}]{1998AJ....116.1009R}
{Riess}, A.~G., {Filippenko}, A.~V., {Challis}, P., {et~al.} 1998, \aj, 116,
  1009

\bibitem[{{Rosati} {et~al.}(2014){Rosati}, {Balestra}, {Grillo}, {Mercurio},
  {Nonino}, {Biviano}, {Girardi}, {Vanzella}, \& {Clash-VLT
  Team}}]{2014Msngr.158...48R}
{Rosati}, P., {Balestra}, I., {Grillo}, C., {et~al.} 2014, The Messenger, 158,
  48

\bibitem[{{Rousseeuw}(1984)}]{Rousseeuw84}
{Rousseeuw}, P.~J. 1984, {Journal of the American Statistical Association}, 79,
  871

\bibitem[{{Schneider} {et~al.}(1992){Schneider}, {Ehlers}, \&
  {Falco}}]{1992grle.book.....S}
{Schneider}, P., {Ehlers}, J., \& {Falco}, E.~E. 1992, {Gravitational Lenses}

\bibitem[{{Schwab} {et~al.}(2010){Schwab}, {Bolton}, \&
  {Rappaport}}]{2010ApJ...708..750S}
{Schwab}, J., {Bolton}, A.~S., \& {Rappaport}, S.~A. 2010, \apj, 708, 750

\bibitem[{{Soucail} {et~al.}(2004){Soucail}, {Kneib}, \&
  {Golse}}]{2004A&A...417L..33S}
{Soucail}, G., {Kneib}, J.-P., \& {Golse}, G. 2004, \aap, 417, L33

\bibitem[{{Steidel} {et~al.}(2000){Steidel}, {Adelberger}, {Shapley},
  {Pettini}, {Dickinson}, \& {Giavalisco}}]{2000ApJ...532..170S}
{Steidel}, C.~C., {Adelberger}, K.~L., {Shapley}, A.~E., {et~al.} 2000, \apj,
  532, 170

\bibitem[{{Suyu} {et~al.}(2013){Suyu}, {Auger}, {Hilbert}, {Marshall}, {Tewes},
  {Treu}, {Fassnacht}, {Koopmans}, {Sluse}, {Blandford}, {Courbin}, \&
  {Meylan}}]{2013ApJ...766...70S}
{Suyu}, S.~H., {Auger}, M.~W., {Hilbert}, S., {et~al.} 2013, \apj, 766, 70

\bibitem[{{Suyu} \& {Halkola}(2010)}]{2010A&A...524A..94S}
{Suyu}, S.~H. \& {Halkola}, A. 2010, \aap, 524, A94

\bibitem[{{Suyu} {et~al.}(2012){Suyu}, {Hensel}, {McKean}, {Fassnacht}, {Treu},
  {Halkola}, {Norbury}, {Jackson}, {Schneider}, {Thompson}, {Auger},
  {Koopmans}, \& {Matthews}}]{2012ApJ...750...10S}
{Suyu}, S.~H., {Hensel}, S.~W., {McKean}, J.~P., {et~al.} 2012, \apj, 750, 10

\bibitem[{{Suyu} {et~al.}(2010){Suyu}, {Marshall}, {Auger}, {Hilbert},
  {Blandford}, {Koopmans}, {Fassnacht}, \& {Treu}}]{2010ApJ...711..201S}
{Suyu}, S.~H., {Marshall}, P.~J., {Auger}, M.~W., {et~al.} 2010, \apj, 711, 201

\bibitem[{{Tollet} {et~al.}(2015){Tollet}, {Macci{\`o}}, {Dutton}, {Stinson},
  {Wang}, {Penzo}, {Gutcke}, {Buck}, {Kang}, {Brook}, {Di Cintio}, {Keller}, \&
  {Wadsley}}]{2015arXiv150703590T}
{Tollet}, E., {Macci{\`o}}, A.~V., {Dutton}, A.~A., {et~al.} 2015, ArXiv
  e-prints [\eprint[arXiv]{1507.03590}]

\bibitem[{{Treu} {et~al.}(2015{\natexlab{a}}){Treu}, {Brammer}, {Diego},
  {Grillo}, {Kelly}, {Oguri}, {Rodney}, {Rosati}, {Sharon}, {Zitrin},
  {Balestra}, {Bradac}, {Broadhurst}, {Caminha}, {Ishigaki}, {Kawamata},
  {Johnson}, {Halkola}, {Hoag}, {Karman}, {Mercurio}, {Schmidt}, {Strolger}, \&
  {Suyu}}]{2015arXiv151005750T}
{Treu}, T., {Brammer}, G., {Diego}, J.~M., {et~al.} 2015{\natexlab{a}}, ArXiv
  e-prints [\eprint[arXiv]{1510.05750}]

\bibitem[{{Treu} {et~al.}(2015{\natexlab{b}}){Treu}, {Schmidt}, {Brammer},
  {Vulcani}, {Wang}, {Brada{\v c}}, {Dijkstra}, {Dressler}, {Fontana},
  {Gavazzi}, {Henry}, {Hoag}, {Huang}, {Jones}, {Kelly}, {Malkan}, {Mason},
  {Pentericci}, {Poggianti}, {Stiavelli}, {Trenti}, \& {von der
  Linden}}]{2015ApJ...812..114T}
{Treu}, T., {Schmidt}, K.~B., {Brammer}, G.~B., {et~al.} 2015{\natexlab{b}},
  \apj, 812, 114

\bibitem[{{Umetsu} {et~al.}(2014){Umetsu}, {Medezinski}, {Nonino}, {Merten},
  {Postman}, {Meneghetti}, {Donahue}, {Czakon}, {Molino}, {Seitz}, {Gruen},
  {Lemze}, {Balestra}, {Ben{\'{\i}}tez}, {Biviano}, {Broadhurst}, {Ford},
  {Grillo}, {Koekemoer}, {Melchior}, {Mercurio}, {Moustakas}, {Rosati}, \&
  {Zitrin}}]{2014ApJ...795..163U}
{Umetsu}, K., {Medezinski}, E., {Nonino}, M., {et~al.} 2014, \apj, 795, 163

\bibitem[{{Zitrin} {et~al.}(2015){Zitrin}, {Fabris}, {Merten}, {Melchior},
  {Meneghetti}, {Koekemoer}, {Coe}, {Maturi}, {Bartelmann}, {Postman},
  {Umetsu}, {Seidel}, {Sendra}, {Broadhurst}, {Balestra}, {Biviano}, {Grillo},
  {Mercurio}, {Nonino}, {Rosati}, {Bradley}, {Carrasco}, {Donahue}, {Ford},
  {Frye}, \& {Moustakas}}]{2015ApJ...801...44Z}
{Zitrin}, A., {Fabris}, A., {Merten}, J., {et~al.} 2015, \apj, 801, 44

\end{thebibliography}

\end{document}